\documentclass[twocolumn]{aastex63}

\received{\today}
\revised{\today}
\accepted{\today}
\submitjournal{ApJL}

\usepackage{amsmath}	
\usepackage{amssymb}
\usepackage[abs]{overpic}
\usepackage{xcolor,varwidth}

\shorttitle{Black hole flares}
\shortauthors{Ripperda, Liska, Chatterjee, et al.}

\graphicspath{{./}{figures/}}

\begin{document}

\title{Black hole flares: ejection of accreted magnetic flux through 3D plasmoid-mediated reconnection}

\correspondingauthor{B. Ripperda \& M. Liska}
\email{bripperda@flatironinstitute.org, \\ matthew.liska@cfa.harvard.edu}

\author[0000-0002-7301-3908]{B. Ripperda}\altaffiliation{Joint Princeton/Flatiron Postdoctoral Fellow}\affiliation{Center for Computational Astrophysics, Flatiron Institute, 162 Fifth Avenue, New York, NY 10010, USA}\affiliation{Department of Astrophysical Sciences, Peyton Hall, Princeton University, Princeton, NJ 08544, USA}
\author{M. Liska}\altaffiliation{Both first authors have contributed equally to this work.}\affiliation{Institute for Theory and Computation, Harvard University, 60 Garden Street, Cambridge, MA 02138, USA; John Harvard Distinguished Science and ITC
Fellow}
\author[0000-0002-2825-3590]{K. Chatterjee}\affiliation{Black Hole Initiative at Harvard University, 20 Garden Street, Cambridge, MA 02138, USA}\affiliation{Anton Pannekoek Institute for Astronomy, University of Amsterdam, Science Park 904, 1098 XH Amsterdam, The Netherlands}
\author[0000-0003-1984-189X]{G. Musoke}\affiliation{Anton Pannekoek Institute for Astronomy, University of Amsterdam, Science Park 904, 1098 XH Amsterdam, The Netherlands}
\author[0000-0001-7801-0362]{A. A. Philippov}\affiliation{Center for Computational Astrophysics, Flatiron Institute, 162 Fifth Avenue, New York, NY 10010, USA}
\author[0000-0001-9564-0876]{S.B. Markoff}\affiliation{Anton Pannekoek Institute for Astronomy, University of Amsterdam, Science Park 904, 1098 XH Amsterdam, The Netherlands}\affiliation{Gravitation \& AstroParticle Physics Amsterdam (GRAPPA), University of Amsterdam, Science Park 904, 1098 XH Amsterdam, The Netherlands}
\author[0000-0002-9182-2047]{A. Tchekhovskoy}\affiliation{Center for Interdisciplinary Exploration \& Research in Astrophysics (CIERA), Physics \& Astronomy, Northwestern University, Evanston, IL 60202, USA}
\author[0000-0001-9283-1191]{Z. Younsi}\affiliation{Mullard Space Science Laboratory, University College London, Holmbury St.~Mary, Dorking, Surrey, RH5 6NT, United Kingdom}

\begin{abstract}
Magnetic reconnection can power bright, rapid flares originating from the inner magnetosphere of accreting black holes. We conduct extremely high resolution {($5376\times2304\times2304$ cells)} general-relativistic magnetohydrodynamics simulations, capturing plasmoid-mediated reconnection in a 3D magnetically arrested disk for the first time. We show that an equatorial, plasmoid-unstable current sheet forms in a transient, non-axisymmetric, low-density magnetosphere within the inner few Schwarzschild radii. Magnetic flux bundles escape from the event horizon through reconnection {at the universal plasmoid-mediated rate} in this current sheet. The reconnection feeds on the highly-magnetized plasma in the jets{ and heats} the plasma that ends up trapped in flux bundles to temperatures proportional to the jet's magnetization. The escaped flux bundles can complete a full orbit as low-density hot spots, consistent with Sgr A$^{*}$ observations by the GRAVITY interferometer. Reconnection near the horizon produces sufficiently energetic plasma to explain flares from accreting black holes, such as the TeV emission observed from M87.
The drop in mass accretion rate during the flare, and the resulting low-density magnetosphere make it easier for very high energy photons produced by reconnection-accelerated particles to escape.
The extreme resolution results in a converged plasmoid-mediated reconnection rate that directly determines the timescales and properties of the flare.


\end{abstract}


\section{\label{sec:level1}Introduction}

Bright flaring from accreting black holes is seen at all wavelengths, but the mechanism powering high-energy flares is still a topic of major debate. Rapid $\gamma$-ray flares have been observed from active galactic nuclei, in the form of very high-energy ($> 100$ GeV) emission (\citealt{Albert2007,Aharonian2007,Aharonian2009,Aleksic2014S}). The variability timescale of the flares can be shorter than the light-crossing time of the event horizon, constraining the emitting region to be of the order of a Schwarzschild radius. Bright TeV flares are also periodically observed from the supermassive black hole M87$^{*}$, in the center of the Messier 87 galaxy (\citealt{Hess2006,Veritas2010,veritas2012,Magic2021}). The flares show a flux rise and decay timescale of 1-3 days, emitting $\gtrsim 10^{41}$ erg/s (\citealt{Hess2012}), which is non-negligible compared to the total jet power of $10^{42}-10^{44}$ erg/s (e.g., \citealt{Prieto2016}). High-energy $\gamma$-rays originating nearby the horizon can be absorbed by background photons to create electron-positron pairs, preventing their escape. Therefore, it is unclear if there is a mechanism that can produce such flares near the horizon and under which conditions the {radiation} can freely escape.
Furthermore, the black hole in the Galactic Center, Sgr A$^{*}$, shows intriguing infrared and X-ray flares on similarly short dynamical timescales (\citealt{baganoff2001,Eckart2004,Neilsen2015}) originating from near the horizon (\citealt{Gravity2018,Gravity2021}).

Magnetically arrested disk (MAD, \citealt{1974Ap&SS..28...45B,1976Ap&SS..42..401B,narayan2003}) accretion is the most plausible scenario for the accretion flow onto active galactic nuclei showing strong jets (see, e.g., \citealt{EHTVII2021} for M87$^{*}$). Sources fed by stellar winds, like Sgr A$^{*}$, are also capable of producing MADs (\citealt{Ressler_2020}). {General-relativistic magnetohydrodynamics (GRMHD) simulations show that a large amount of poloidal (pointing in the $R$- and $z$-directions) magnetic flux (proportional to the square root of the mass accretion rate) is forced into the black hole by the accreting gas, until the flux becomes dynamically important and strong enough to push the accreting gas away (\citealt{2003ApJ...592.1042I,2008ApJ...677..317I,Tchekhovskoy2011}).} The MAD state is accompanied by large-amplitude fluctuations, caused by quasi-periodic accumulation and escape of the magnetic flux bundles in the vicinity of the black hole (\citealt{2008ApJ...677..317I,Tchekhovskoy2011,dexter2020sgr,Porth2020flares}). 

{Recently, extreme resolution two-dimensional (2D) GRMHD simulations showed that escape of magnetic flux bundles from the black hole, resulting in the decay of magnetic flux on the horizon, occurs through {plasmoid-mediated} reconnection (\citealt{Ripperda2020}{, hereafter RBP20}). The magnetic flux decay is accompanied by the ejection of the accretion disk (\citealt{Proga2003}). The ejection results in the formation of a magnetosphere, consisting of an equatorial plasmoid-unstable} current sheet of oppositely directed magnetic field, that separates two highly magnetized jet regions. {Reconnection in the current sheet releases energy that can power a flare and the tension of the reconnected flux can push gas away and suppress the mass accretion rate.} The jets, which supply matter in the current sheet, are highly magnetized because their large-scale magnetic field serves as a barrier to ions within the accretion disk. Pair discharges can generate ample electron-positron plasma to fill the magnetospheric region \citep[e.g.,][]{Crinquand2020}. The collisional mean free path of particles is much larger than the characteristic length scale of the system. As a result, the magnetospheric electron-positron plasma is collisionless, and can be accelerated in a reconnecting current sheet into a power-law distribution, and subsequently power high-energy flares. In magnetized and collisionless plasma conditions, reconnection occurs in the {plasmoid-mediated} regime {at a universal reconnection rate of $v_{\rm rec} / v_{\rm A} \sim 0.1$, where $v_{\rm rec}$ is the inflow velocity into a current sheet, and $v_{\rm A} \sim c$ is the Alfv\'{e}n speed (\citealt{sironi2014,Guo_2014,Werner_2015}).}

{In {collisional systems as described by GRMHD, the reconnection rate in the plasmoid-mediated regime at high Lundquist numbers (and at sufficiently high resolution to resolve the spatial scales associated to that Lundquist number)  converges to a universal value of $v_{\rm rec} / v_{\rm A} \sim 0.01$, becoming independent of the resistivity} (\citealt{bhattacharjee2009,uzdensky2010,ripperda2019}; RBP20) \footnote{{Note that the reconnection rate in the plasmoid-mediated regime in collisionless systems is approximately ten times faster than in collisional systems described by GRMHD (\citealt{sironi2014,Guo_2014,Werner_2015,Bransgrove2021}). At low resolutions, GRMHD simulations show higher reconnection rates, which are however a result of large numerical diffusion instead of plasmoid-mediated reconnection.}}. Resolving plasmoid-mediated reconnection, and hence a converged universal reconnection rate, in global black hole simulations requires resolutions higher than $\sim 2000$ cells in the {$\theta$}-direction to capture thin current sheets liable to the plasmoid instability (RBP20; \citealt{Bransgrove2021}).}  The flare time-scale is governed by the flux decay which is directly set by the reconnection rate (\citealt{Bransgrove2021}){, which makes it particularly important to resolve the plasmoid instability in thin current sheets}.

Our goal here is to understand if a macroscopic reconnecting current sheet can form and power a flare in 3D GRMHD simulations {with a converged universal reconnection rate, $v_{\rm rec} \sim 0.01c$, for the largest current sheets in the system}, despite the excitation of non-axisymmetric effects like a Rayleigh–Taylor-type instability (RTI) preventing the complete arrest of accretion (\citealt{Tchekhovskoy2011,Papadopoulos2019}).
{In this Letter we conduct the highest-resolution global 3D GRMHD simulations to-date to show that plasmoid-mediated magnetic reconnection in transient, non-axisymmetric current sheets can power flares from accreting black holes and that the magnetic flux decay on the black hole event horizon is governed by the universal reconnection rate.}

{Throughout the manuscript we use geometrized units with gravitational constant, black-hole mass, and speed of light $G = M = c = 1$; such that length scales are normalized to the gravitational radius $r_{\rm g} = GM/c^2$ and times are given in units of $r_{\rm g}/c$.
We employ spherical Kerr-Schild coordinates, where $r$ is the radial coordinate, $\theta$ and $\phi$ are the poloidal and toroidal angular coordinates, respectively, and $t$ is the temporal coordinate.}

\section{Numerical setup}
{Reconnecting current sheets are plasmoid-unstable for Lundquist numbers (\citealt{bhattacharjee2009,uzdensky2010})}
\begin{equation}
    S=v_{\rm A} w / \eta_{\rm num} \geq S_{\rm{crit}} = 10^4,
    \label{eq:lundquist}
\end{equation}
{assuming the Alfv\'{e}n speed $v_{\rm A} \sim c$, and the length of a current sheet $w \sim r_{\rm g}$. Here, we assume that the numerical resistivity proportional to the cell size $\eta_{\rm num} \propto \Delta x^p$, where $p \approx 2$ depending on the details of the second order accurate algorithm. Thus, the constraint on $S$ (Eq. \ref{eq:lundquist}) directly determines the required resolution.} In the plasmoid-mediated regime, the reconnection rate converges to the asymptotic $v_{\rm rec} \sim 0.01c$ in GRMHD (RBP20; \citealt{Bransgrove2021}), directly determining the (converged) rate of magnetic flux decay on the horizon. To achieve the resolution required to capture the plasmoid-mediated reconnection {and, hence, achieve long-sought convergence in the reconnection rate,} we employ our GPU-accelerated GRMHD code H-AMR (\citealt{liska2019}). We set the effective numerical resolution to $N_r \times N_\theta \times N_\phi =  5376\times2304\times2304$ (dubbed ``extreme resolution'' from here onward) to {ensure} that we capture thin plasmoid-unstable current sheets (RBP20).  To study convergence of the reconnection rate and the rate at which magnetic flux can escape {from the black hole}, we also conduct three {lower resolution} runs at $N_r \times N_\theta \times N_\phi =  2240\times1056\times1024$ (``high resolution''); $580\times288\times256$ (``standard resolution''); and $288\times128\times128$ (``low resolution''). {The resolution in the $r-$ and $\theta-$ dimensions is satisfied throughout the domain. To keep the cell aspects ratio approximately uniform in our spherical grid, we use 3 internal and 4 external derefinement levels \citep{liska2019} in $\phi$ to reduce the resolution from the full $N_{\phi}=128-2304$ at $30^{\circ}<\theta<150^{\circ}$ to $N_{\phi}=16-18$ within $0.5^{\circ}-7.5^{\circ}$ of each pole.} {In all of these runs we fix the radial domain to $[1.2, 2000] r_g$ and we use a minimum $10000 r_{\rm g}/c$ integration time. We use outflow boundary conditions in $r$, transmissive boundary conditions in $\theta$, and periodic boundary conditions in $\phi$ as described in \cite{liska2018}.}
{We initialize our simulation to obtain a prograde MAD around a Kerr black hole with dimensionless spin $a=0.9375$, starting from a torus threaded by a single weak poloidal magnetic field loop, defined by the vector potential $A_{\phi} \propto \max\left[{\rho}/{\rho_{\rm max}}\left({r}/{r_{\rm in}}\right)^3\sin^3\theta\exp\left(-{r}/{400}\right)-0.2, 0\right]$, normalized to the gas-to-magnetic-pressure ratio $\beta = 2p/b^2 = 100$.} We replenish gas density $\rho$ in low-density regions to maintain $\sigma_{\rm max}=25$ where the magnetization $\sigma=b^2/(4 \pi \rho c^2)$ is defined using the magnetic field strength $b$ co-moving with the fluid, and fluid-frame rest-mass density $\rho$. We adopt an equation of state for a relativistic ideal gas with an adiabatic index of $\hat{\gamma} = 13/9$, in between a fully relativistic gas $\hat{\gamma}=4/3$ and a fully non-relativistic gas $\hat{\gamma}=5/3$. We employ dimensionless temperature units $T=p/\rho$ with {thermal gas} pressure $p$, where $T=1$ corresponds to $k_{\rm B} T=m_{\rm i} c^2$ with ion mass $m_{\rm i}$ and Boltzmann's constant $k_{\rm B}$ such that $T>1$ indicates relativistic {ion} temperatures.

\section{Reconnection-powered flares}
\begin{figure*}
    \centering
    \includegraphics[width=0.353\textwidth,trim= 0.85cm 2.3cm 13.4cm 1.3cm, clip=true]{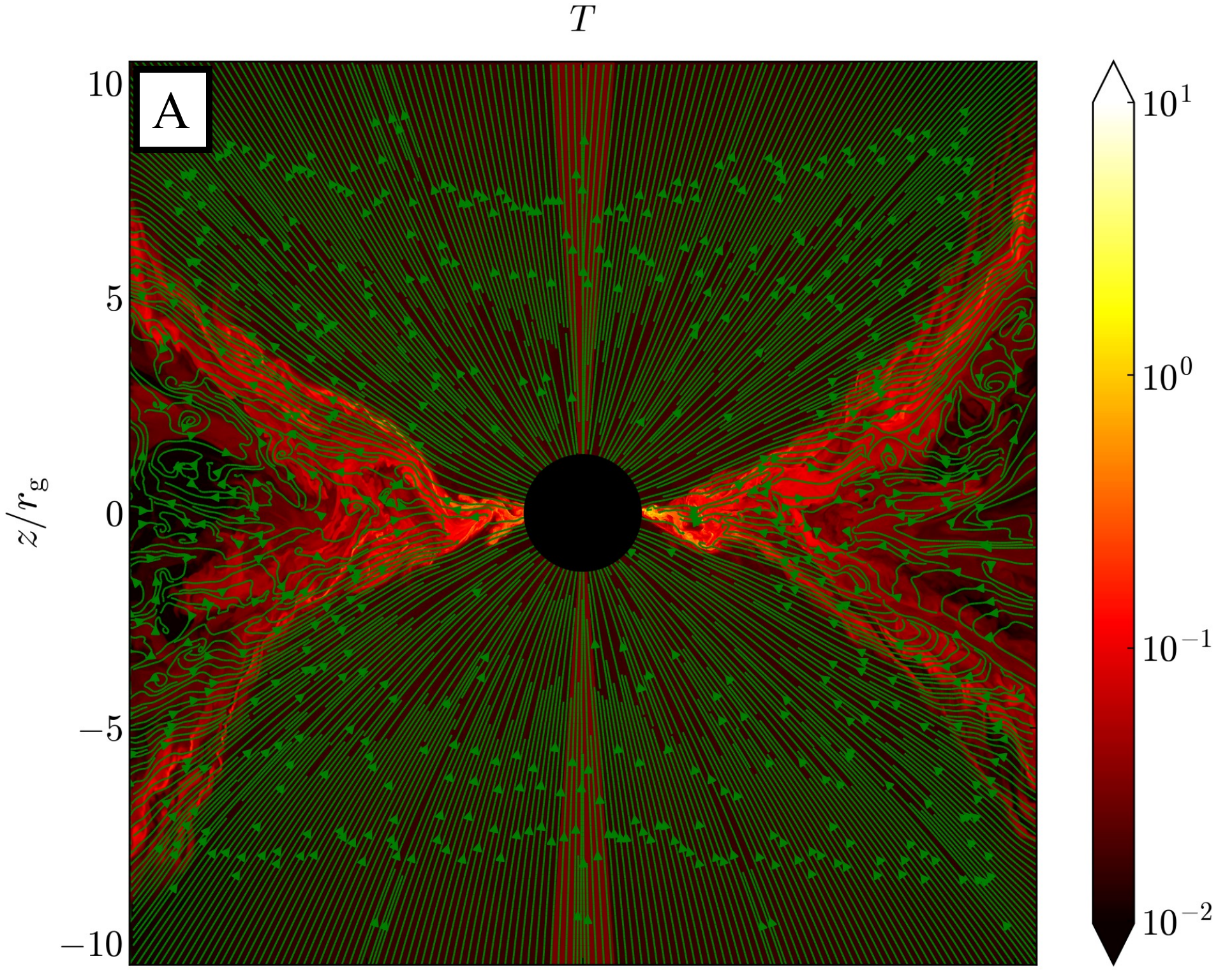}
    \includegraphics[width=0.318\textwidth,trim= 2.8cm 2.3cm 13.4cm 1.3cm, clip=true]{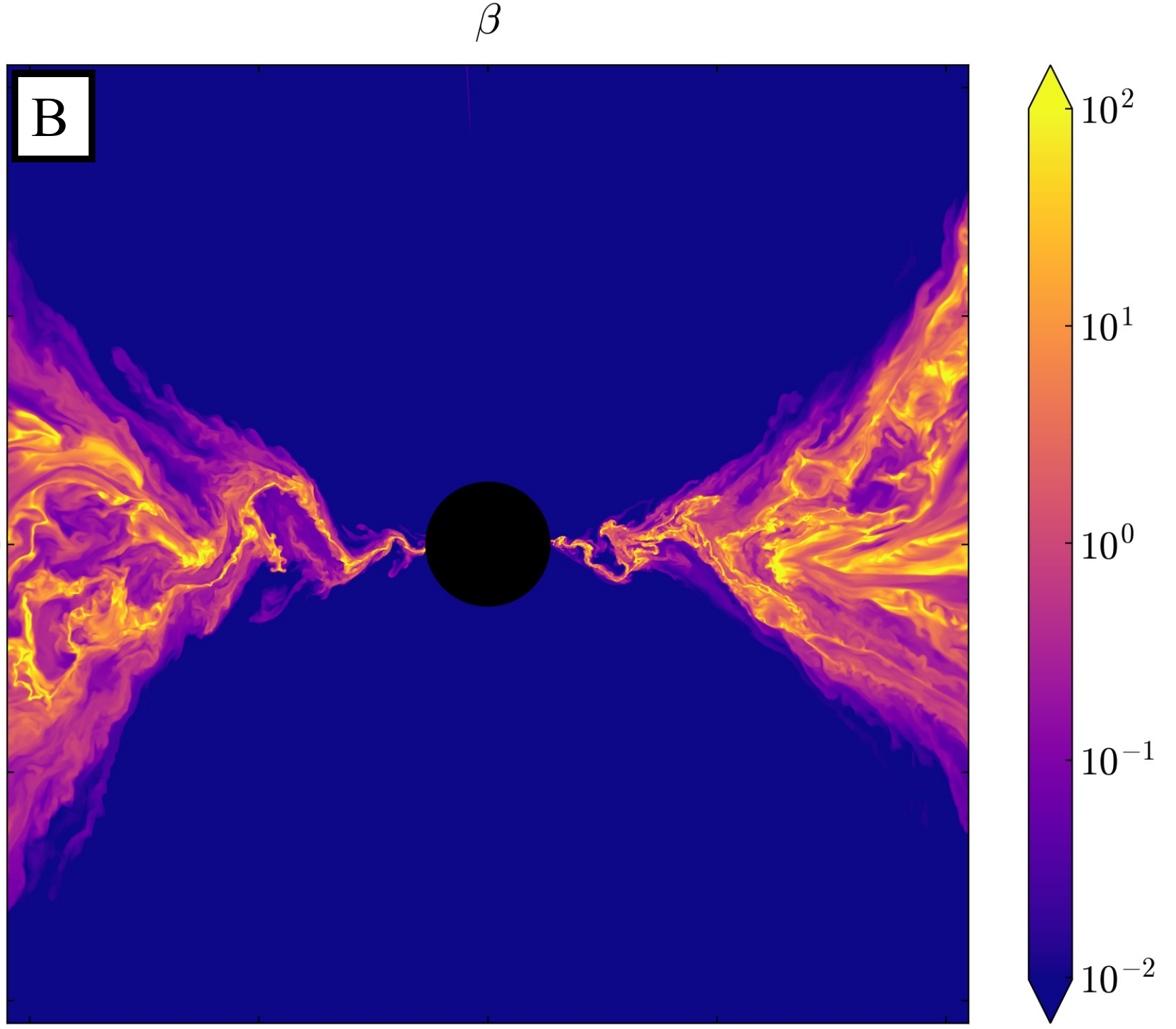}
    \includegraphics[width=0.318\textwidth,trim= 2.8cm 2.3cm 13.4cm 1.3cm, clip=true]{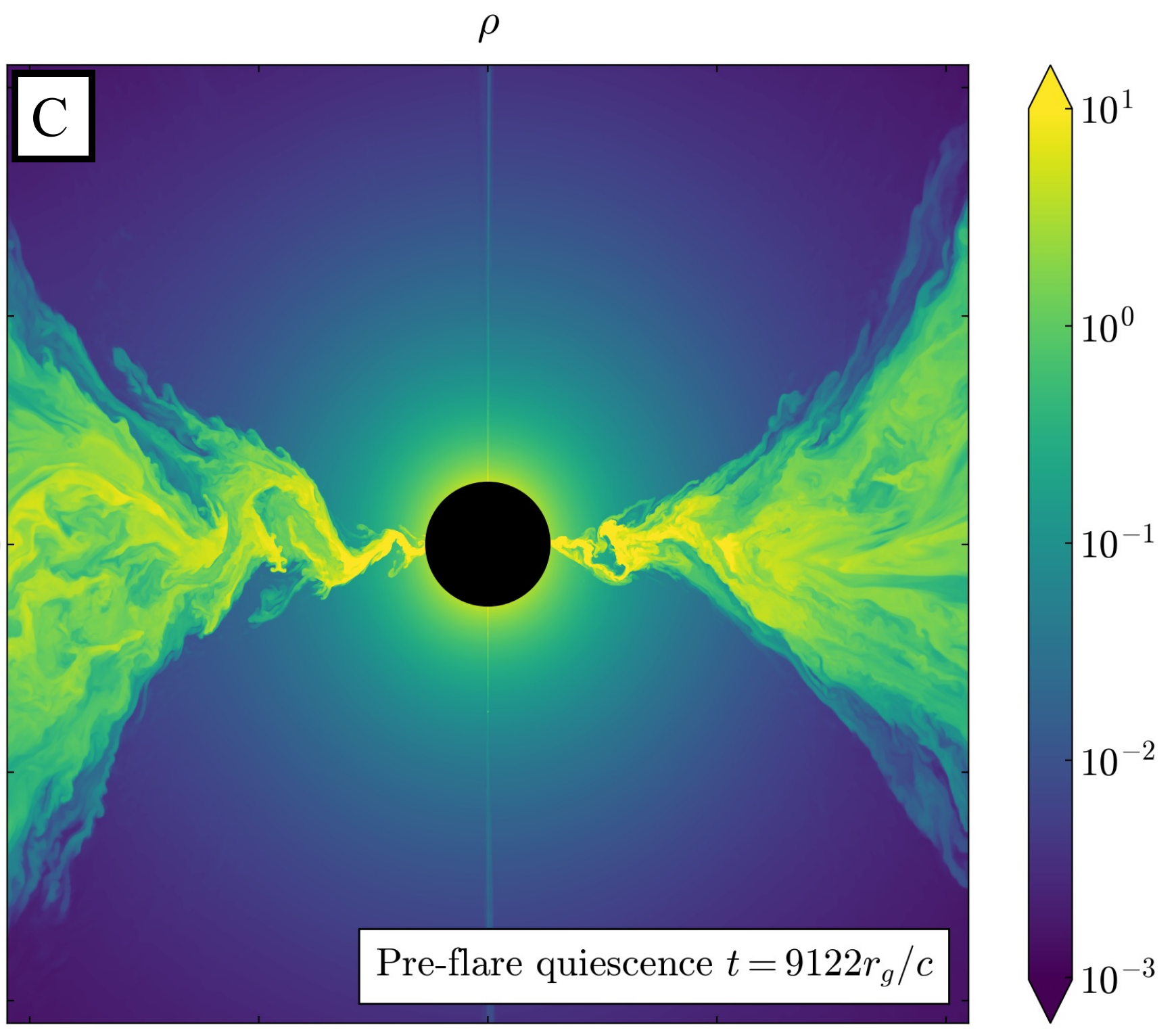}

    \includegraphics[width=0.353\textwidth,trim= 0.85cm 0.785cm 13.4cm 1.95cm, clip=true]{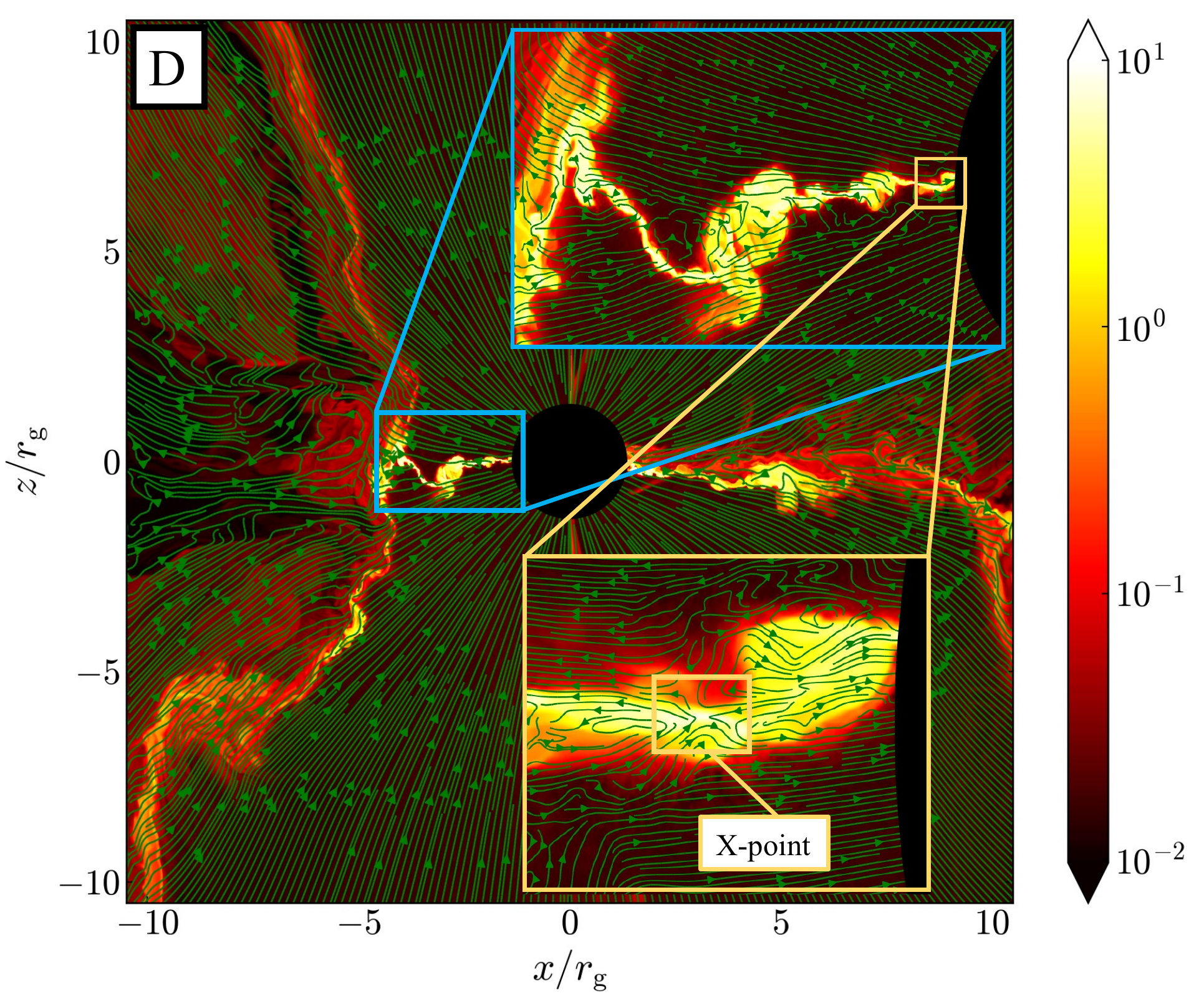}
    \includegraphics[width=0.318\textwidth,trim= 2.8cm 0.785cm 13.4cm 1.95cm, clip=true]{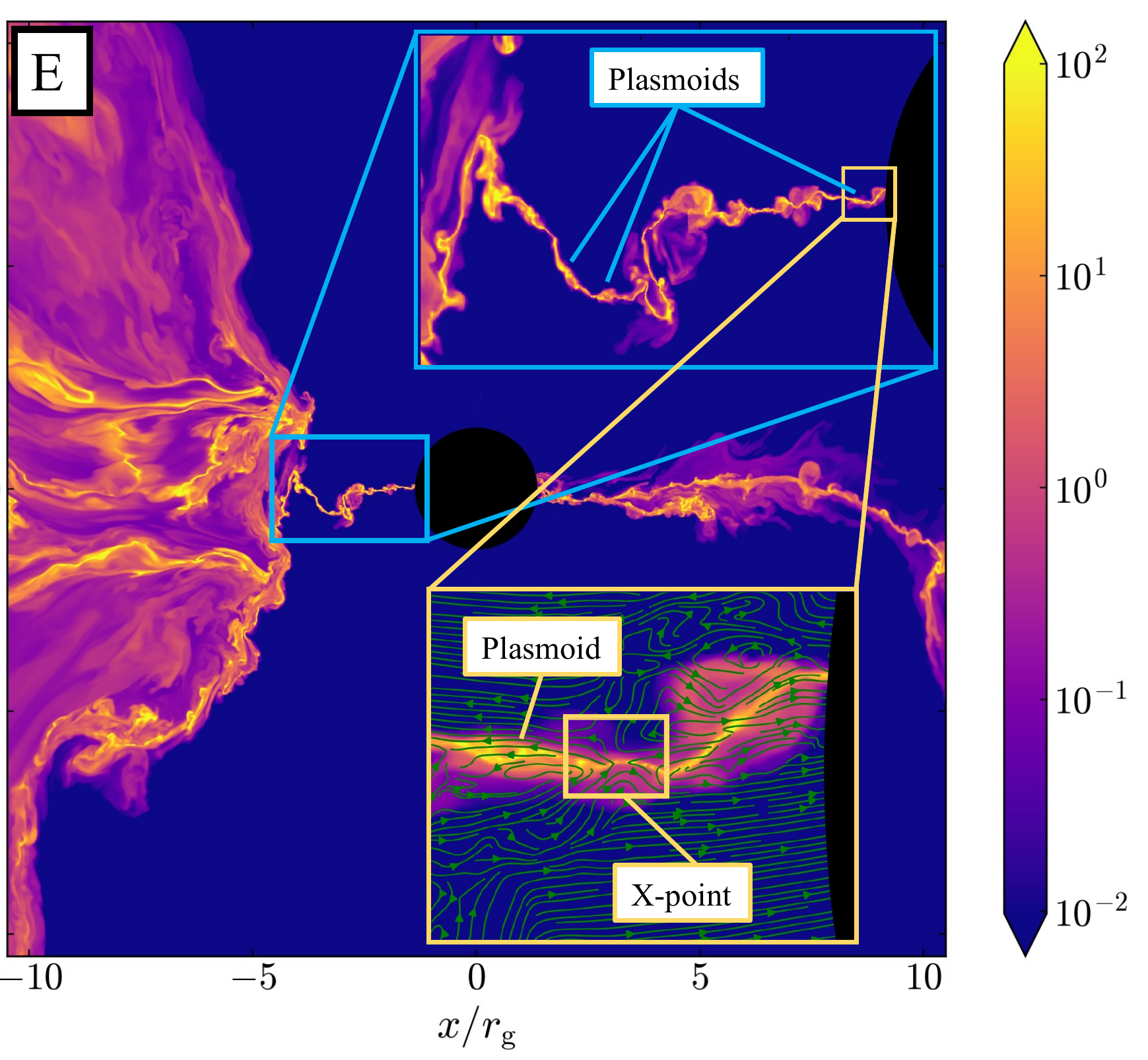}
    \includegraphics[width=0.318\textwidth,trim= 2.8cm 0.785cm 13.4cm 1.95cm, clip=true]{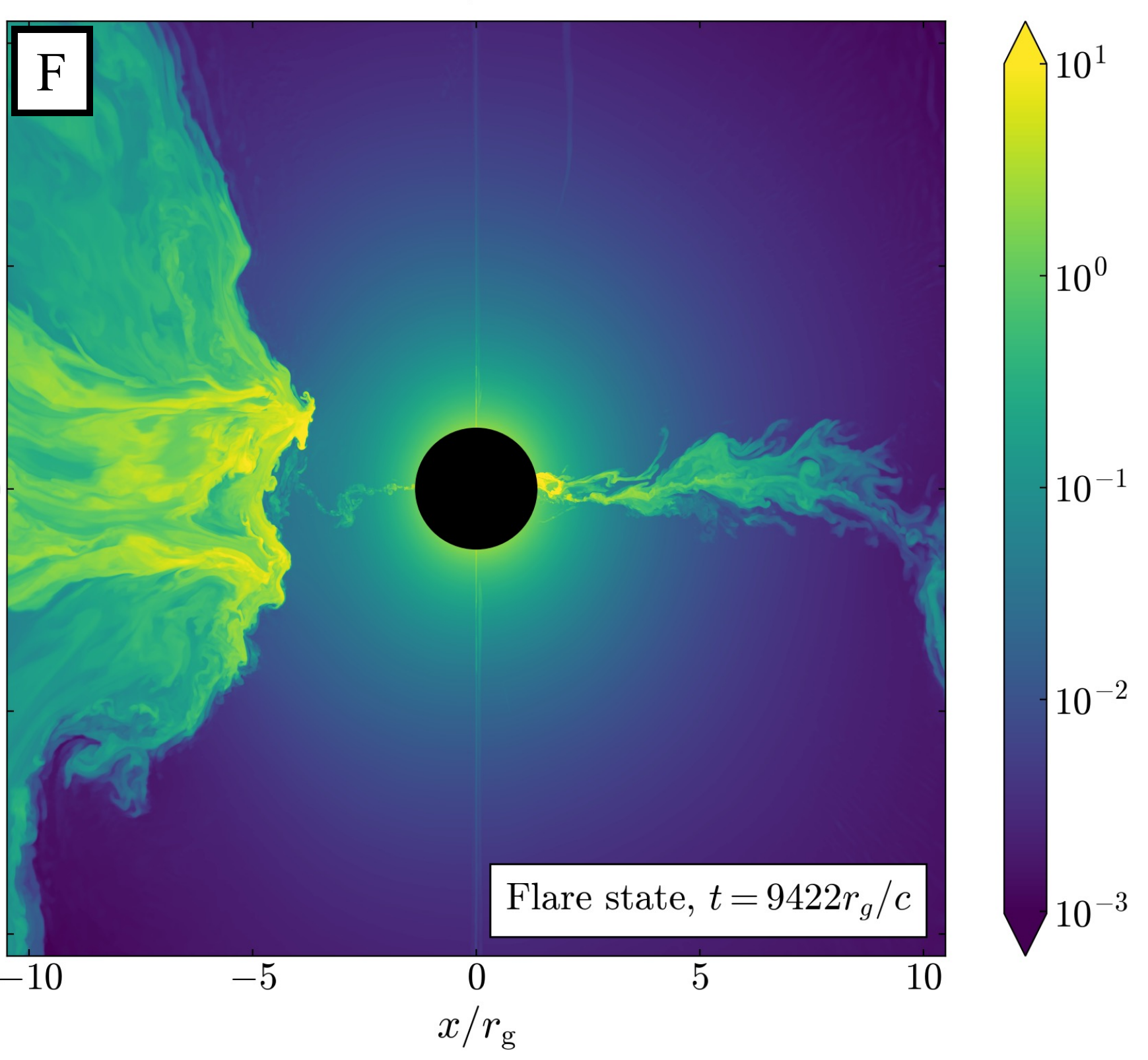}
    
    \includegraphics[width=0.353\textwidth,trim= 0.85cm 0.785cm 13.4cm 2.15cm, clip=true]{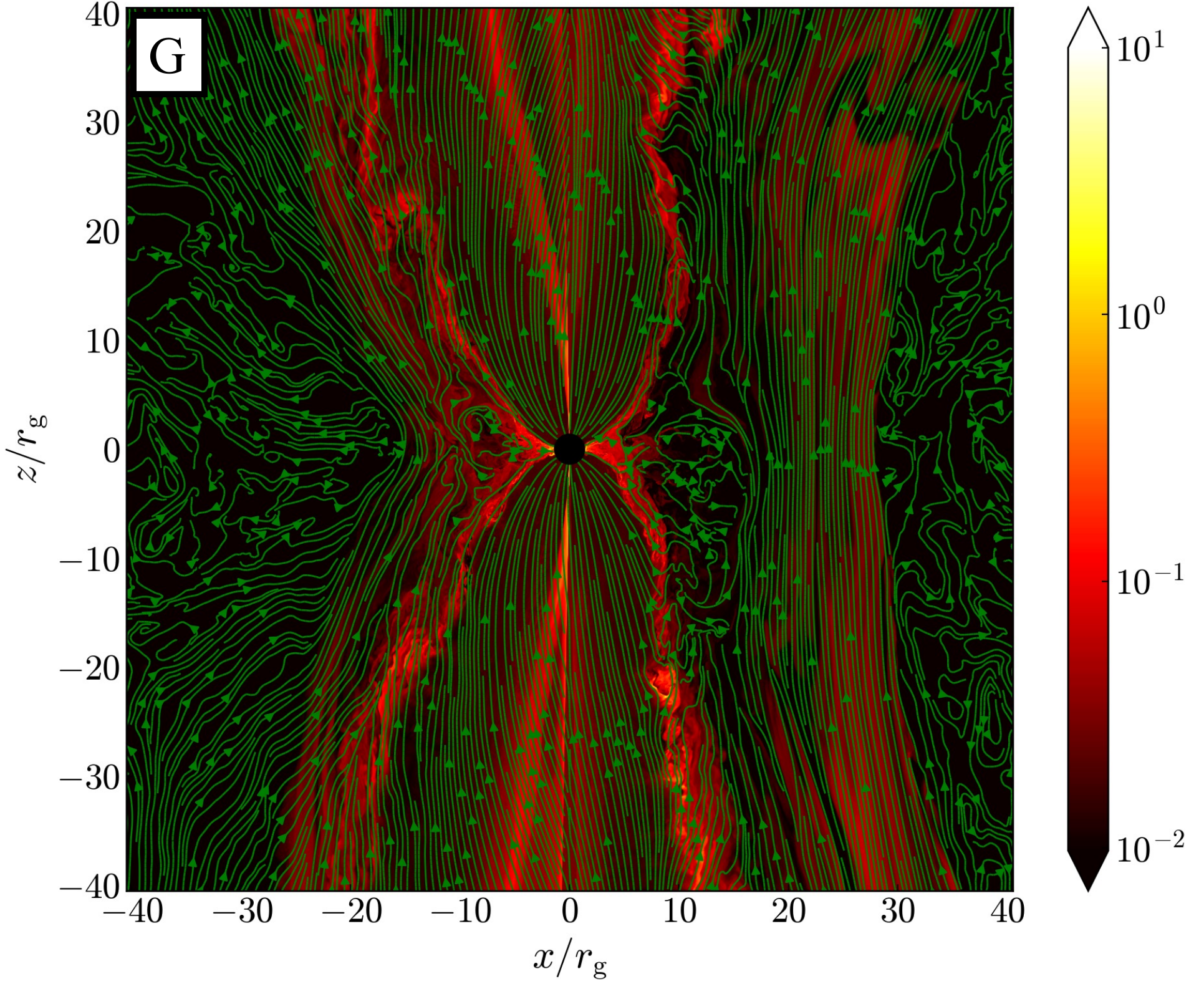}
    \includegraphics[width=0.318\textwidth,trim= 2.8cm 0.785cm 13.4cm 2.15cm, clip=true]{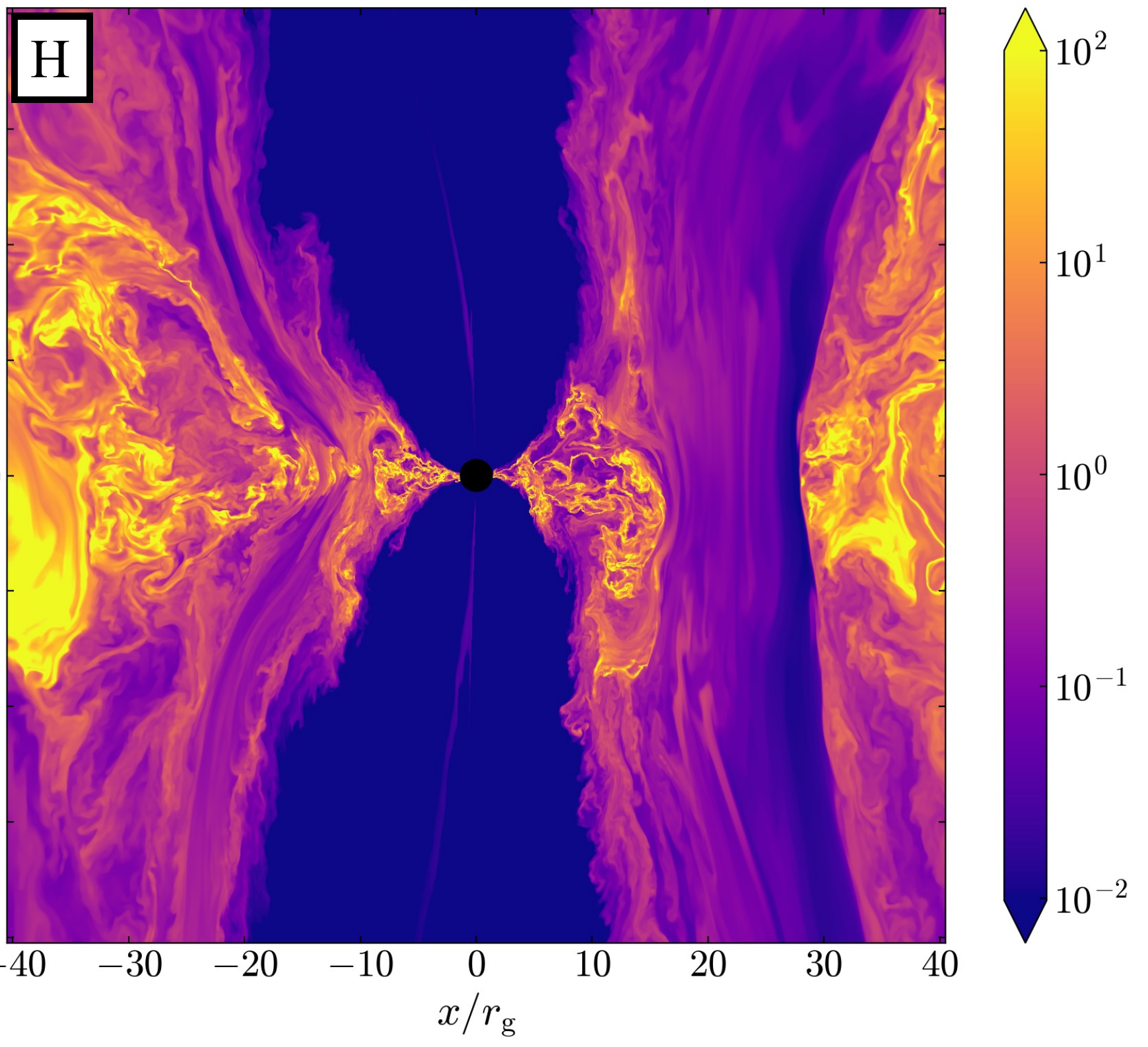}
    \includegraphics[width=0.318\textwidth,trim= 2.8cm 0.785cm 13.4cm 2.15cm, clip=true]{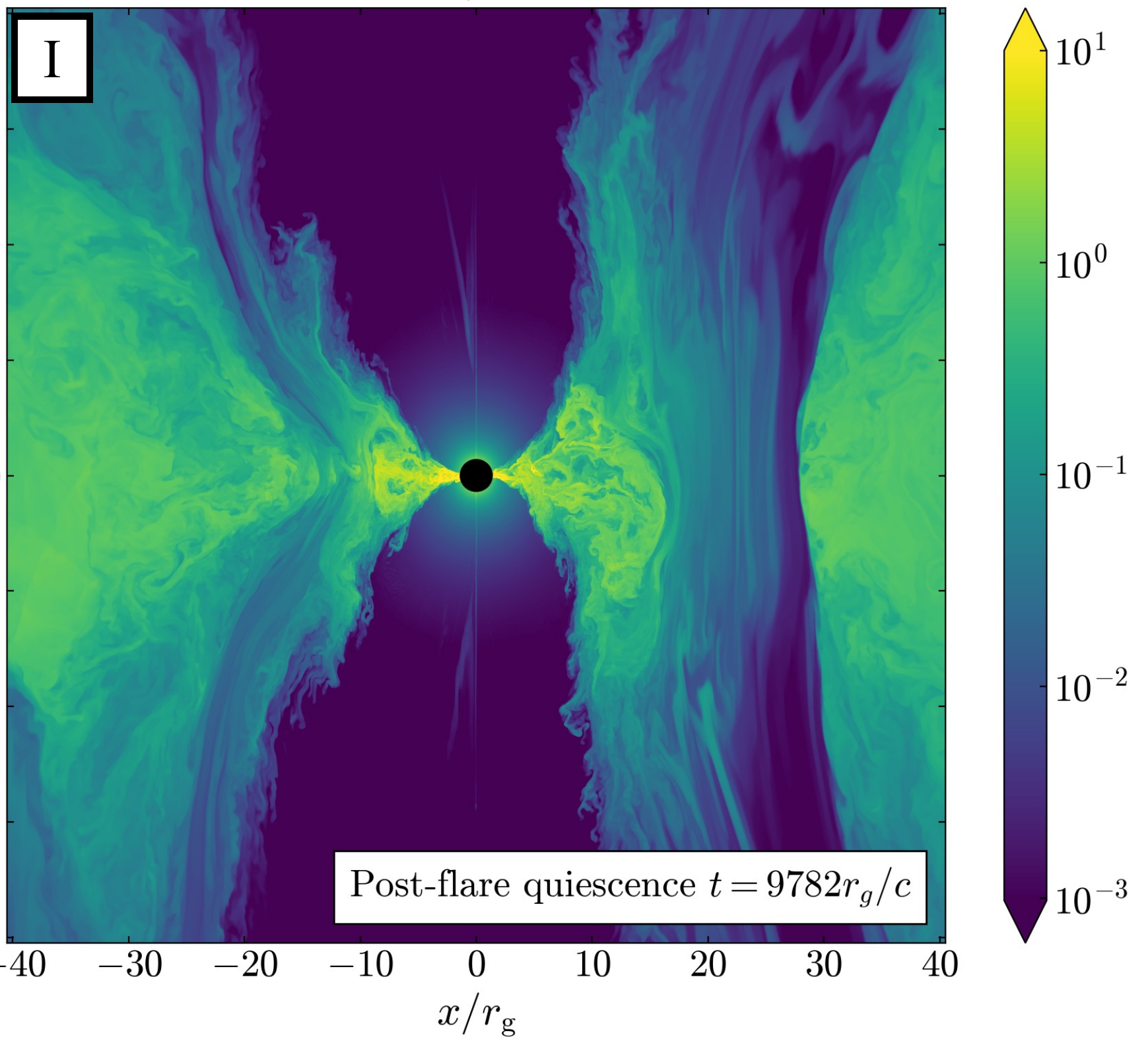}    
    \caption{{Plasmoid-mediated reconnection, which takes place at sufficiently high resolutions in MHD, is seen in a 3D GRMHD simulation for the first time. Resolving the dynamics of X-points and plasmoids in the current sheet can be the key to understanding the source of black hole non-thermal emission, e.g., high-energy flares.} 
    Dimensionless temperature $T=p/\rho$, plasma-$\beta$, and density $\rho$ (from left to right) in the {meridional plane before (top row), during (middle row) in the inner $10 r_{\rm g}$ and after (bottom row) the large magnetic flux eruption} in the inner $40 r_{\rm g}$. During the {magnetic flux eruption}, the accretion disk is ejected and the broad accretion inflow is reduced to a thin {plasmoid-unstable} current sheet, indicated by {X-points and} magnetic nulls shown by the antiparallel in-plane field lines (in green{, see inset in panel D) and the high $\beta$ (inset panel E)}. The hot ($T \sim \sigma_{\rm max}$) exhaust of the reconnection layer heats the jet sheath. Reconnection transforms the horizontal field in the current sheet to vertical field that is ejected in the form of hot coherent flux tubes (panel G) at low $\beta$ and density (panels H,I).}
    \label{fig:panelXZ}
\end{figure*}

\begin{figure*}
    \centering
    \includegraphics[width=0.353\textwidth,trim= 0.85cm 2.3cm 13.4cm 1.3cm, clip=true]{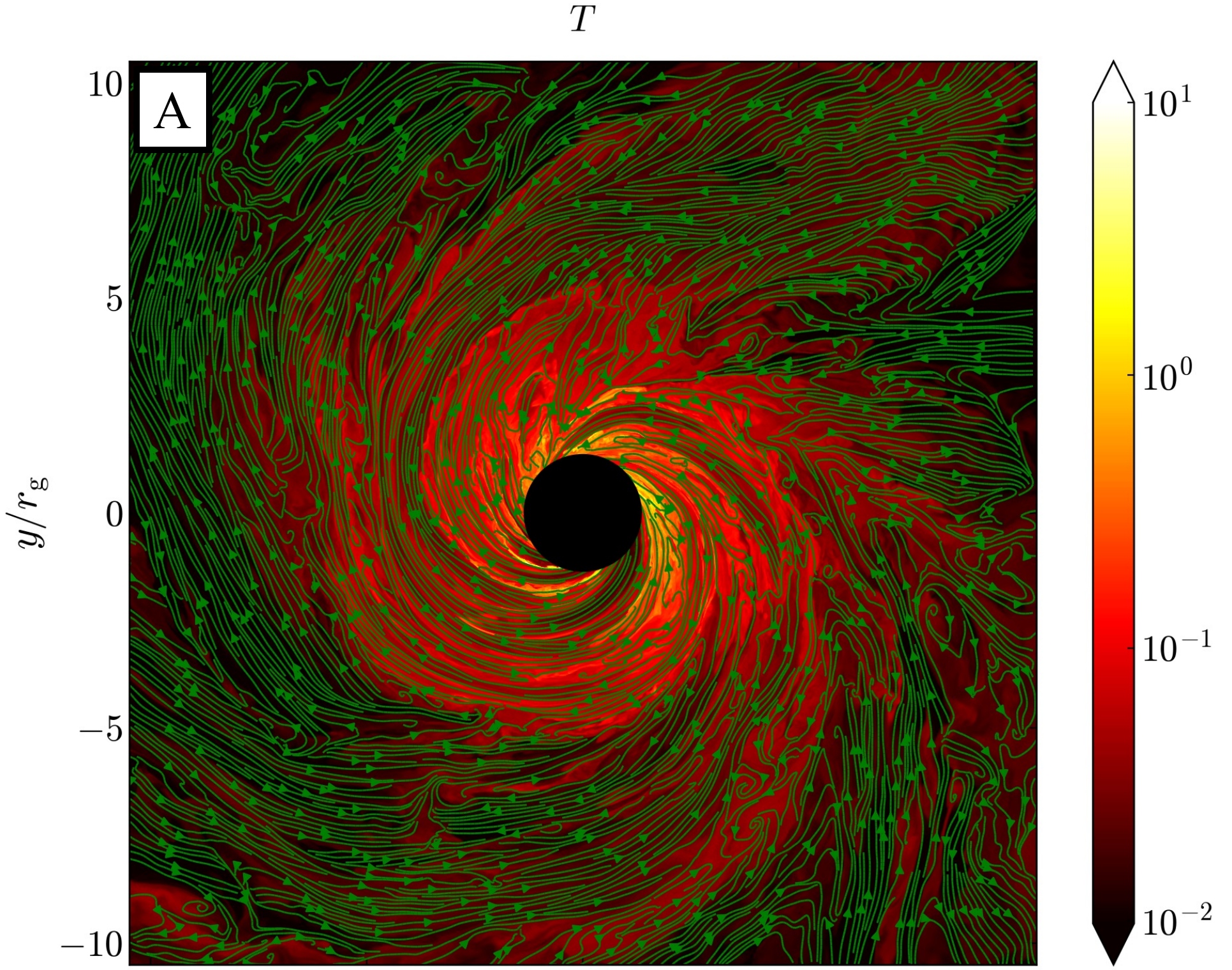}
    \includegraphics[width=0.318\textwidth,trim= 2.8cm 2.3cm 13.4cm 1.3cm, clip=true]{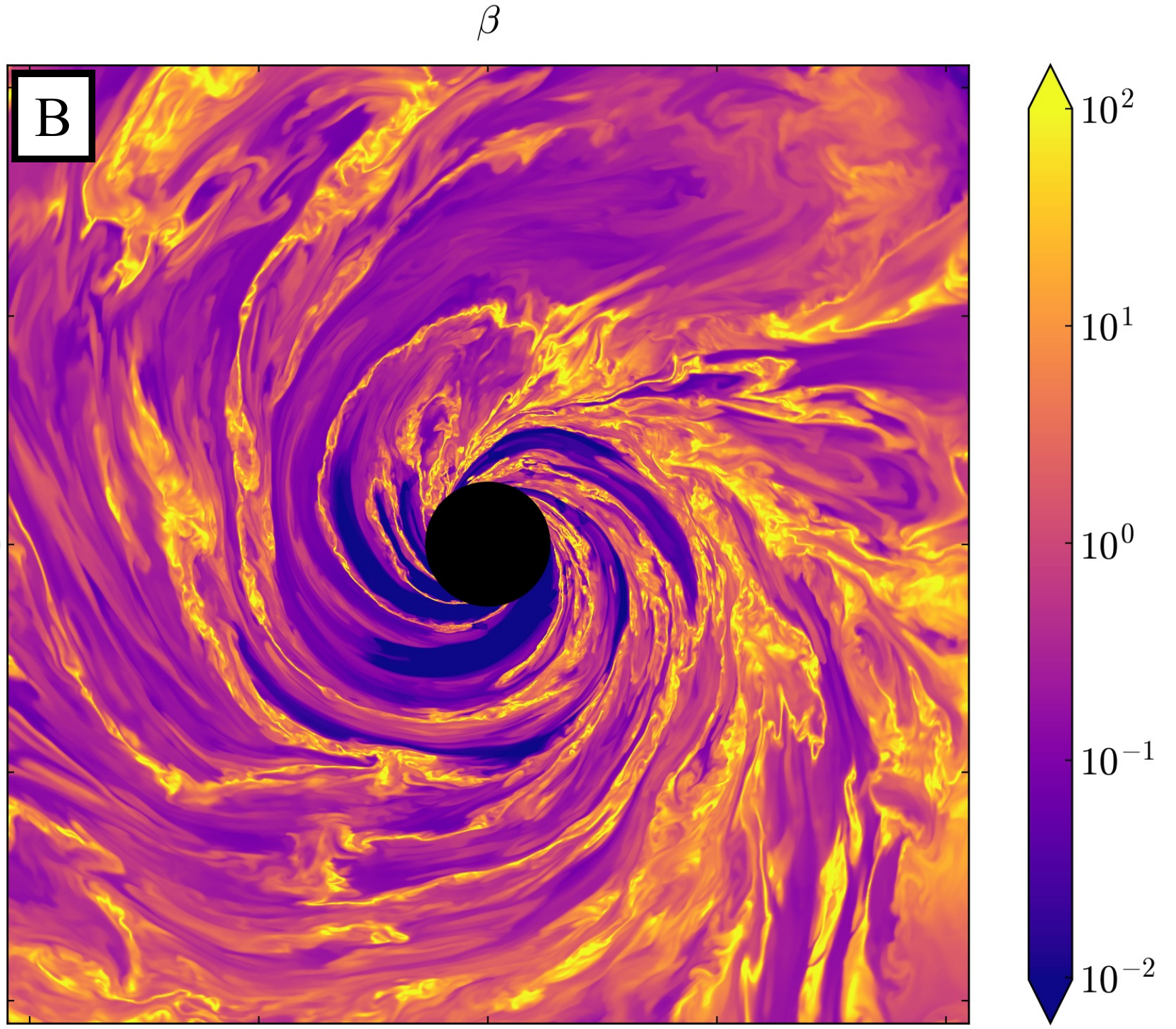}
    \includegraphics[width=0.318\textwidth,trim= 2.8cm 2.3cm 13.4cm 1.3cm, clip=true]{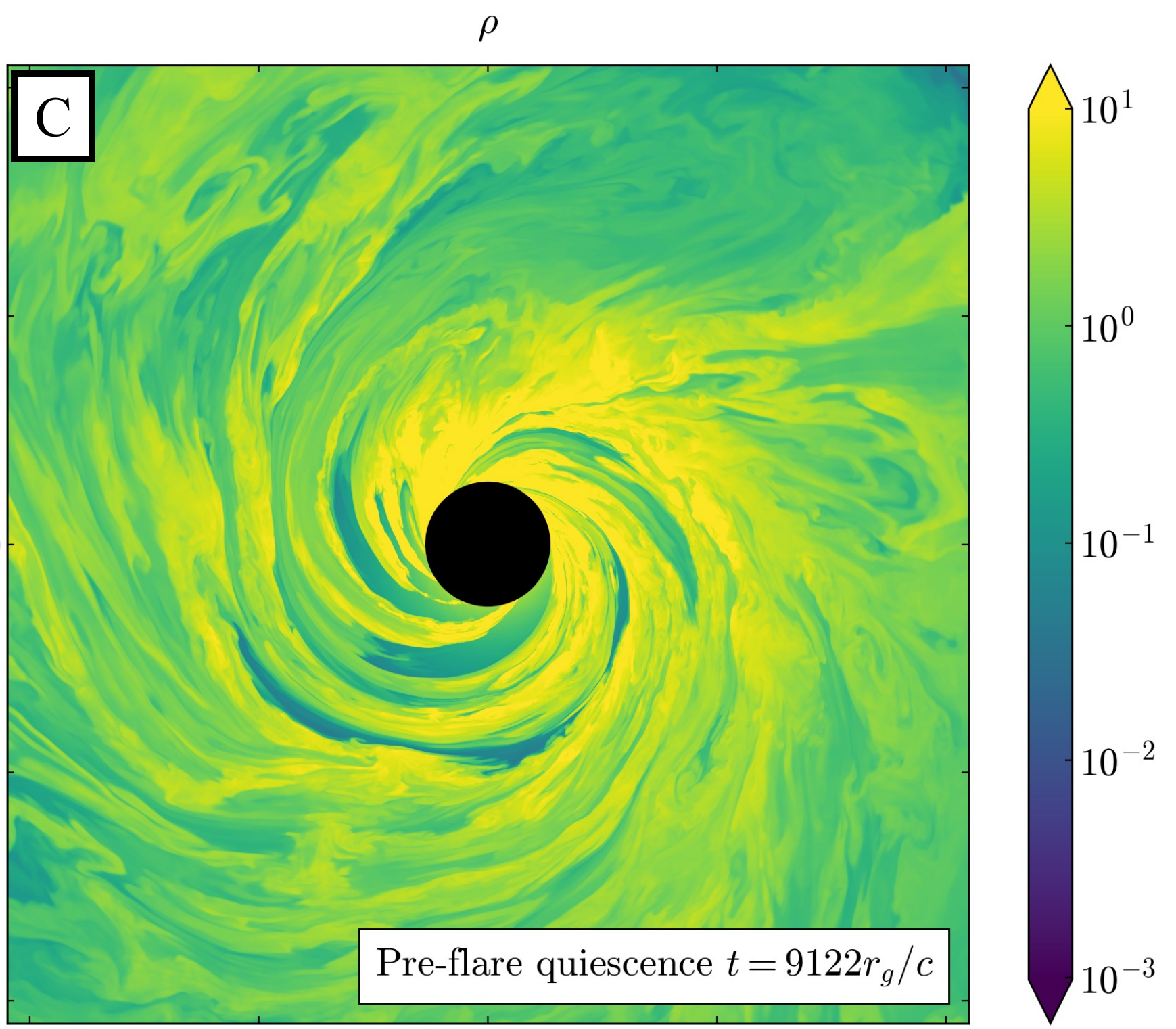}
    
    \includegraphics[width=0.353\textwidth,trim= 0.85cm 0.785cm 13.4cm 1.95cm, clip=true]{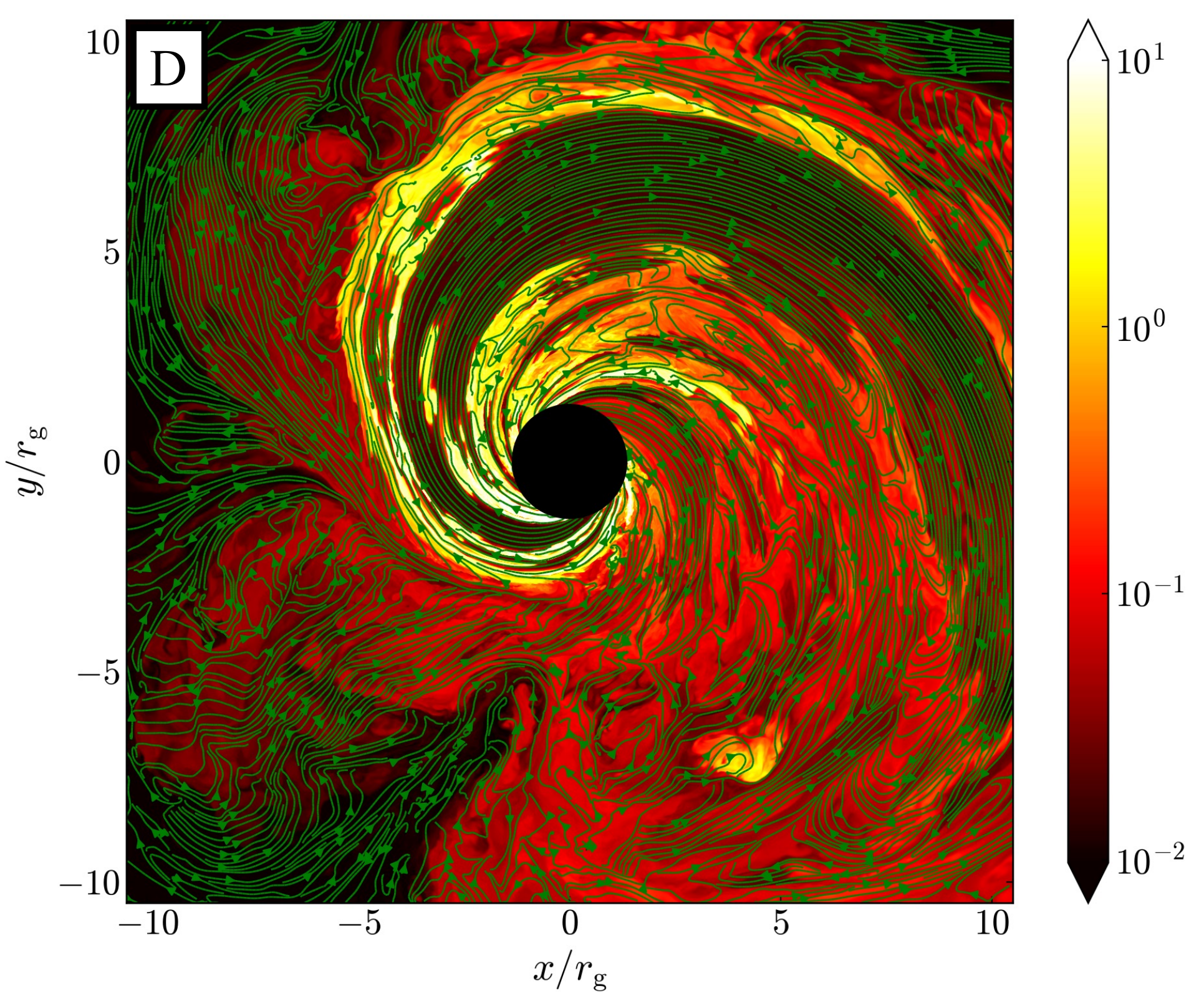}
    \includegraphics[width=0.318\textwidth,trim=  2.8cm 0.785cm 13.4cm 1.95cm, clip=true]{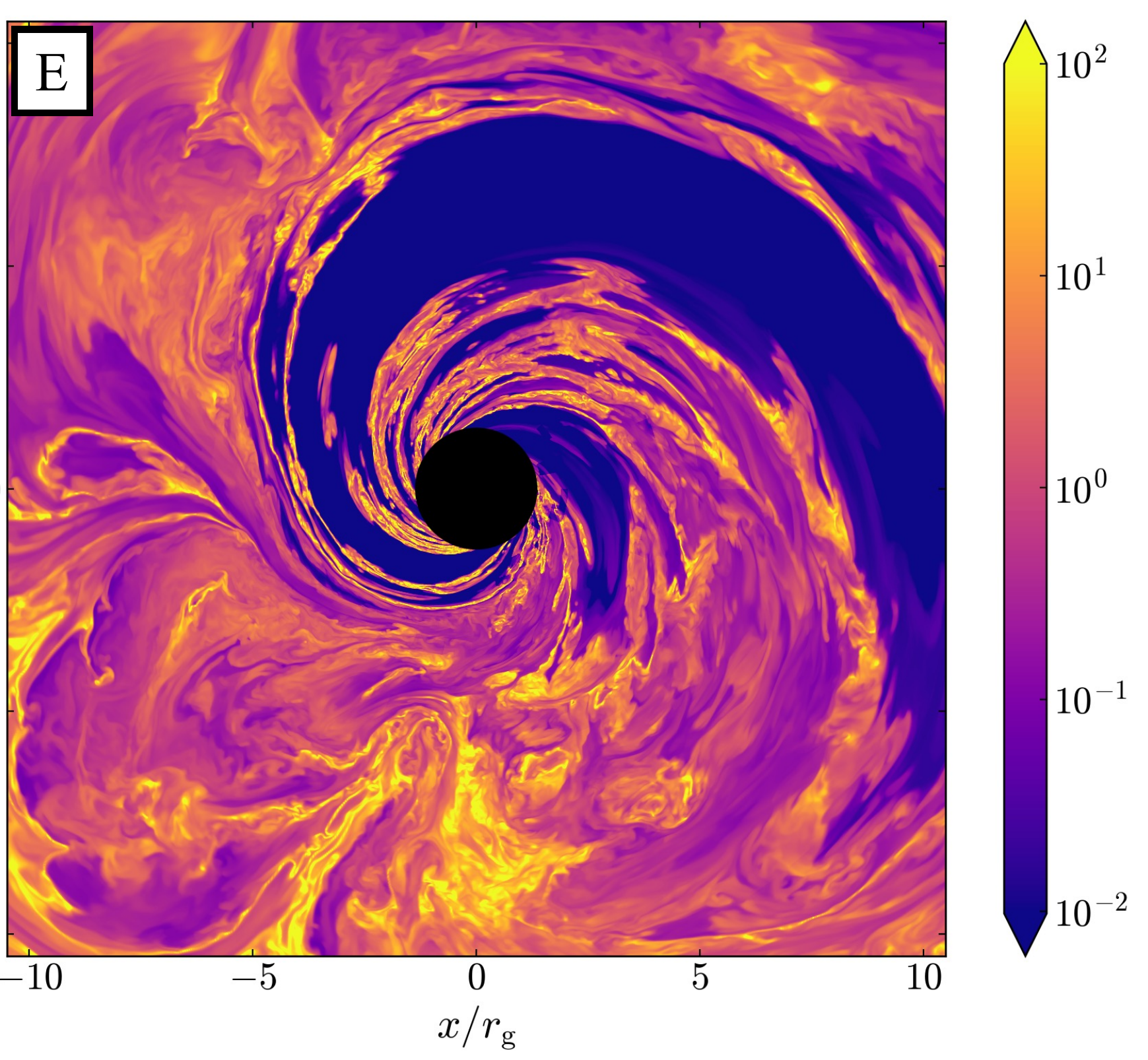}
    \includegraphics[width=0.318\textwidth,trim= 2.8cm 0.785cm 13.4cm 1.95cm, clip=true]{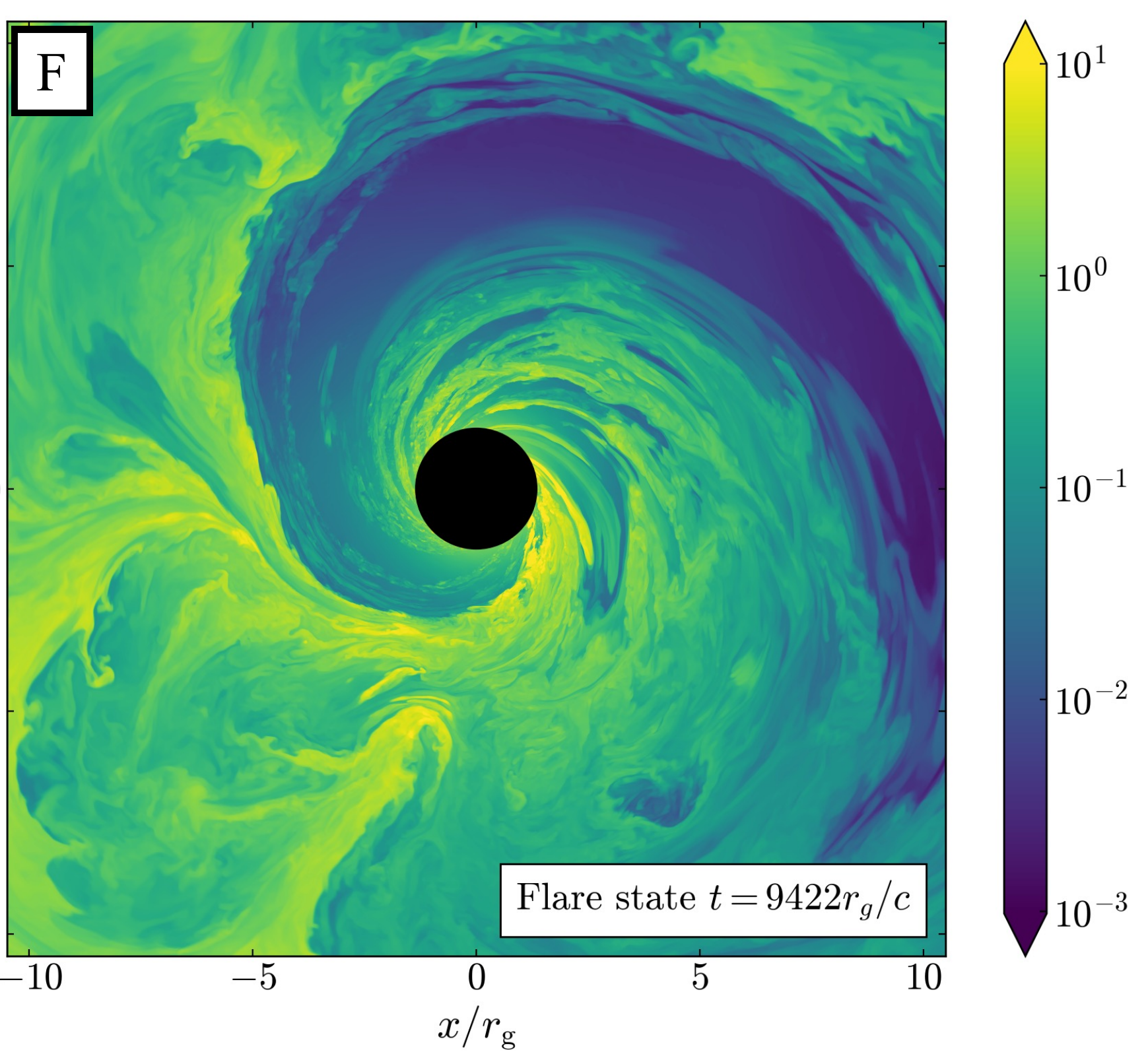}
    
    \includegraphics[width=0.353\textwidth,trim= 0.85cm 0.785cm 13.4cm 2.15cm, clip=true]{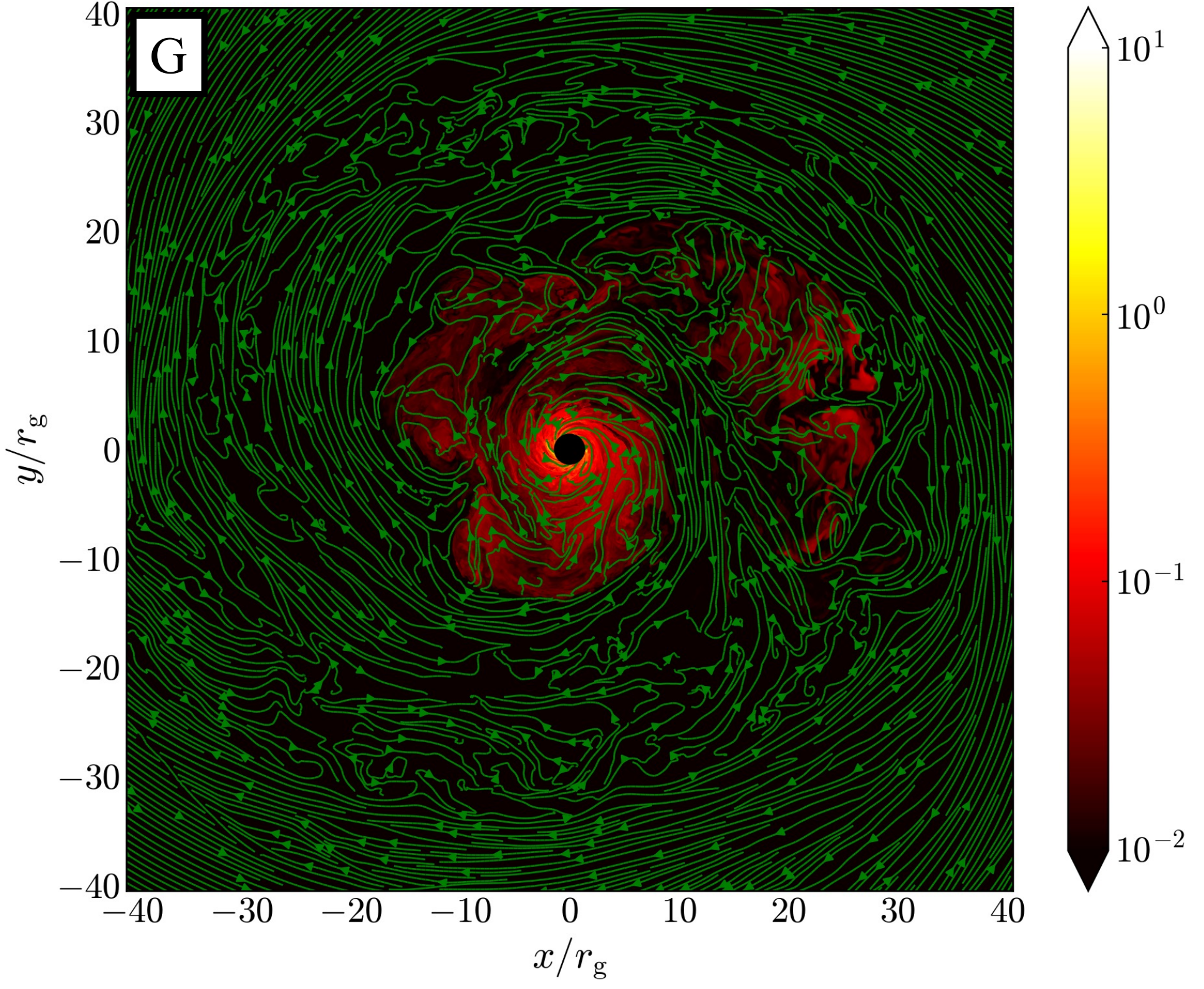}
    \includegraphics[width=0.318\textwidth,trim= 2.8cm 0.785cm 13.4cm 2.15cm, clip=true]{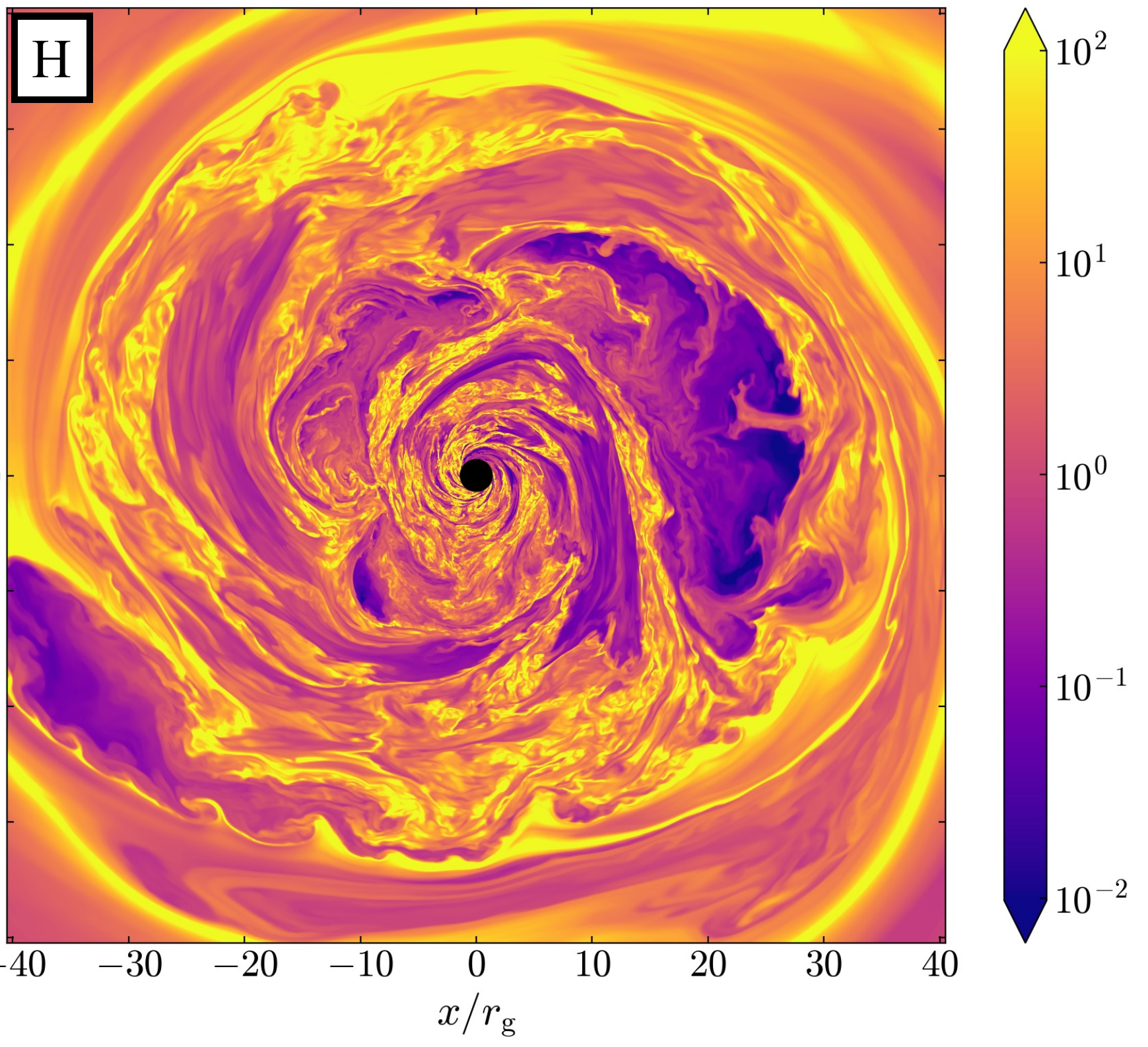}
    \includegraphics[width=0.318\textwidth,trim= 2.8cm 0.785cm 13.4cm 2.15cm, clip=true]{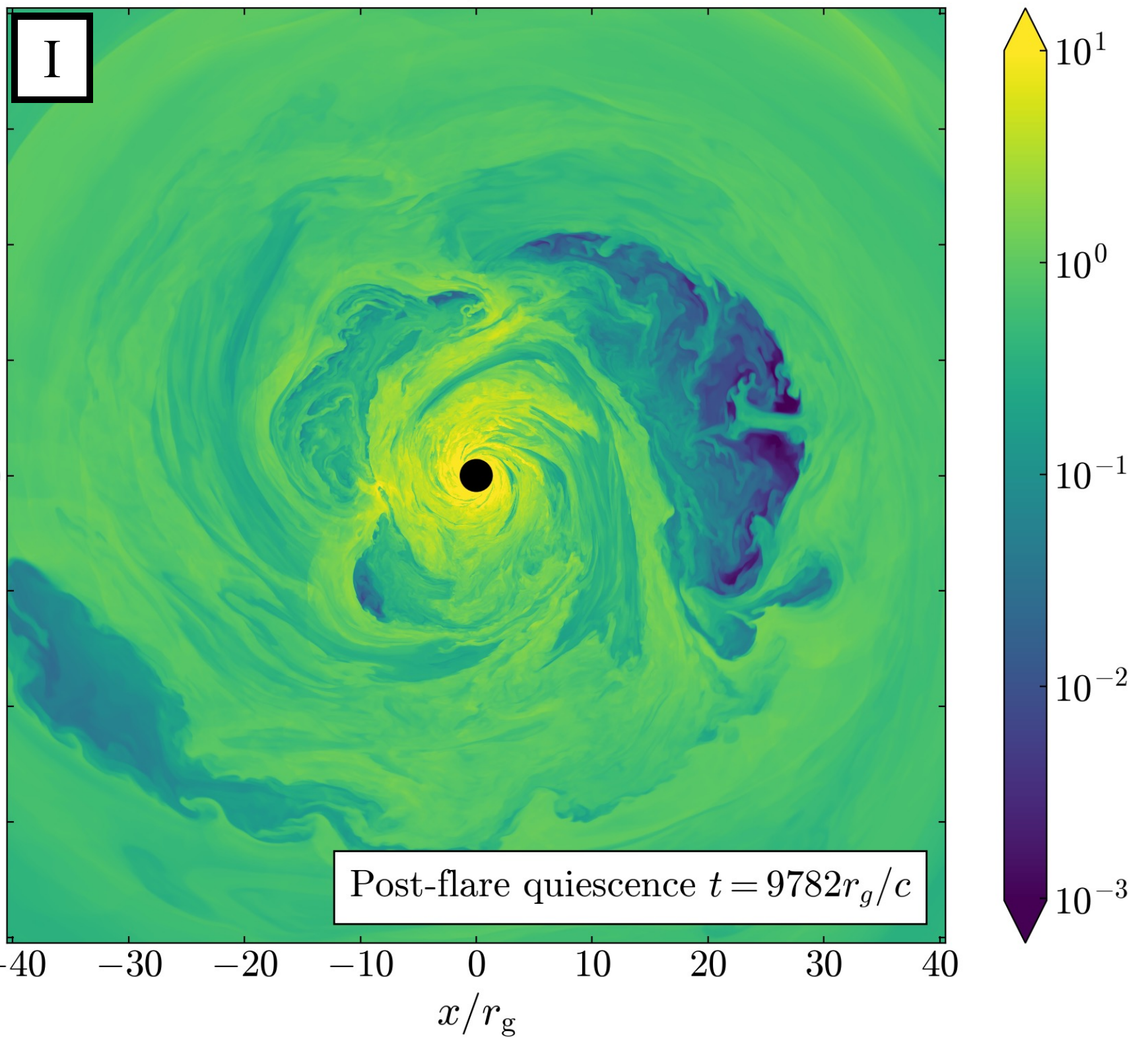}    
    \caption{{Our extreme resolution simulation reveals small-scale structure and interface instabilities of magnetic flux bundles escaping from the black hole, in an equatorial slice through the system.} Dimensionless temperature $T=p/\rho$, plasma-$\beta$, and density $\rho$ (from left to right) in the equatorial plane before a large {magnetic flux eruption} (top row), during the {magnetic flux eruption} (middle row) in the inner $10 r_{\rm g}$ and after the {magnetic flux eruption} (bottom row) in the inner $40 r_{\rm g}$. Gaps of low $\beta$ and density form during the {pre-eruption} quiescence while many azimuthal RTI modes accrete.
    During the {magnetic flux eruption} a single large $T>1$ spiral forms with a gap where the sheet moved out of the equatorial plane. Magnetic flux escapes through the spiral current sheet, while accretion continues over a small angle $\phi<2\pi$ at $x \approx 2 r_{\rm g}$ and $y \approx -1$ to $y \approx -2$. In the bottom row the inner $10 r_{\rm g}$ is in quiescent accretion state, and a hot flux tube that is ejected from the reconnection layer is in orbit at $x \approx 10 r_{\rm g}$ to $x \approx 30 r_{\rm g}$ and $y \approx -10 r_{\rm g}$ to $y \approx 20 r_{\rm g}$. The low $\beta$ flux tube shows clear signatures of instabilities at its boundaries mixing low density plasma into the disk.}
    \label{fig:panelXY}
\end{figure*}
{We analyze the flaring mechanism and its properties in the {MAD} after $t \approx 5000 r_{\rm g}/c$ {when} the accretion flow has settled into a quasi-steady state of a constant mass accretion rate and magnetic flux {on the black hole event horizon} (see Figure \ref{fig:mdot} in {Appendix C})}.
The accumulation of magnetic flux on the horizon cannot continue beyond the limit in which {the outward magnetic force balances the inward gravitational force}. When the magnetic flux reaches this limit in axisymmetry (2D), accretion is halted completely and a low density magnetosphere with an equatorial current sheet can form transiently (RBP20). In 3D, a large spectrum of RTI modes develops in the turbulent inner edge of the disk, steadily driving accretion. 
{The magnetic flux periodically erupts from the black hole into the disk. These eruptions are made possible by near-event-horizon reconnection, which converts the magnetic energy into the energy of emitting particles and can naturally power a flare.}
Figures \ref{fig:panelXZ} (at $\phi=0$, i.e., the meridional plane) and \ref{fig:panelXY} (at $\theta=\pi/2$, i.e., the equatorial plane) show the gas temperature $T=p/\rho$ with magnetic field lines plotted as green lines, the gas-to-magnetic-pressure ratio $\beta=8 \pi p/B^2$, and rest-mass density $\rho$ {around the time of one such flares at $t\sim 9500 r_{\rm g}/c$. Namely, we show the quantities in the quiescent period (i.e., a period of quasi-constant magnetic flux at the horizon) before, during, and after the large magnetic flux eruption, respectively, at $t=9122 r_{\rm g}/c$, $t=9422 r_{\rm g}/c$, and $t=9782 r_{\rm g}/c$ (where we zoom out to show large-scale effects).} 
Shortly before and during a flare, accretion only occurs through large{-scale} (i.e., low azimuthal mode-number) spiral RTI modes (see also \citealt{Takasao2019} for a very similar scenario explaining protostellar flares) creating a transient, non-axisymmetric (i.e., over an angle $\phi<2\pi$), magnetized (i.e., low plasma-$\beta$), low-density magnetosphere (top and middle rows in Figures \ref{fig:panelXZ} and \ref{fig:panelXY}) pushing the accretion disk outward and resulting in a drop in mass accretion rate. A macroscopic equatorial current sheet forms in the magnetosphere, extending from the horizon to the disk at $x=r\sin\theta\cos\phi\approx-5 r_{\rm g}$ at $z=r\cos\theta\approx 0$ shown by the antiparallel magnetic field lines ({inset in panel D}, green lines).
Reconnection pinches off the horizontal magnetic field in the sheet, transforming it into vertical ($z$) magnetic field, reminiscent of the 2D results of RBP20.
The {flux eruption} originates from the inner magnetosphere where the highly magnetized plasma in the jet directly feeds the current sheet.
The plasma density in the jet is determined by the density floor at $\sigma_{\rm max}=25$ in our simulations, whereas in reality it is much more strongly magnetized ($\sigma \gg \sigma_{\rm max}$) pair plasma.  {Reconnection occurs locally in X-points where a field line breaks and reconnects to other field line (see insets in Figures~\ref{fig:panelXZ}D and \ref{fig:panelXZ}E). In these X-points, reconnection heats the plasma up to $T \sim \sigma_{\rm max} = 25$ (left panels) after which it is expelled from the layer at Lorentz factors up to $\Gamma \propto \sqrt{\sigma_{\rm max}} = 5$ (\citealt{lyubarsky2005}, see also {Appendix B} for an exploration of different $\sigma_{\rm max}$ in 2D).} The flux is expelled through reconnection into the low-density region in between the large low-mode-number RTI modes accreting spirals. {Electrons and positrons accelerated to non-thermal energies through reconnection at the X-points in the macroscopic equatorial current sheet can power high-energy flares that may reach a distant observer during the drop in the mass accretion rate.} 

Small plasmoids are visible close to the horizon and a larger hot plasmoid is detected at $x=-3 r_{\rm g}$ (middle row in Figure \ref{fig:panelXZ}) as a result of the merger of smaller escaping plasmoids. The plasmoids that escape the gravitational pull of the black hole interact with the disk and jet sheath resulting in significant heating up to at least $z \gtrsim \pm 40 r_{\rm g}$. The bottom row of Figure \ref{fig:panelXZ} shows a large magnetic flux tube at $x \approx 20-30 r_{\rm g}$: a low density region of strong vertical field (low plasma-$\beta$) heated to medium temperature $T \sim 0.1-1$. {The flux tube forms as a result of the reconnection that converts horizontal magnetic field into vertical field that is ejected from the reconnection layer. Filled with heated plasma, the flux tube can appear as a hot spot. The accumulated vertical magnetic flux in this hot spot} can remain coherent for approximately one orbital time scale {between $10$ and $30 r_{\rm g}$} (bottom row in Figure \ref{fig:panelXY} between $y \approx -20 r_{\rm g}$ and $y\approx 20 r_{\rm g}$), while the inner $10 r_{\rm g}$ is already in the quiescent accretion state at $t=9782 r_{\rm g}/c$. RTIs develop at the boundary of the hot spot, which mix the hot low density plasma into the surrounding accreting gas. The hot spots are expected to be filled with positrons and electrons energized by the reconnection, which in this way can end up in the accretion disk.
After the flaring episode, magnetic flux builds up on the horizon and the quasi-steady-state accretion cycle develops again. Smaller and less hot current sheets where $B^{\phi}$ changes sign also exist in the inner $\sim 20 r_{\rm g}$ of the turbulent accretion disk during the quiescent period, indicated by thin high-$\beta$ layers of anti-parallel field lines (top and bottom rows in Figures \ref{fig:panelXZ} and \ref{fig:panelXY}).
\begin{figure*}
    \centering
    \includegraphics[width=1.0\textwidth, clip=true]{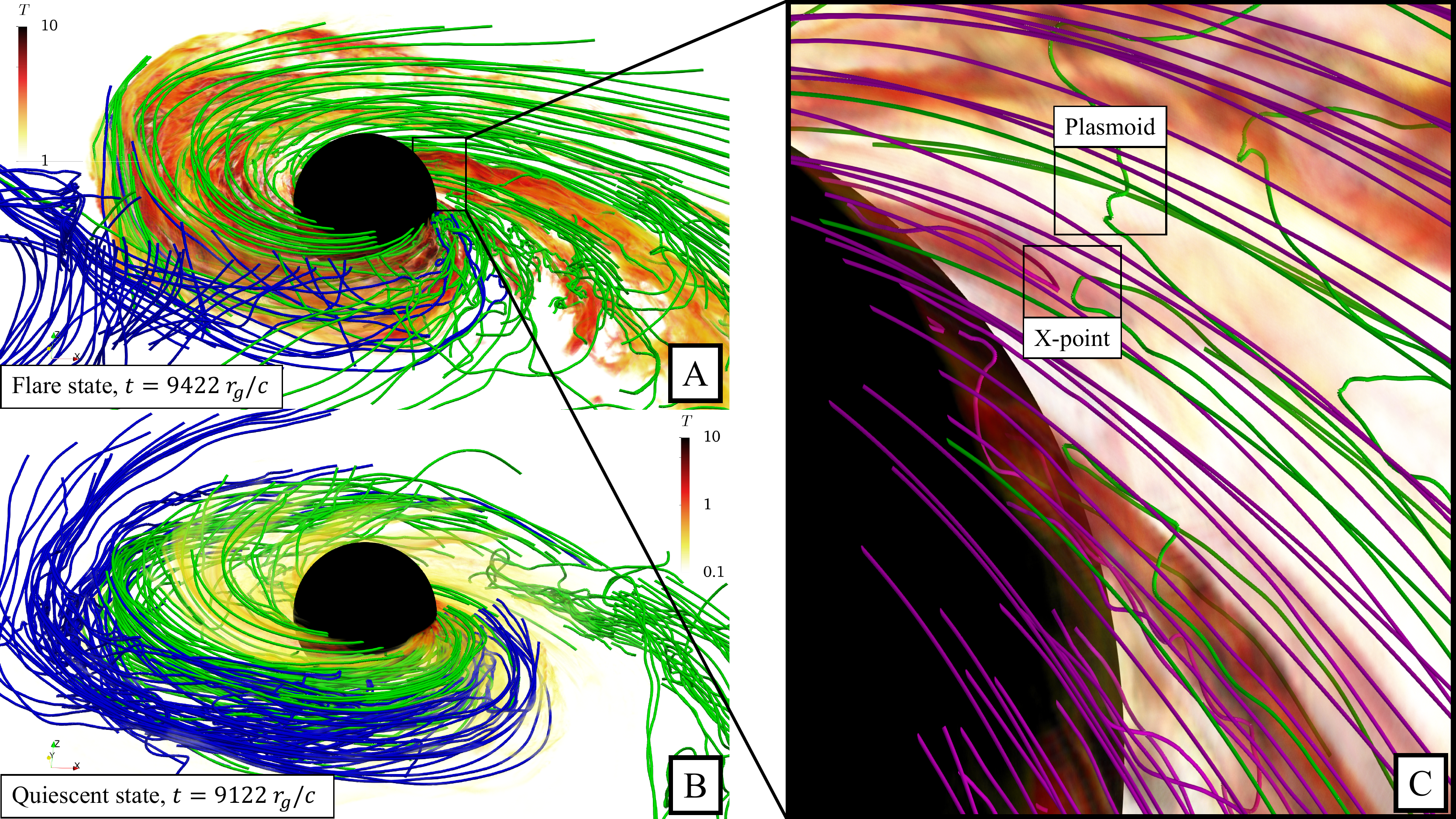}
    \caption{
    Volume rendering of the temperature $T=p/\rho$ shows plasmoids and hot current sheets. Extreme resolution allows the current sheets to become thinner and hotter than typically seen in GRMHD simulations. 
    (Panel A:) During a large flare a relativistically hot $T>1$ spiral current sheet forms. Accretion occurs over a small azimuthal angle $\phi < 2\pi$ in the $T<1$ (white) regions. The green field lines, seeded in the current sheet ($T>1$), remain in the current sheet and are mostly attached to the black hole. Blue field lines are seeded in the disk, where some disk field lines are accreting onto the black hole in the $T<1$ region. (Panel B:) In the quiescent state $T\leq 1$ everywhere, and both green and blue field lines (with the same seeds as in panel A) are in the disk, accreting onto the black hole. The inset (C) shows a zoom into {the inner $r_{\rm g}$ in} the flare state with multiple escaping flux loops (green field lines). In the small black box we highlight an escaping flux tube with vertical field as the result of reconnection (green) and an infalling flux tube (purple). We also show a plasmoid, indicated by the helical field line (green) in the second small black box.}
    \label{fig:3D}
\end{figure*}

Figure \ref{fig:3D}{A} visualizes the 3D nature of the hot current sheet by showing the temperature and magnetic field line structure in the inner $10 r_{\rm g}$ during the flare at $t=9422 r_{\rm g}/c$. The current sheet has a relativistic temperature $T>1$, whereas shortly before the flare at $t=9122 r_{\rm g}/c$ ({\ref{fig:3D}B}) there are no structures at $T>1$. During flare, the (green) field lines in the current sheet {(i.e., seeded in the $T>1$ region in \ref{fig:3D}{A})} have a clear spiral structure and are separated from the more vertical field lines in the disk (blue). During the quiescence before the flare {(Figure \ref{fig:3D}B)} no such distinction is visible and all field lines (green and blue{, which are seeded at the same points as in panel \ref{fig:3D}A}) are part of the disk. {The extreme resolution allows to capture multiple plasmoids identified as 3D helical field line structures in the sheet (Figure~\ref{fig:3D}C) during the magnetic flux eruption}. We highlight a typical X-point as the manifestation of reconnection, separating an infalling (purple field line) and escaping flux tube (green field line) in the hot current sheet. Similar X-points can be detected in {e.g., the inset in Figure \ref{fig:panelXZ}D}.
\begin{figure*}
    \centering
    \includegraphics[width=0.476\textwidth,trim= 2.45cm 6.25cm 8.5cm 0cm, clip=true]{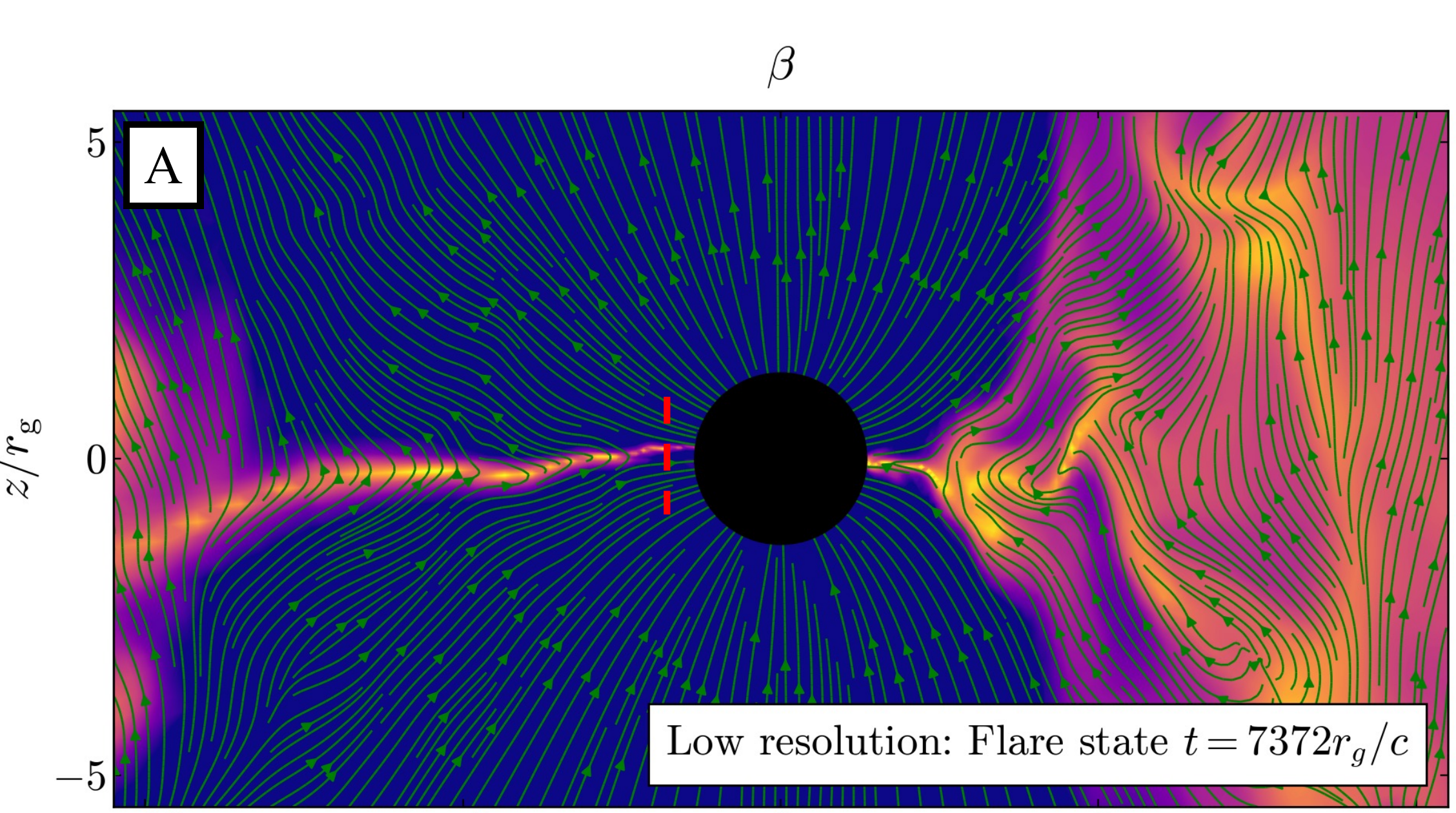}
    \includegraphics[width=0.51\textwidth,trim= 4.2cm 6.25cm 5.1cm 0cm, clip=true]{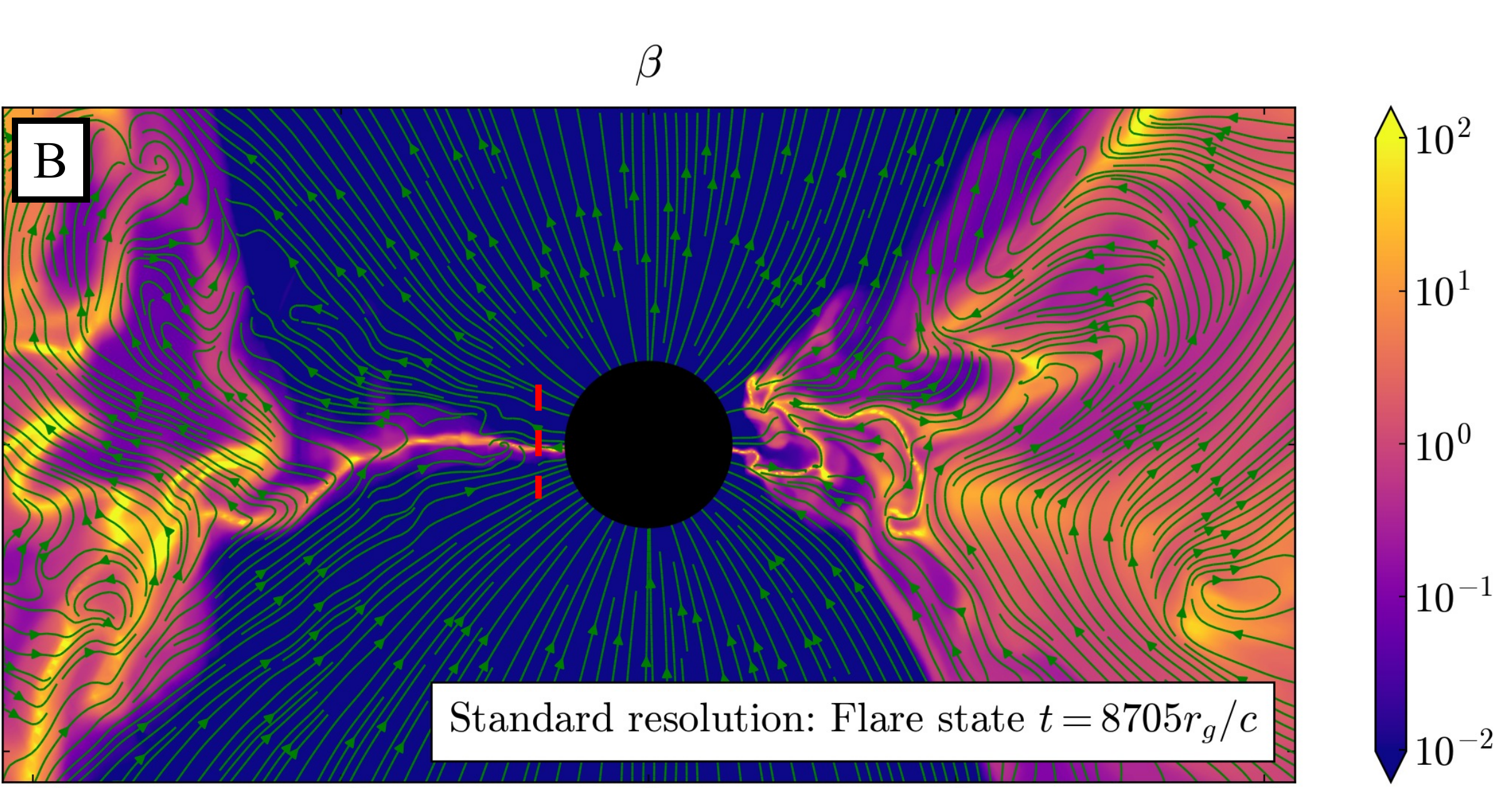}
    
    \includegraphics[width=0.476\textwidth,trim= 2.45cm 4.7cm 8.5cm 1.5cm, clip=true]{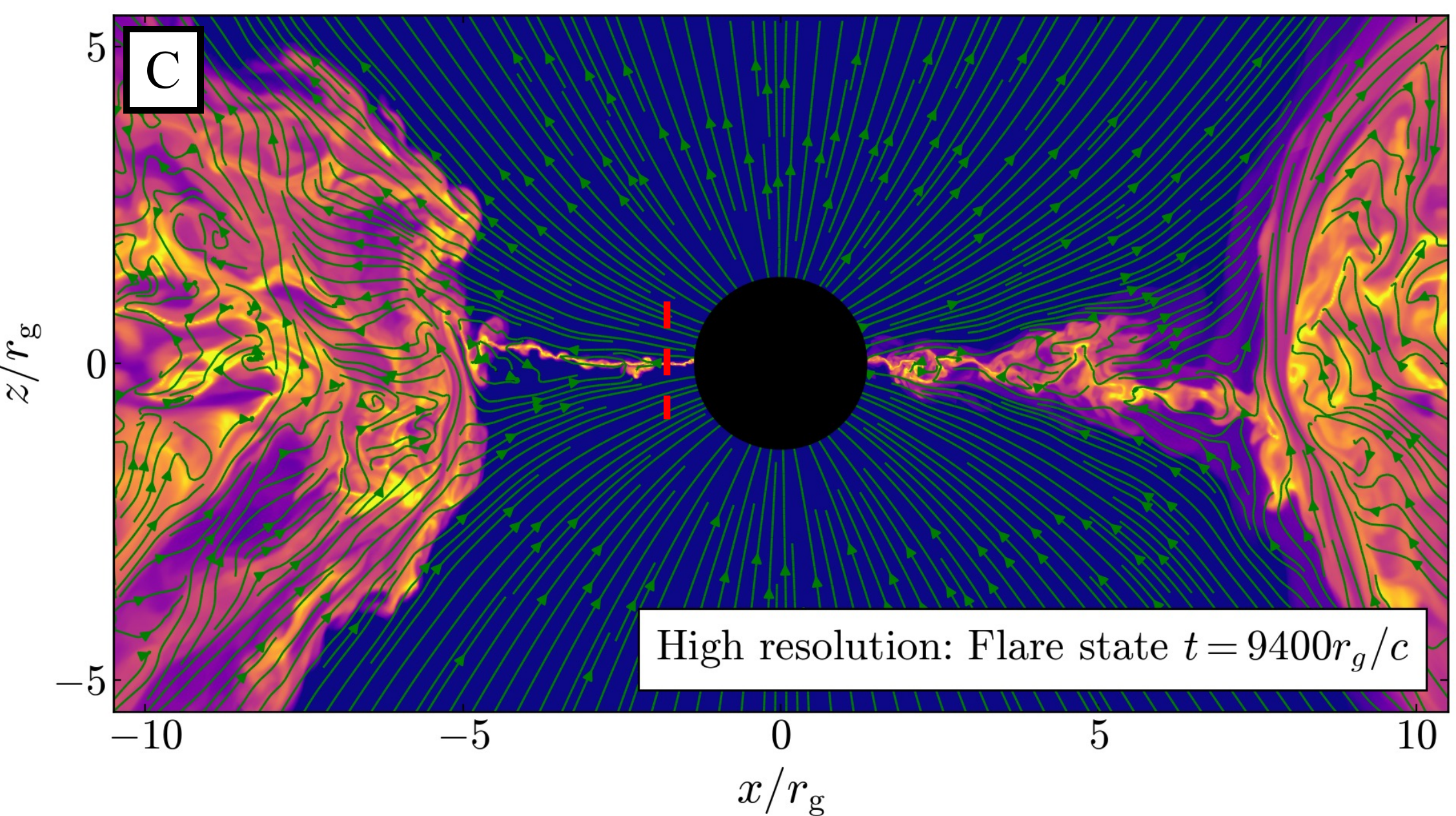}
    \includegraphics[width=0.51\textwidth,trim= 4.2cm 4.7cm 5.1cm 1.5cm, clip=true]{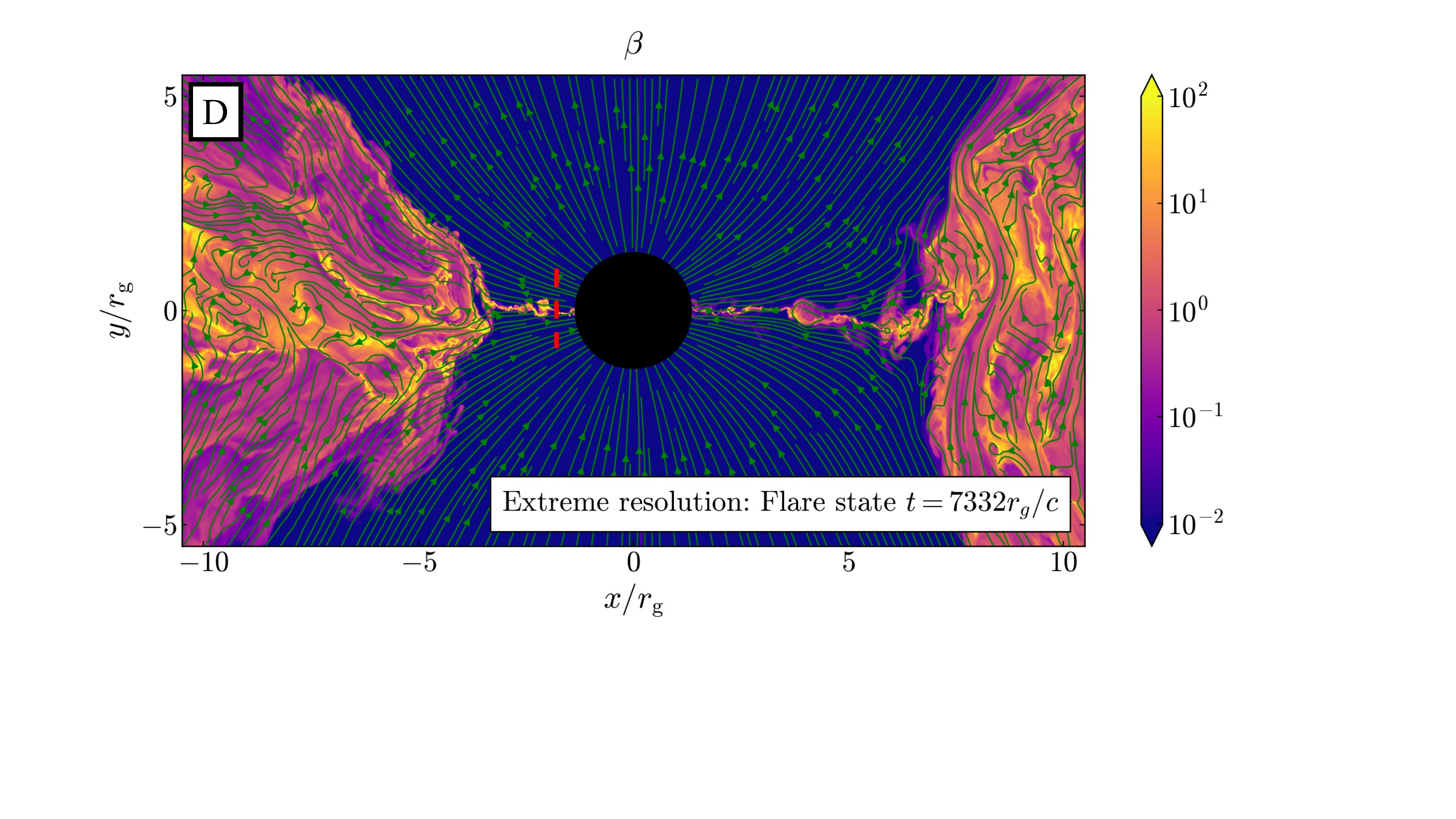} 
    
    \includegraphics[width=0.498\textwidth, trim= 0.3cm 10.5cm 14.3cm 0.4cm, clip=true]{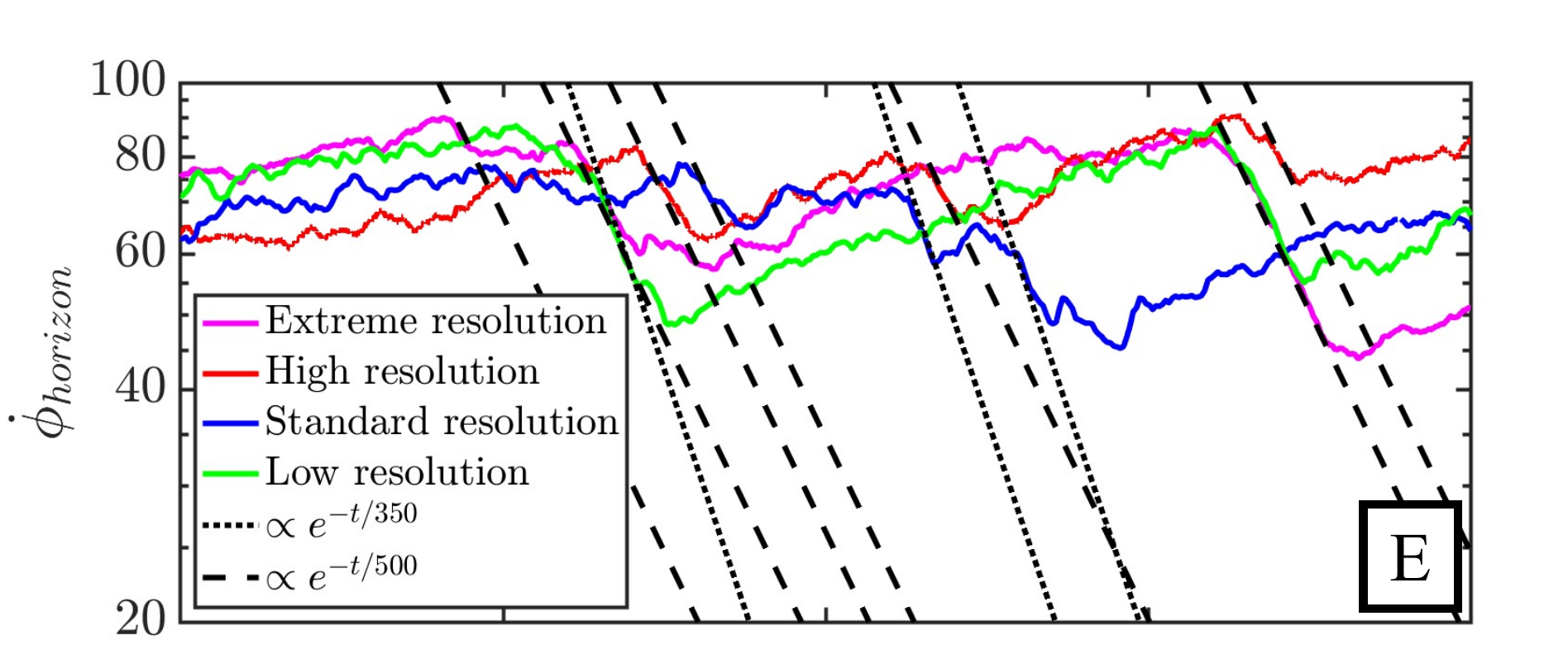}
    \includegraphics[width=0.496\textwidth, trim= 0.3cm 10.5cm 14.5cm 0.4cm, clip=true]{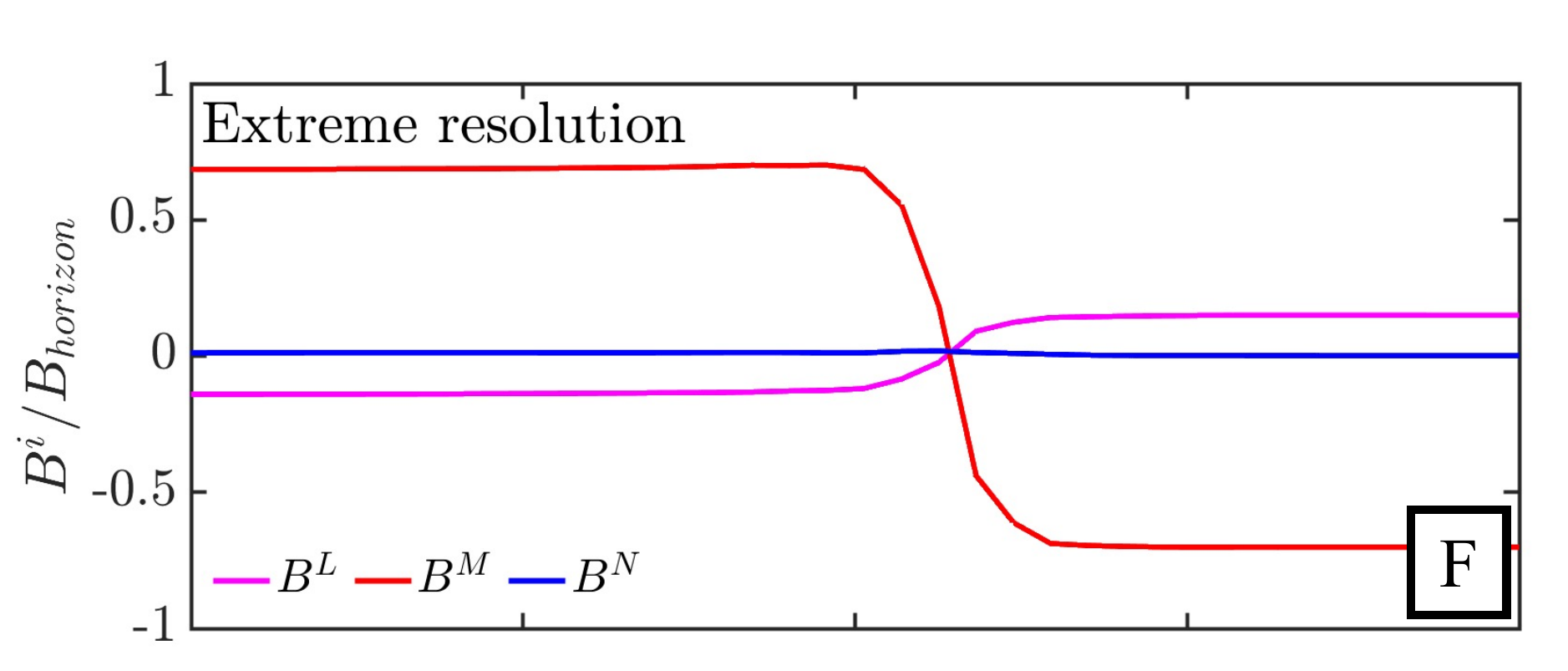}
    
    \includegraphics[width=0.498\textwidth, trim= 0.3cm 0.3cm 14.3cm 9.9cm, clip=true]{mdot_phidot_paper.pdf}
    \includegraphics[width=0.496\textwidth,  trim= 0.3cm 0.3cm 14.5cm 9.9cm, clip=true]{recrate.pdf}
    \caption{{The equatorial current sheet that forms during the magnetic flux eruption is unresolved at low and standard resolutions (panels A,B) such that magnetic field lines (green lines) diffuse through the current sheet and do not reconnect, due to the high numerical resistivity. At high and extreme resolutions (C,D), the field lines are antiparallel in the current sheet, and they reconnect in well-defined X-points. Smaller current sheets are resolved in the accretion disk at high and extreme resolutions, potentially heating the plasma through reconnection.} {Panel E shows the} magnetic flux on the horizon for the four numerical resolutions. The extreme and high resolution runs show two and three large flare periods, respectively, indicated by flux decay at a rate $\propto e^{-t/500}$ governed by the reconnection rate (dashed black lines). A mini-flare is indicated by the small flux drop at $t\approx 6800 r_{\rm g}/c$ in the extreme resolution run. The standard and low resolution runs show a faster flux decay $\propto e^{-t/350}$ governed by the enhanced reconnection rate due to an increased numerical resistivity. Flares in the extreme resolution run are accompanied by clear drops in the mass accretion rate {(panel G)}, due to the expulsion of the disk over a large azimuthal angle. {Panel F shows a cut through the equatorial current sheet at $x\approx1.5 r_{\rm g}$ during the flare state (indicated by the red dashed line in panels A-D). Both the (nearly) radial field $B^L$ and the (nearly) toroidal field $B^M$ {components (see definition in the text) change sign in the equatorial current sheet}, while $B^N$ is (close to) zero, indicating zero-guide field reconnection. Panel H shows the flow speeds left and right of the current sheet. After correcting for the bulk flow, we measure the reconnection rate to be $v_{\rm rec}\approx0.01c$. We confirm this measurement at 10 radial cuts during separate flare periods.} 
    }
    \label{fig:recrate}
\end{figure*}

{Figure \ref{fig:recrate}A-D shows zooms into the current sheet during large magnetic flux eruptions for the four numerical resolutions employed. The drop in magnetic flux at low and standard resolutions (panels A,B) is not accompanied by a large drop in mass accretion rate (see panels E,G), due to the large numerical diffusion. The magnetic field diffuses through the thick current sheet and does not reconnect, due to the large numerical resistivity. This results in a too high reconnection rate and a large heated area (see {Appendix C}, Figure \ref{fig:panelXZlowres} for more properties of the large magnetic flux eruption at low resolution). The current sheet is in these cases not plasmoid-unstable. The high resolution flux eruption (panel C) behaves similarly to the extreme resolution result (panel D) from Figure \ref{fig:panelXZ}, indicating that the plasmoid instability is resolved on the grid, and that the reconnection rate is converged to a universal value of $0.01c$ (panel H).}
In Figure \ref{fig:recrate} we {also} analyze the magnetic flux $\dot{\phi}_{\rm BH} := \frac{1}{2}\int_{0}^{2 \pi} \int_{0}^{\pi} |^{*} F^{rt}| \sqrt{-g} \, d\theta d\phi$ on the horizon {(\ref{fig:recrate}E)} and the mass accretion rate $\dot{m} := -\int_{0}^{2 \pi} \int_{0}^{\pi} \rho u^r \sqrt{-g} d\theta d\phi$ through the inner $5 r_{\rm g}$ {(\ref{fig:recrate}G)}, where $g$ is the metric determinant, $u^\mu$ is the fluid 4-velocity, $^* F^{\mu\nu}$ is the dual of the Faraday tensor, and $\rho$ is the fluid-frame rest-mass density. After $\sim 5000 r_{\rm g}/c$ the flow sets into a quasi-steady state which is globally converged for all resolutions. For the extreme resolution run (magenta line {Figure \ref{fig:recrate}E)} we observe two major flux decays, which we associate with large flares, at $t\approx 7300 r_{\rm g}/c$ and $t\approx 9300 r_{\rm g}/c$, both lasting for a few $\sim 100 r_{\rm g}/c$. We also observe a small flux decay at $t\approx 6800 r_{\rm g}/c$, associated with a smaller flare, or ``mini-flare''. For all flares, the magnetic flux on the event horizon decays quasi-exponentially with time with characteristic timescale $\tau\approx 500$~$r_{\rm g}/c$ (indicated by the black dashed lines), implying that the decay is governed by reconnection at a universal rate of $0.01c$, consistent with the decay observed for a split monopole magnetic field on the event horizon (\citealt{Bransgrove2021}). All three events are accompanied by a large drop in mass accretion rate (Figure \ref{fig:recrate}G) that is related to the ejection of the accretion disk such that the accretion is funneled through a small azimuthal angle $\phi < 2\pi$ and nearly halts. 

For {the high resolution run (red line, Figure \ref{fig:recrate}E), }similar flare episodes can be observed at $t\approx 7500 r_{\rm g}/c$, $t\approx 8300 r_{\rm g}/c$, and $t\approx 9400 r_{\rm g}/c$, with flux decaying on the same timescale $\tau\approx 500$~$r_{\rm g}/c$. For lower resolutions (blue and green line) there is a clearer distinction: large flares show (e.g., at $t\approx 7300 r_{\rm g}/c$ for low resolution, and $t\approx 8300 r_{\rm g}/c$ and $8600 r_{\rm g}/c$ at standard resolution) a faster decay rate $\tau\approx 350$~$r_{\rm g}/c$, implying a faster reconnection rate $> 0.01c$. Mini-flares (e.g., at $t\approx 9300 r_{\rm g}/c$ for low resolution and $t\approx 7500 r_{\rm g}/c$ for high resolution) instead show a flux decay at a rate of $\tau\approx 500$~$r_{\rm g}/c$ implying a reconnection rate of $\sim 0.01c$. At low and standard resolution{s}, these mini-flares are typically {\it not} accompanied by a clear drop in $\dot{m}_{5r_{\rm g}}$ {(Figure \ref{fig:recrate}G)}, while large flares are showing a clear drop in $\dot{m}_{5r_{\rm g}}$ implying a large ($\gtrsim 5 r_{\rm g}$) current sheet. {This can be explained by the large numerical diffusion of the thinning current sheet in both the $z$ and $y$ directions, resulting in a too broad accretion funnel at low and standard resolution (Figure \ref{fig:recrate}A,B). Mini-flares are better captured at lower resolutions than large flares due to the shorter length of the current sheet and the higher effective resolution of the spherical grid at small radii (see {Appendix D}).}

The reconnection rate can be determined directly by selecting a current sheet during a flare episode and measure the inflow speed of the plasma into the reconnection layer. {To do so, we first transform the Eulerian velocity and magnetic fields into a locally Minkowski frame (see e.g., \citealt{white2016})} to apply standard reconnection analysis in flat spacetime (RBP20).
The fields are expressed in minimum variance coordinates (\citealt{Howes_2016}), with $B^L$ projected in the flat frame along the poloidal direction parallel to the current sheet, $B^M$ along the toroidal direction and $B^N$ perpendicular to the current sheet, to determine the upstream geometry, showing a typical Harris-type sheet structure in {Figure \ref{fig:recrate}F}. Both the toroidal and poloidal components switch sign in the sheet, indicating that zero-guide-field reconnection occurs. {The total vertical velocity of the flow consists of the inflow of the fluid into the current sheet due to reconnection, $v_{\rm rec}$,  and the advection of the current sheet with the bulk velocity, $v_{\rm bulk}$. In Figure {\ref{fig:recrate}H} we then measure the flow speeds from left and right of the current sheet as $v_{\rm in, left}=(v_{\rm bulk} + v_{\rm rec}) / (1 + v_{\rm bulk} v_{\rm rec}/c^2)$ and $v_{\rm in, right}=(v_{\rm bulk} - v_{\rm rec}) / (1 - v_{\rm bulk} v_{\rm rec}/c^2)$ and solve for $v_{\rm rec}$, where we account for the relativistic speed of the bulk flow. {We determine the profile of the upstream magnetic field projected along the current sheet and find the location where the profile becomes flat (Figure \ref{fig:recrate}F).} We then} select 10 cuts of the current sheet at different radii and consistently find a reconnection rate of $\sim 0.01c$, indicating a Lundquist number of at least $S = v_{\rm A} w / \eta_{\rm num} = (v_{\rm rec}/c)^{-2} = 10^4$. {Reconnection thus occurs in the asymptotic {plasmoid-mediated} regime where $S \geq S_{\rm{crit}} = 10^4$ (\citealt{bhattacharjee2009}) for our extreme resolution run, where the length of the sheet $w \gtrsim r_{\rm g}$, Alfv\'{e}n speed {in the upstream, $v_{\rm A} = \sqrt{\sigma_{\rm up} / (\sigma_{\rm up}+1)}c \sim c$, for $\sigma_{\rm up} =25$}, and numerical resistivity $\eta_{\rm num}$. The reconnection rate is consistent with 2D resistive GRMHD simulations of plasmoid-dominated reconnection in MAD flows (RBP20) for Lundquist numbers $S = L_{\rm{sheet}} c / \eta \gtrsim \mathcal{O}(10^5)$ with explicit resistivity $\eta=5\times 10^{-5}$.} In {Appendix D} we show the same analysis for the lower resolution simulations, concluding that the {extreme and} high resolution results are in the {plasmoid-mediated} regime, whereas the standard and low {resolution} runs show reconnection rates {a factor two to three larger than} $0.01c$, and do not display plasmoids. The enhanced reconnection rate due to larger numerical resistivity at lower resolutions manifests itself as an increased flux decay rate and hence directly affects the flare time scale.
\begin{figure*}
    \centering
    \includegraphics[width=0.3527\textwidth,trim= 0.85cm 2.3cm 13.4cm 1.3cm, clip=true]{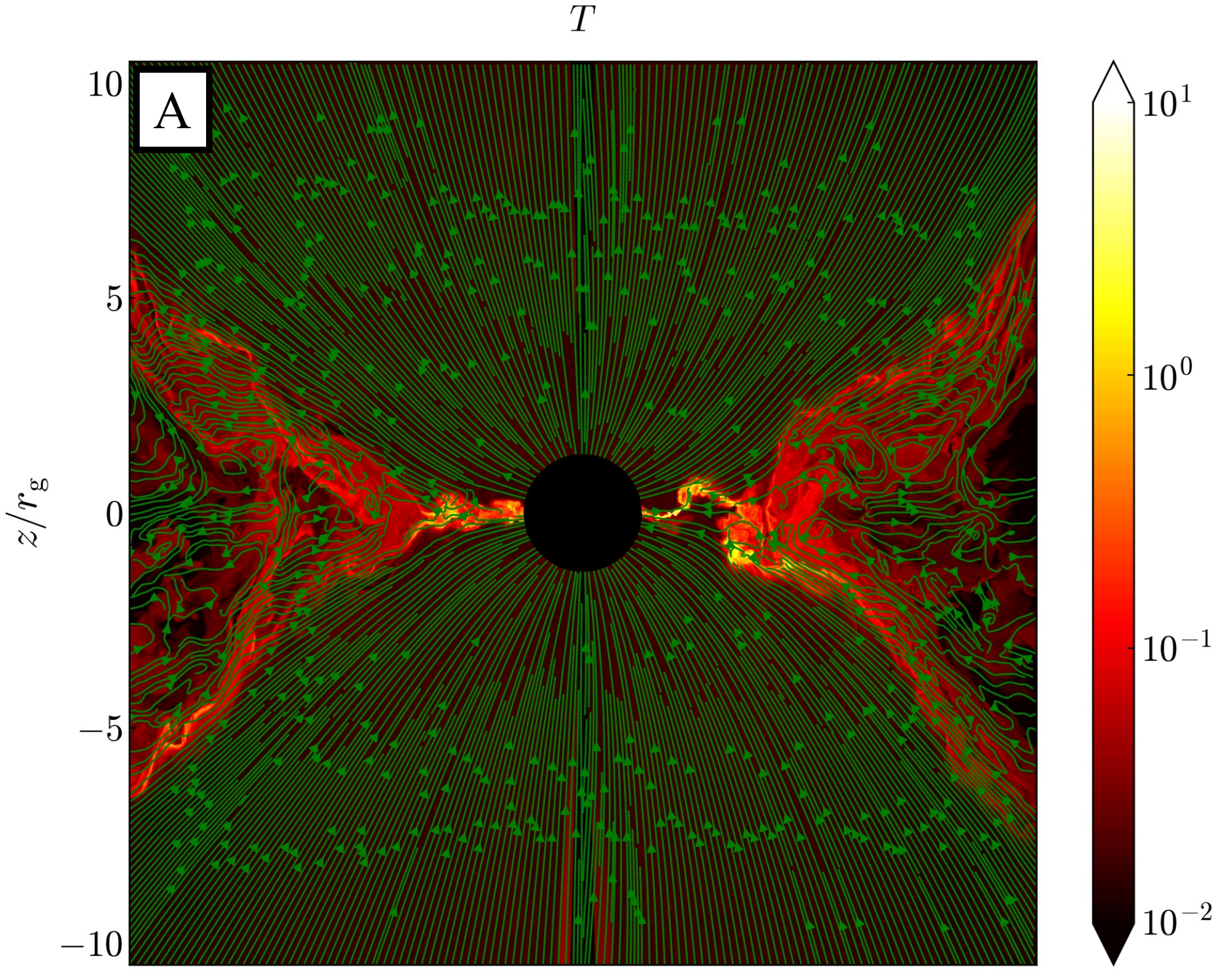}
    \includegraphics[width=0.3183\textwidth,trim= 2.8cm 2.3cm 13.4cm 1.3cm, clip=true]{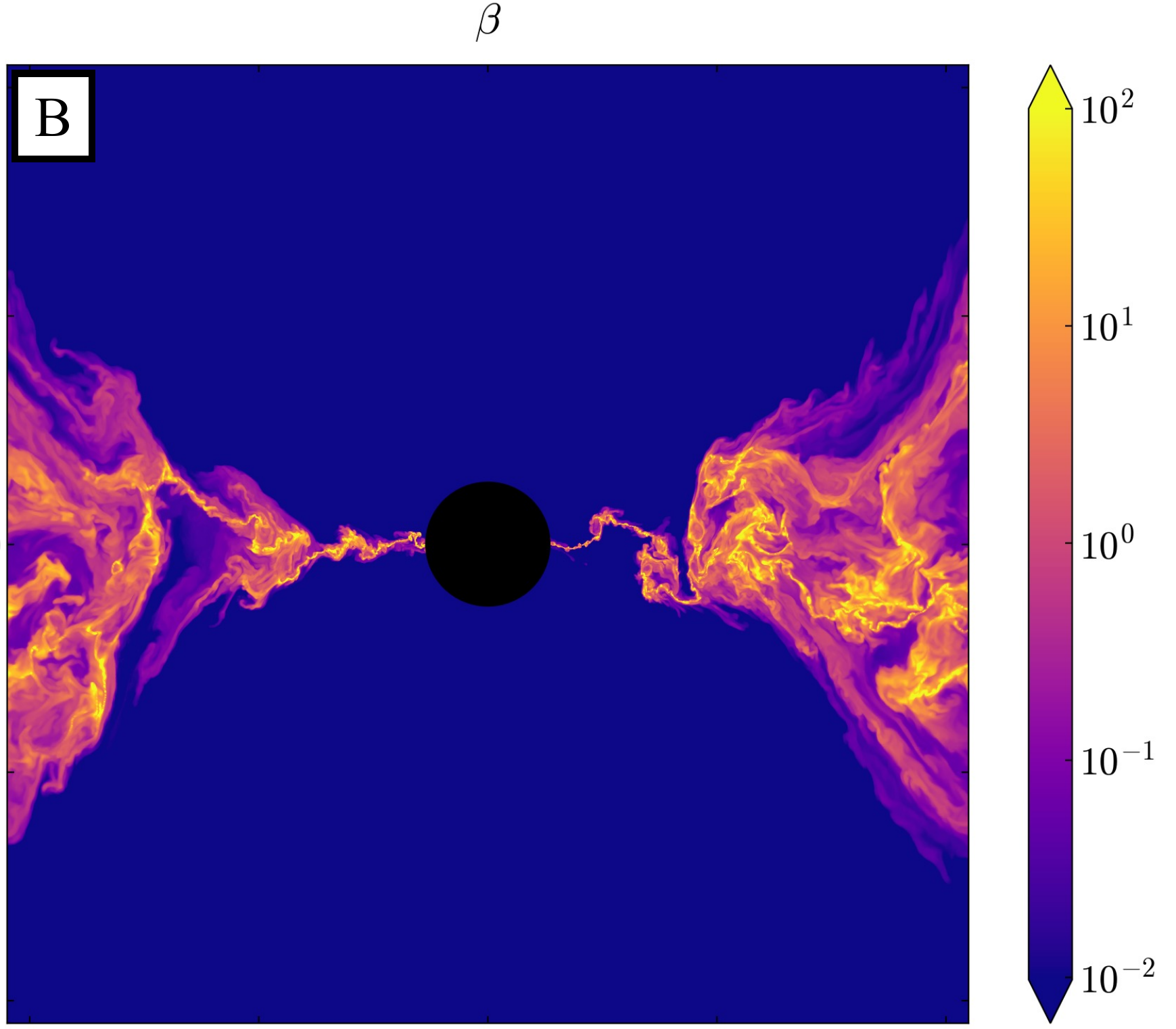}
    \includegraphics[width=0.3183\textwidth,trim= 2.8cm 2.3cm 13.4cm 1.3cm, clip=true]{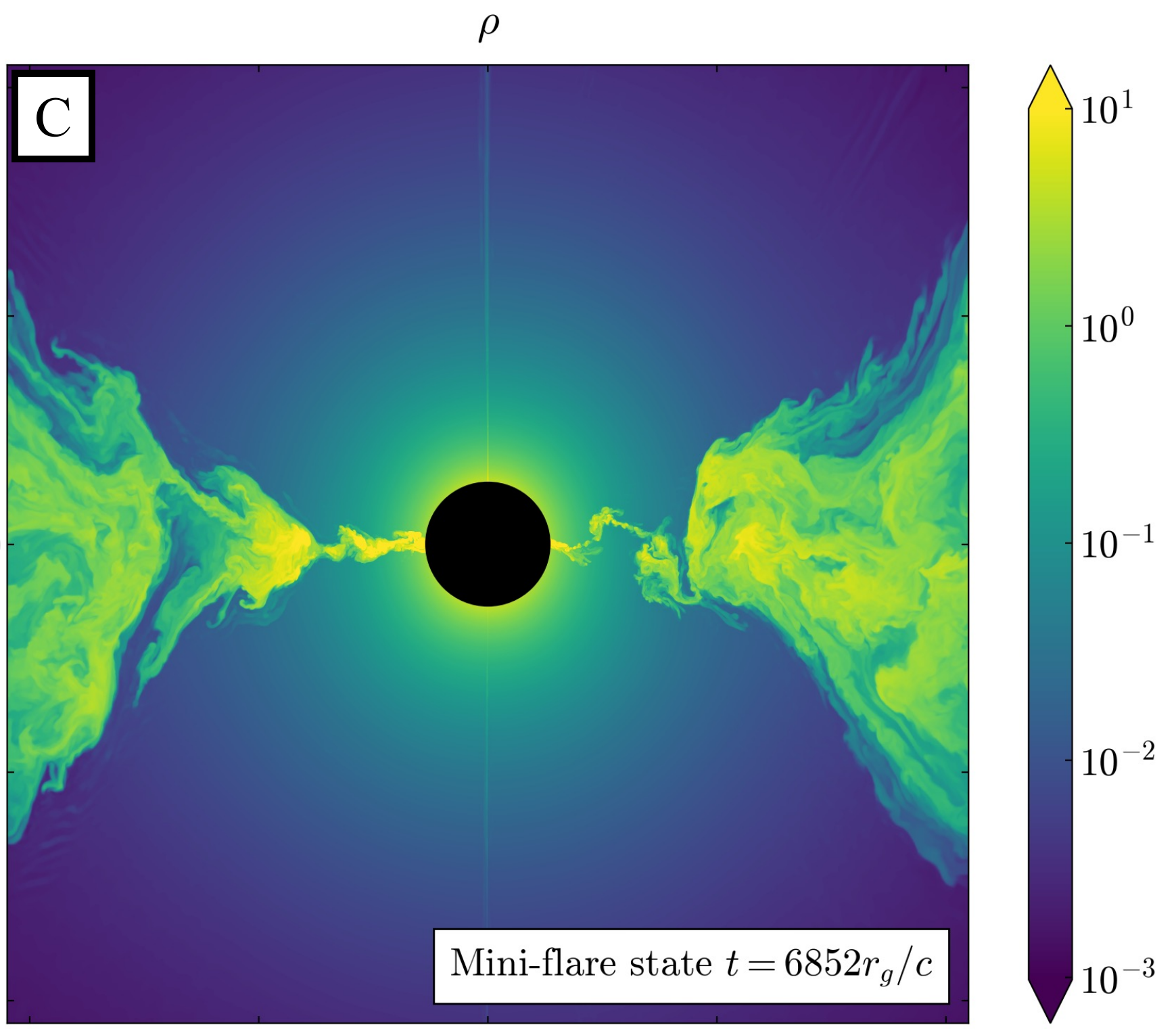}
    
    \includegraphics[width=0.354\textwidth,trim= 0.85cm 0.785cm 13.4cm 2.15cm, clip=true]{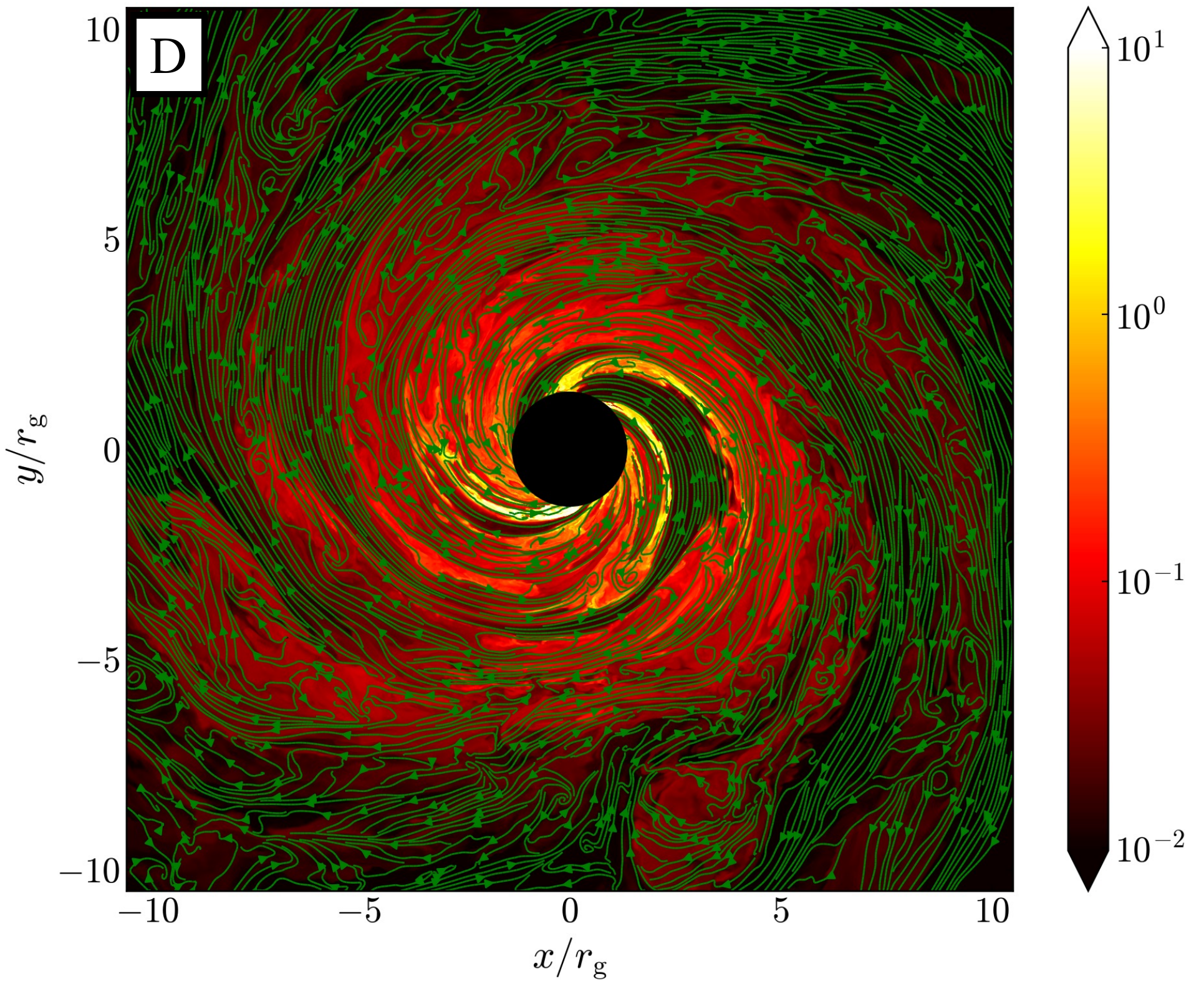}
    \includegraphics[width=0.3177\textwidth,trim= 2.8cm 0.785cm 13.4cm 2.15cm, clip=true]{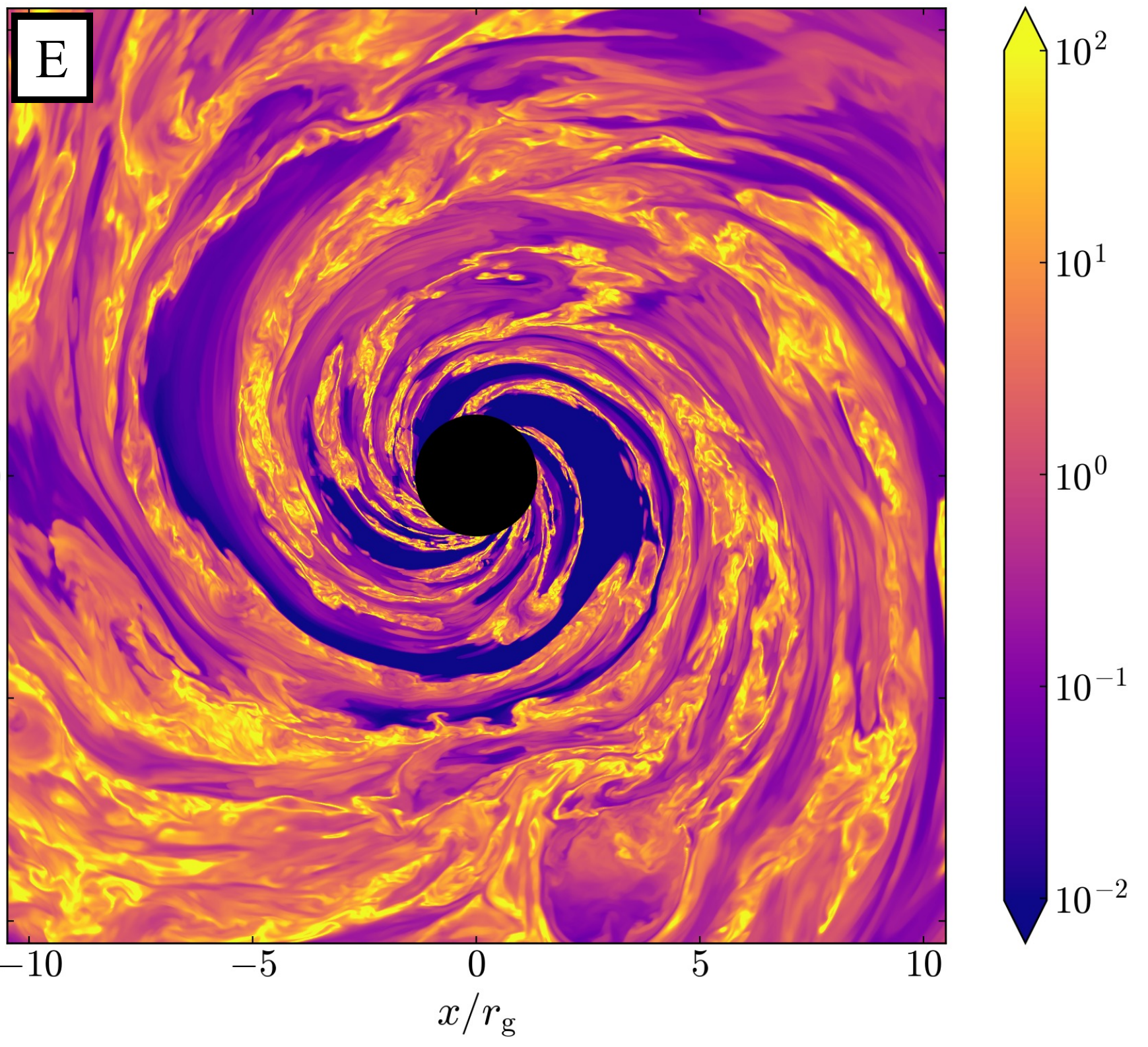}
    \includegraphics[width=0.3177\textwidth,trim= 2.8cm 0.785cm 13.4cm 2.15cm, clip=true]{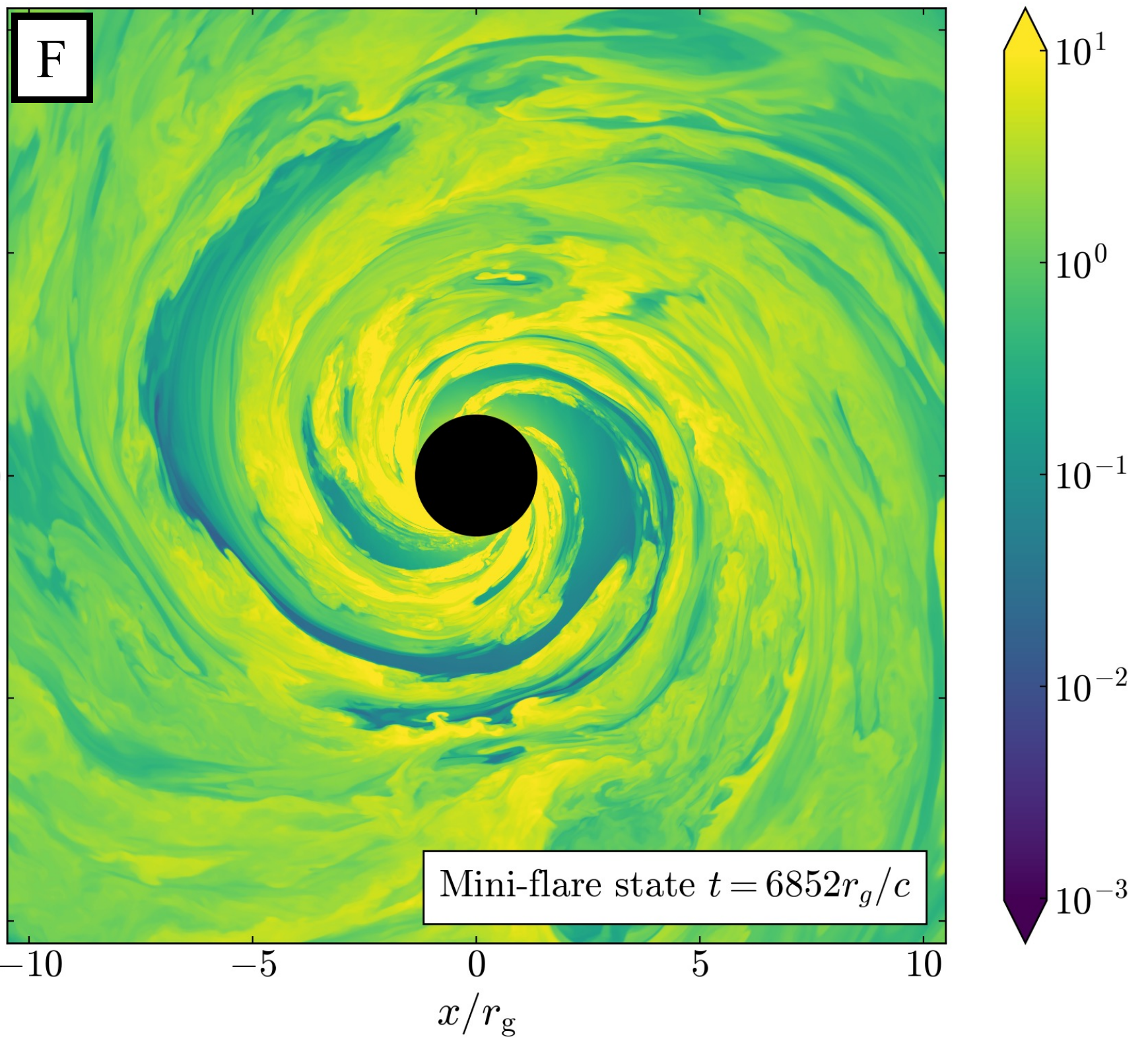}    
    \caption{
    {Smaller flux eruptions show shorter current sheets, potentially powering mini-flares that are not accompanied by a large-scale evacuation of the accretion disk.}
    Meridional (top row) and equatorial (bottom row) cuts of temperature $T=p/\rho$ (left), plasma-$\beta$ (middle) and density $\rho$ (right) during the mini-flare at $t=6852 r_{\rm g}/c$. The magnetic flux is expelled through a smaller ($w \lesssim 3 r_{\rm g}$) current sheet, close to the horizon, in a short time $\ll 100 r_{\rm g}/c$. The accretion disk is not expelled over a large azimuthal angle, yet the flare is accompanied by a significant drop in mass accretion rate {(see Figure \ref{fig:recrate}G)} and clear gaps in the density (F). Multiple small current sheets are visible in the accretion disk at $x\geq 3 r_{\rm g}$ indicated by the high plasma-$\beta$ (B).}
    \label{fig:panelminiflare}
\end{figure*}

In Figure \ref{fig:panelminiflare} we show temperature (left column), plasma-$\beta$ (middle column), and rest-mass density (right column) in both the {meridional} (top row) and {equatorial} (bottom row) plane for the mini-flare in the extreme resolution run at $t\approx 6800 r_{\rm g}/c$. In this case, the accretion disk is not ejected far beyond $5 r_{\rm g}$, still creating a spiral density gap causing the mass accretion rate to drop significantly. Reconnection occurs in a shorter, $\lesssim 5 r_{\rm g}$ plasmoid-unstable current sheet, very close to the horizon, and this is also the main area that is heated to relativistic temperatures $T>1$. These mini-flares could potentially result in smaller very high energy {flares} and shorter variability time scales (\citealt{Hess2012}). 

\section{Radiative properties of the reconnection layer}
To probe the radiation emitted by accelerated particles in the reconnection layer a self-consistent radiative kinetic approach is necessary (\citealt{Hakobyan2019,Crinquand2020,Crinquand2020b}). {Motivated by the results in the previous Sections, we assume the flaring is associated with the formation of a transient macroscopic current sheet in a magnetospheric region near the event horizon without further relying on the GRMHD results.  We then use }the well-constrained parameters for M87$^{*}$ and Sgr A$^{*}$ to estimate the expected emission properties due to reconnection occurring in the radiative regime.

\subsection{M87$^{*}$ flares powered by radiative reconnection}
In our simulations, the current sheet is fed by plasma in the jet at the floor density with a magnetization $\sigma_{\rm max}=25$. In reality, the reconnection powering the flare close to the event horizon is fed by collisionless pair plasma from the jet with a rate of $v_{\rm rec}/c = 0.1$ (\citealt{Bransgrove2021}) at magnetization $\sigma_{\rm up} = B^2_{\rm up} / (4\pi n m_{\rm e} c^2) = 2U_{\rm B}/ (n m_{\rm e} c^2)$, where $n$ is the number density of electrons with mass $m_{\rm e}$, $B_{\rm up}$ is magnetic field strength upstream from the current sheet and magnetic energy density $U_{\rm B} = B^2_{\rm up}/8\pi$\footnote{\cite{Scepi2021} find a typical $\sigma_{\rm up}=100$ in the upstream, which is due to the floor $\sigma_{\rm max}=100$ that they set. However, for realistic funnel densities that are not limited by floors in GRMHD simulations, the magnetization parameter in the upstream, $\sigma_{\rm up}$, can be much higher.}. The plasma particles are impulsively accelerated by non-ideal electric fields at the X-points \citep{sironi2014}. When they encounter plasmoids, they experience strong synchrotron losses. To parametrize the effect of the radiation backreaction, we define the particle Lorentz factor $\gamma_{\rm rad}^{\rm sync}$ for which the radiation drag force is comparable to the force due to the accelerating electric field $E \sim B_{\rm up} v_{\rm rec} / c$ (\citealt{Uzdensky2011a}):
\begin{equation}
    2\sigma_{\rm T} U_{\rm B} (\gamma^{\rm sync}_{\rm rad})^2 = v_{\rm rec} e B_{\rm up} / c,
    \label{eq:1}
\end{equation}
where $\sigma_{\rm T} = (8\pi/3)r_{\rm e}^2$ is the Thomson cross-section, $r_{\rm e} = e^2 / (m_{\rm e} c^2)$ is the classical electron radius, and $e$ is the electron's charge. We then find $(\gamma_{\rm rad}^{\rm sync})^2 = 3 v_{\rm rec} B_{\rm cl} / (2 c B_{\rm up})$, where $B_{\rm cl} = m_{\rm e}^2 c^4/e^3 \simeq 6 \times 10^{15}$ G is the classical magnetic field. The global magnetic field strength at $5 r_{\rm g}$ is estimated to be $1-30$G (\citealt{EHTVII2021}), resulting in $5-150$G at the horizon, assuming a $1/r$ dependence (RBP20). We can compare this to the magnetic field strength in the jet, feeding the current sheet close to the event horizon of M87$^{*}$ by equating the observed limits on the total jet power $L_{\rm jet}\sim 10^{42} - 10^{44}$ erg/s (\citealt{Prieto2016}) to the Blandford-Znajek jet power $L_{\rm BZ} = \kappa \Omega^2_{\rm BH} \dot{\phi}^2_{\rm BH}/(4\pi c)$, where $\kappa \approx 0.044$ for a parabolic field geometry, $\Omega_{\rm BH} = ac / 2r_{\rm H} \simeq c/2r_{\rm g}$ is the black hole's angular frequency, $M\approx 6\cdot 10^9 M_{\odot}$ for M87$^{*}$, and $r_{\rm H} = r_{\rm g}(1+\sqrt{1-a^2})$ is the horizon radius (\citealt{BZ1977,Tchekhovskoy2011}), resulting in a range $B_{\rm horizon} \sim 20 - 200$ G at the horizon. By normalizing to a fiducial $B_{\rm up} = 100$ G in this range, we then obtain 
\begin{equation}
    \gamma_{\rm rad}^{\rm sync} \approx 3 \cdot 10^6 \left(\frac{B_{\rm up}}{100 {\rm G}}\right)^{-1/2}
    \label{eq:2}
\end{equation}
The magnetization $\sigma_{\rm up}$ sets the available magnetic energy per particle and determines the typical particle Lorentz factor, $\gamma \sim \sigma_{\rm up}$ {(which in GRMHD corresponds to the temperature of reconnection-heated fluid, whereas the bulk Lorentz factors of reconnection outflows scale as $\Gamma \sim \sqrt{\sigma_{\rm up}}$)}, for the acceleration at X-points, if cooling were negligible (\citealt{sironi2014,Guo_2014,Werner_2015}). We can rewrite $\sigma_{\rm up} = \omega_{\rm B} / (2\Omega_{\rm BH} \lambda)$, where we plugged in the nominal electron gyrofrequency $\omega_{\rm B} = e B_{\rm up} / (m_{\rm e} c)$ and defined plasma density with respect to the Goldreich-Julian density, $n = \lambda n_{\rm GJ} = \lambda \Omega_{\rm BH} B_{\rm up} / (2\pi c e)$, where $\lambda$ is the multiplicity of the pair cascade in the charge-starved gap in the funnel region $\lambda \lesssim 10^3$ (\citealt{chen2019,Crinquand2020}) or of collisions of photons from the disk, if that process is more efficient (\citealt{Moscibrodzka2011}). The typical ratio between the electron gyrofrequency and the angular frequency of M87$^{*}$ is $\omega_{\rm B} / \Omega_{\rm BH} \sim 10^{14} ({M}/{6 \cdot 10^9 M_{\odot}}) ({B_{\rm up}}/{100 {\rm G}})$, such that $\sigma_{\rm up} \sim 10^{14} ({M}/{6 \cdot 10^9 M_{\odot}}) ({B_{\rm up}}/{100 {\rm G}}) / 2\lambda$. For these parameters, $\gamma^{\rm sync}_{\rm rad} \ll \sigma_{\rm up}$ such that leptons impulsively accelerated at X-points are quickly cooled in plasmoids \citep{Hakobyan2019}. Thus, the reconnection occurs in the radiative regime \citep{uzdensky2011}.


To understand the radiative efficiency of reconnection, we determine the magnetic {\it compactness} $\ell_{\rm B} = U_{\rm B} \sigma_{\rm T} w / (m_{\rm e} c^2)$ \citep{Beloborodov2017}. Using Eq.~\ref{eq:1} and the $\omega_{\rm B} / \Omega_{BH}$ relation, we can rewrite $\ell_{\rm B} = v_{\rm rec} w \omega_{\rm B} / (c^2 (\gamma^{\rm sync}_{\rm rad})^2)$ and obtain
\begin{equation}
    \ell_{\rm B} \sim 1 \left(\frac{w}{1 r_{\rm g}}\right)\left(\frac{M}{6 \cdot 10^9 M_{\odot}}\right)\left(\frac{B_{\rm up}}{100 {\rm G}}\right)^2,
    \label{eq:3}
\end{equation}
so $\ell_{\rm B} \sim 1$, suggesting potentially efficient pair production, but negligible annihilation \citep{Beloborodov2017}. In this regime the cooling time of accelerated particles, $c t_{\rm sync} / w \sim 1/(\ell_{\rm B} \gamma)$, is much shorter than the light-crossing time of the current sheet. Inverse Compton (IC) cooling of accelerated particles on the $\sim 10^{41}$ ${\rm erg/s}$ low-energy photons with energy density $U_{\rm rad}^{\rm soft} \sim 0.003$ ${\rm erg}$ ${\rm cm}^{-3}$ in the inner $10 r_{\rm g}$ results in $\gamma_{\rm rad}^{\rm IC} \sim \gamma_{\rm rad}^{\rm sync} \sqrt{U_{\rm B}/U_{\rm rad}^{\rm soft}} \sim 10^9$ \citep{Broderick2015,MWL2021}, which is well above $\gamma_{\rm rad}^{\rm sync}$. The jet's magnetic field reconnects with a rate of $0.1c$ in the collisionless radiative regime, after which all reconnected power is directly radiated such that the higher energy density of photons produced by accelerated particles, $U_{\rm rad}^{\rm rec} \sim 0.1 U_{\rm B}$ and hence $L_{\rm rad} \sim 0.1 L_{\rm jet}$ (\citealt{Beloborodov2017,Bransgrove2021}), can lead to very efficient IC cooling. The exact result depends on the spectral shape and reduction by Klein-Nishina effects.



The peak of the synchrotron radiation spectrum is expected to be at the synchrotron burnoff limit $\mathcal{E}_{\rm ph} \sim (\gamma_{\rm rad}^{\rm sync})^2 \hbar \omega_{\rm B} \sim 200 {\rm MeV}$ (\citealt{Uzdensky2011a}), which is independent of the magnetic field strength. The highest energy photons will be produced by IC scattering. Conservatively, the characteristic photon energy that can be produced is ${\rm max}(\mathcal{E}_{\rm ph}) = m_{\rm e} c^2 \gamma_{\rm rad}^{\rm sync} \sim 0.511 {\rm MeV} \cdot \gamma_{\rm rad}^{\rm sync} \sim$ few {\rm TeV}. Additionally, particles can be accelerated beyond $\gamma > \gamma_{\rm rad}^{\rm sync}$ because synchrotron cooling is suppressed in X-points (\citealt{Uzdensky2011a,Cerutti2014}).
For photons with energy above the electron rest-mass energy $m_{\rm e}c^2=0.5 {\rm MeV}$, $e^{\pm}$ pairs are created if there are enough photon-photon collisions with  seed photons with low energy $\mathcal{E_{\rm s}} \sim (m_{\rm e} c^2)^2 / \mathcal{E}_{\rm ph}$.  High-energy photons of energy $\mathcal{E}_{\rm ph,TeV}$ produced in the magnetospheric region around the current sheet will interact most efficiently with seed photons of energy $\mathcal{E}_{\rm s} \sim (1 {\rm TeV} / \mathcal{E}_{\rm ph,TeV})$ ${\rm eV}$. 
Given the uncertainties about the density of a $1 {\rm eV}$ photon field near the event horizon during the flaring state, the escape of TeV photons from the region is an open question \citep{Levinson2011,MWL2021}. Conservatively, if $\sim 1 \%$ of the reconnection dissipated power $U_{\rm rad} \sim 0.1 U_{\rm B}$, $L_{\rm rad} \sim 0.1 L_{\rm jet} \sim 10^{41} - 10^{43}$, is emitted in very high-energy photons, a $\gamma$-ray flux of $10^{39} - 10^{41}$ erg/s can be emitted as a flare.

Our {extreme resolution} GRMHD simulation shows transient flaring periods where the mass accretion rate drops (and, thus, luminosity of seed  photons) significantly, by a factor $\sim 5-10$, resulting in large low density regions, such that opacity constraints for the escape of $\gamma$-ray photons from the equatorial current sheet are less strict than during a quiescent state.
The decrease of the mass accretion rate and the local soft photon field can also create favorable conditions for the activation of pair discharges on the jet's magnetic field lines and the potential escape of TeV emission from spark gaps, if the opacity becomes prohibitive during the quiescent state (\citealt{Levinson2011,Crinquand2020}).
The flaring state is distinctively different from the quiescent state observed by \cite{EHTpaper1}, implying that observations during a mass accretion rate drop/flare may result in different $230 {\rm GHz}$ images (Chatterjee et al., in prep.).
The magnetic flux decay and mass-accretion drop lasts for a period of $\sim 100 r_{\rm g}/c$ $\sim$ 1 month for M87$^{*}$, which is longer than the typical observed $\sim 1-3$ day TeV flux rise and decay timescale (\citealt{Hess2012}). However, in a collisionless plasma, the magnetic flux decay period is typically $\sim 3-10$ times shorter due to the faster reconnection rate of $v_{\rm rec} \approx 0.1c$ (\citealt{Bransgrove2021}) compared to $v_{\rm rec} \approx 0.01c$ in GRMHD models \footnote{{Note that the higher reconnection rate in collisionless models is caused by kinetic plasma effects, e.g., gradients of the anisotropic pressure tensor of electrons and positrons in pair plasma (\citealt{Bessho2005}), and is unrelated to the increased reconnection rate due to large numerical diffusion in low resolution GRMHD models.}}, resulting in a flare timescale of $\sim$ few days. 
We find that pair production in the current sheet can efficiently mass load the jet with electrons and positrons with energies $\gamma \sim 1-1000$, that can emit synchrotron photons with energies ranging from the radio to optical wavelengths (see {Appendix A}).

\subsection{Sgr A$^{*}$ flares powered by radiative reconnection}
Sgr A$^{*}$ shows daily near-infrared and X-ray flares from the inner $10 r_{\rm g}$, on average every 6 and 12 hours, lasting for 30-80 minutes, respectively (\citealt{baganoff2001,eckart2006,Gravity2018,Witzel2020,Murchikova2021}). The flare periods in our simulation last for $\sim 100 r_{\rm g}/c \sim 30$ minutes, and the subsequent quiescent period for $\sim 2000 r_{\rm g}/c \sim 10$ hours for Sgr A$^{*}$. The resulting hot spot orbits for $\sim 500 r_{\rm g}/c \sim 150$ minutes in the inner $20 r_{\rm g}$ until it diffuses due to mixing instabilities. The magnetic field strength in quiescence is well constrained in the range of $10-50$ G in the inner $10 r_{\rm g}$ for Sgr A$^{*}$ with black hole mass $4 \cdot 10^6 M_{\odot}$ (\citealt{Dodds_Eden2009}). Using Eq. \ref{eq:2}, this results in $\gamma_{\rm rad}^{\rm sync} \approx 9 \cdot 10^6 (B_{\rm up}/10 {\rm G})^{-1/2}$, limiting the energy of accelerated particles by synchrotron cooling for a typical magnetization $\sigma_{\rm up} \sim 10^{10} (M/4 \cdot 10^6 M_{\odot}) (B_{\rm up}/10 {\rm G}) / 2\lambda \gg \gamma_{\rm rad}^{\rm sync}$.
Using Eq. \ref{eq:3}, the compactness is $\ell_{\rm B} \sim 10^{-5} (w/1 r_{\rm g})(M/4 \cdot 10^6 M_{\odot})(B_{\rm up}/10 {\rm G})^2$.
Synchrotron photons emitted by the particles accelerated to the highest energies in the reconnection layer, up to $\gamma_{rad}^{\rm sync} \sim 10^7$, should extend in the hard X-ray range. The energy of particles accumulated in the orbiting hot spot will be constrained by the synchrotron cooling time which has to be larger than the lightcrossing time of the current sheet, $c t_{\rm sync} / w \sim 1/(\ell_{\rm B} \gamma) \geq 1$, or $\gamma \lesssim 1 / \ell_{\rm B} \sim \gamma_{\rm {cool}}=10^4$ for the hot spot at $\sim 10r_{\rm g}$. These particles are likely to emit in the (near-)infrared range, $(\gamma_{\rm cool})^2 \hbar \omega_{\rm B} \sim 1 {\rm eV}(B/10 {\rm G})$. Thus, reconnection near the event horizon can power a multi-wavelength flare solely by synchrotron emission from reconnection-accelerated particles. Mini-flares are a potentially viable route to produce only near-infrared emission without strong enough X-rays to be detected as flares, as they do not produce a long-lasting extended current sheet, which would be the source of highest energy particles. {Mini-flares could be distinguishable from large flares with concurrent multi-wavelength observations of Sgr A$^{*}$ (e.g., \citealt{ponti2017})}. The characteristic power of the X-ray emission can be estimated from the total dissipated power in reconnection, $\sim 0.1 L_{\rm BZ}\sim 10^{35} (B_{\rm horizon}/10{\rm G})^2$ erg/s. Thus, reconnection in the magnetospheric current sheet provides enough energy to power the observed X-ray flares from Sgr A$^{*}$ with typical luminosities in a range $10^{34}-10^{35}$ erg/s \citep{Neilsen2015}.



%
%
%
%

\section{Conclusions}
By conducting extreme resolution 3D GRMHD simulations we have shown that during periods of magnetic flux decay at the horizon, MAD flows form transient and non-axisymmetric magnetospheres that possess special qualities revealed only at such high resolutions. Namely, these eruptions lead to a substantial, order-of-magnitude drop in the mass accretion rate and the formation of a thin equatorial current sheet that extends from the horizon out to $\sim 5-10 r_{g}$  into the disk and separates the two polar jets. This current sheet is filled with the electron-positron plasma from the jets and reconnects in the {plasmoid-mediated} regime. The formation of plasmoids is revealed here for the first time in 3D thanks to the unusually high resolutions achieved in this work, $N_r \times N_\theta \times N_\phi =  5376\times2304\times2304$. Reconnection-heated to relativistic temperatures, the plasma in the current sheet escapes the black hole's gravitational pull through the exhaust of the reconnection layer: this injects magnetic flux tubes filled with the low-density pair plasma into the accretion disk, and hot plasma along the jet-disk boundary. This reconnection-heated plasma can produce a multiwavelength flare. 
Hot flux tubes orbit in the accretion disk and can remain coherent for one to a few orbital periods. The time scales of the flare are directly governed by the reconnection rate in the equatorial current sheet. We have shown that this rate \emph{decreases} with increasing numerical resolution until the critical resolution beyond which it reaches the \emph{universal {converged} value} that no longer changes when the resolution is increased any further.  
Importantly, only at such high resolutions, the structure of the current sheet -- X-points and plasmoids -- are {resolved for the first time with our extreme resolution 3D GRMHD simulations.}

The universal reconnection rate directly sets the magnetic flux decay rate at the horizon. Other studies have related flux decay at the horizon with flares (\citealt{Ball2018b,dexter2020sgr,Chashkina2021,Scepi2021}) or observed orbiting flux tubes in retrograde disks (those rotating in the opposite sense to their black hole; \citealt{Porth2020flares}). However, due to limited numerical resolution they did do not capture {plasmoid-mediated} reconnection as the power source and did not identify a direct link between the magnetic flux decay at the event horizon and its origin in reconnection in the equatorial magnetospheric current.

We note that the trigger behind such large flux eruption events is still not understood. Large flares occur when the accretion is governed by large, low azimuthal mode-number spiral RTI modes. It is as of yet unclear why the accretion state switches from a large spectrum of RTI modes in quiescence to a single azimuthal spiral RTI mode during the flare.

{In reality, t}he reconnection powering the flare is fed by highly magnetized pair plasma that eventually ends up in the hot flux tube, buoyantly rising in and mixing with the electron-ion plasma that makes up the accretion disk. {Additionally, matter originating from the jet that enters into the equatorial current sheet and gets heated by reconnection, can travel along the jet sheath for large distances, during and shortly after a flux eruption}. 
{The temperature of this reconnection-heated matter is proportional to the magnetization in the jet, which in GRMHD simulations is set by the density floor. Therefore, the temperature in the parts of the jet sheath that are causally connected to the exhaust of the reconnection layer cannot be used during a flare period to determine its emission properties. The main uncertainty is the electron temperature, that is unknown in GRMHD simulations.} Commonly used parametrized relations connecting the temperatures of ions and electrons based on local plasma-$\beta$ or $\sigma$ values in the accretion flow \citep[e.g.,][and references therein]{Moscibrodzka2016, Davelaar2019, EHTpaper5, Chatterjee2020b, dexter2020sgr, Yoon2020} or two-temperature GRMHD approaches (\citealt{Ressler2015,Chael2019}) therefore cannot describe the non-thermal emission from these events which involves reconnection in high-$\sigma$ collisionless pair plasma regime, the transport and cooling of non-thermal lepton distributions, as well as efficient pair production.

We note that while the reconnection rate in the equatorial current sheet is converged in GRMHD at the extremely high numerical resolutions used in this work, it converges to $v_{\rm rec}/v_{\rm A} \sim 0.01 $, which is an order of magnitude lower than the converged value of $\sim 0.1$ in kinetic simulations \citep{Bransgrove2021}. This difference comes from GRMHD simulations being unaware of the collisionless plasma microphysics, which is important at scales where reconnection happens, i.e., electron skin depth. Incorporating non-ideal effects beyond scalar resistivity (e.g., \citealt{ripperda2019b}) into GRMHD simulations, such as electron inertia and anisotropic electron pressure tensor effects in the Ohm’s law {\citep{most2021}}, holds promise of matching the (collisional) GRMHD and collisionless reconnection rates \citep{NG2020}. {General relativistic (radiative)} kinetic simulations (e.g., \citealt{parfrey2019,Crinquand2020b,Crinquand2020}) are crucial for probing the non-thermal effects and the impact of the higher reconnection rate in collisionless plasma on the flare properties. 

In upcoming work we will investigate the radiative properties of the flares, and the consequences for the image variability as observed by the Event Horizon Telescope (Chatterjee et al., in prep.). {During large flux eruptions, the accretion disk is ejected over a large fraction of the azimuthal angle. The very hot current sheet will then emit non-thermal emission powered by reconnection in the inner $10 r_{\rm g}$, that may not be in the radio band. This may result in an observable dimming of the radio image, potentially distinguishable with concurrent Event Horizon Telescope and multi-wavelength flare observations for Sgr A$^{*}$. Flares are less frequent for M87$^{*}$ and, hence, observing potential dimming requires much longer Event Horizon Telescope observation windows, or several observations for separated periods.}

{In this work we have for the first time reached a numerical Lundquist number above the plasmoid instability threshold for the largest current sheets close to the event horizon in MAD flows with 3D GRMHD simulations. The formation of these macroscopic plasmoid-unstable current sheets is similar in 2D resistive GRMHD simulations with a resolved explicit resistivity (RBP20). We robustly find that reconnection in the largest current sheets in MAD flows can act as the powering mechanism for large, bright, rapid flares originating from near the event horizon. MHD turbulence is known to intermittently form plasmoid-unstable current sheets at smaller scales (\citealt{Zhdankin_2013,Zhdankin_2017,dong2018,Chernoglazov2021}) that are not resolved in our simulations and that may substantially modify the turbulent cascade and dissipation at even higher Lundquist numbers $S \gtrsim 10^6$ (\citealt{boldyrev2017,Comisso_2018}), potentially heating the accretion disk. Additionally, the ideal GRMHD approach taken here does not describe the dynamics of non-ideal electric fields and resistive dissipation, and relies on a numerical resistivity that is only controlled by numerical resolution. To analyze the formation of non-ideal electric fields and probe heating through turbulent reconnection in the accretion disk on smaller scales, even higher resolution resistive GRMHD simulations are required (RBP20; \citealt{Chernoglazov2021}).}

The robust formation of a macroscopic plasmoid-unstable current sheet close the event horizon that can heat and accelerate plasma, and eject flux tubes as low density hot spots into an orbiting disk in our extreme resolution GRMHD simulation, suggests that  flux eruptions powered by magnetic reconnection are a widespread phenomenon that can potentially explain observations of bright, rapid TeV flares from M87$^{*}$ and flaring hot spots from Sgr A$^{*}$.

\section*{Acknowledgements}
We would like to thank Ashley Bransgrove, Alexander Chernoglazov, Luca Comisso, Doosoo Yoon, Hayk Hakobyan, Gabriele Ponti, Benjamin Crinquand, Elias Most, Amir Levinson and Yuri Levin for useful discussions. B.R. and M.L. contributed equally to this work. This research was enabled by support provided by grant no. NSF PHY-1125915 along with a INCITE program award PHY129, using resources from the Oak Ridge Leadership Computing Facility, Summit, which is a US Department of Energy office of Science User Facility supported under contract DE-AC05- 00OR22725, as well as Calcul Quebec (http://www.calculquebec.ca) and Compute Canada (http://www.computecanada.ca). The computational resources and services used in this work were partially provided by facilities supported by the Scientific Computing Core at the Flatiron Institute, a division of the Simons Foundation. This research is part of the Frontera (\citealt{Frontera}) computing project at the Texas Advanced Computing Center (LRAC-AST20008). Frontera is made possible by National Science Foundation award OAC-1818253.
B.R. is supported by a Joint Princeton/Flatiron Postdoctoral Fellowship. M.L. was supported by John Harvard Distinguished Science Fellowship and ITC
Fellowship. K.C. is supported by a Black Hole Initiative Fellowship at Harvard University, which is funded by grants from the Gordon and Betty Moore Foundation, John Templeton Foundation and the Black Hole PIRE program (NSF grant OISE-1743747). The opinions expressed in this publication are those of the authors and do not necessarily reflect the views of the Moore or Templeton Foundations. G.M. is supported by a Netherlands Research School for Astronomy (NOVA), Virtual Institute of Accretion (VIA) postdoctoral fellowship. A.P. acknowledges support by the National Science Foundation under Grants No. AST-1910248 and PHY-2010145. Research at the Flatiron Institute is supported by the Simons Foundation. K.C. and S.M. are thankful for support by Dutch Research Council (NWO) VICI award, grant Nr. 639.043.513. A.T. acknowledges support by Northwestern University
and by the National Science Foundation grants AST-1815304, AST-1911080. Z.Y. is supported by a UK Research \& Innovation (UKRI) Stephen Hawking Fellowship.

\bibliography{sample63,mylib3}{}
\bibliographystyle{aasjournal}

\clearpage

\section*{{Appendix A.} Mass loading of the jet by pair production in the reconnection layer}
Pair production in the current sheet near M87$^{*}$ can significantly contribute to the mass loading of the jet. The optical depth for photons of energy $\mathcal{E}_{\rm ph}$ is 
$\tau_{\gamma\gamma} \sim 0.1 \sigma_{\rm T} U_{\rm rad}^{\rm s} w /\mathcal{E_{\rm s}}$, where $U_{\rm rad}^{\rm s}$ is the energy density of photons at energy $\mathcal{E_{\rm s}} \sim (m_{\rm e} c^2)^2 / \mathcal{E}_{\rm ph}$ such that the photon-photon pair production opacity is maximal for $\mathcal{E_{\rm s}} \mathcal{E_{\rm ph}} \sim (m_{\rm e}c^2)^2$. Most of the reconnected power, $L_{\rm rad} \sim 0.1 L_{\rm jet}$, is radiated around the burnoff limit, $\sim 200 {\rm MeV}$, and the peak can be quite broad \citep{Hakobyan2019}. Estimating $U_{\rm rad}^{\rm s} \sim L_{\rm rad}/4\pi w^2 c$, we get $\tau_{\gamma\gamma} \sim 0.1 \sigma_{\rm T} L_{\rm rad} / (4\pi c w\mathcal{E}_{\rm s})$. For $w \sim r_{\rm g}$, and for typical photon energies $\mathcal{E}_{\rm ph} \sim \mathcal{E}_{\rm s} \sim {\rm MeV}$, we find $\tau_{\gamma\gamma} \sim 10^{-2} \sigma_{\rm T} L_{\rm jet} / (4\pi c r_{\rm g}\mathcal{E}_{\rm s}) \sim 10^{-4}$ for a jet power $L_{\rm jet} \sim 10^{43}$ erg/s. The rate of pair creation is then $\dot{N}_{\rm pair} \sim \tau_{\gamma \gamma} \cdot (0.1 L_{\rm jet} / \mathcal{E}_{ph}) \sim 10^{44} s^{-1}$. We can compare this estimate to the Goldreich Julian number flux $\dot{N}_{\rm GJ} = I_{\rm GJ} / e$, where $I_{\rm GJ} \simeq (c L_{\rm jet})^{1/2}$ is the Goldreich Julian current, such that $\dot{N}_{\rm GJ} \sim (c L_{\rm {\rm jet}})^{1/2} / e \sim 10^{36} s^{-1}$. The resulting multiplicity $\lambda = \dot{N}_{\rm pair} / \dot{N}_{\rm GJ} \sim 10^8$ indicates that pairs produced in the current sheet with energies $\gamma \sim 1-1000$ can significantly contribute to mass loading of the jet, and emit synchrotron photons with energies $\sim \hbar \omega_{\rm B} \gamma^2 \sim 10^4 {\rm GHz} ({\gamma}/{400})^2$, ranging from the radio to optical wavelengths.

\section*{{Appendix B.} Influence of mass loading on plasma heating due to reconnection}
\begin{figure*}
    \centering
    \includegraphics[width=0.329\textwidth,trim=0.8cm 0.785cm 16.4cm 1.3cm, clip=true]{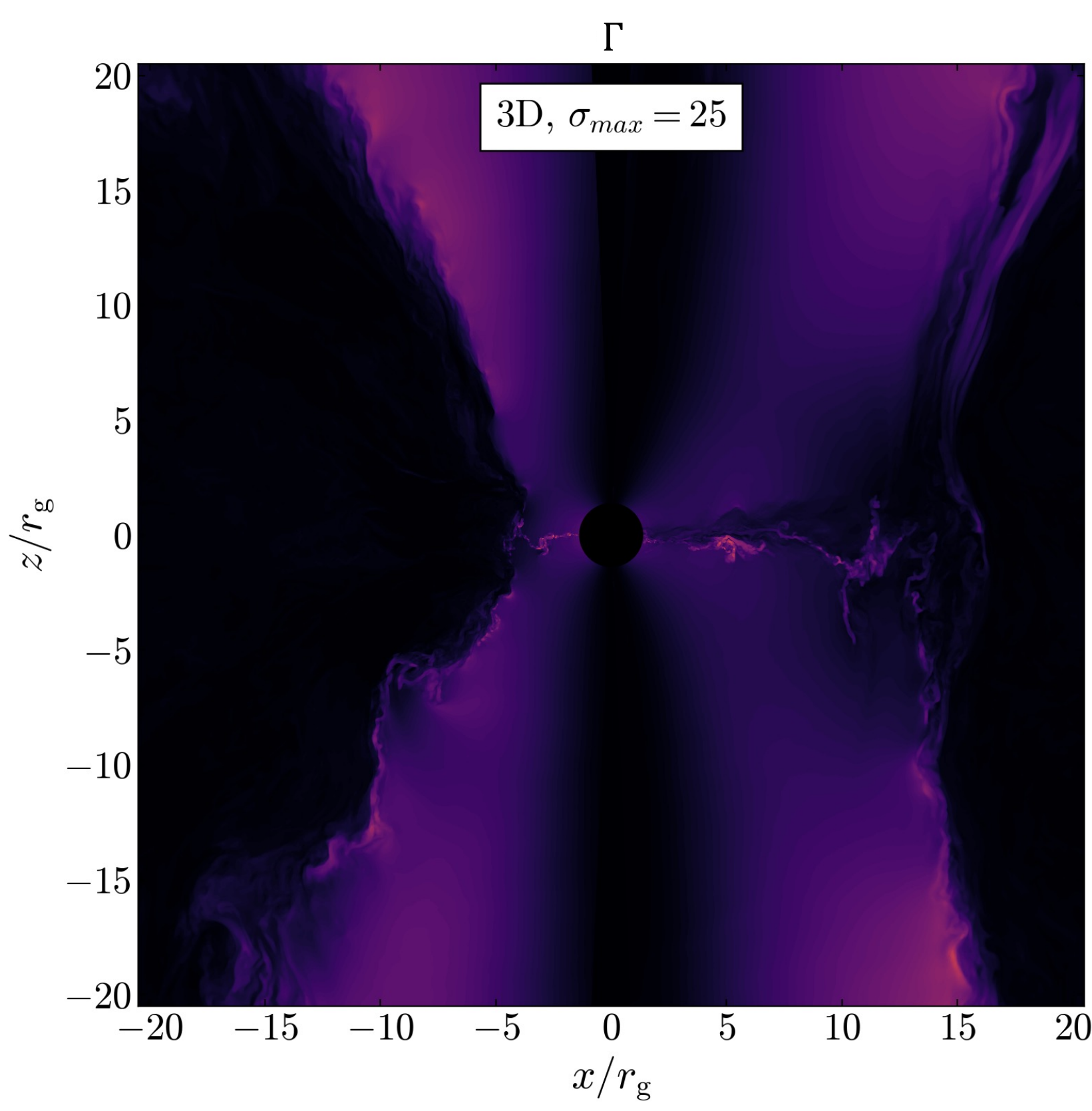}
    \includegraphics[width=0.29\textwidth,trim= 2.8cm 0.785cm 16.4cm 1.3cm, clip=true]{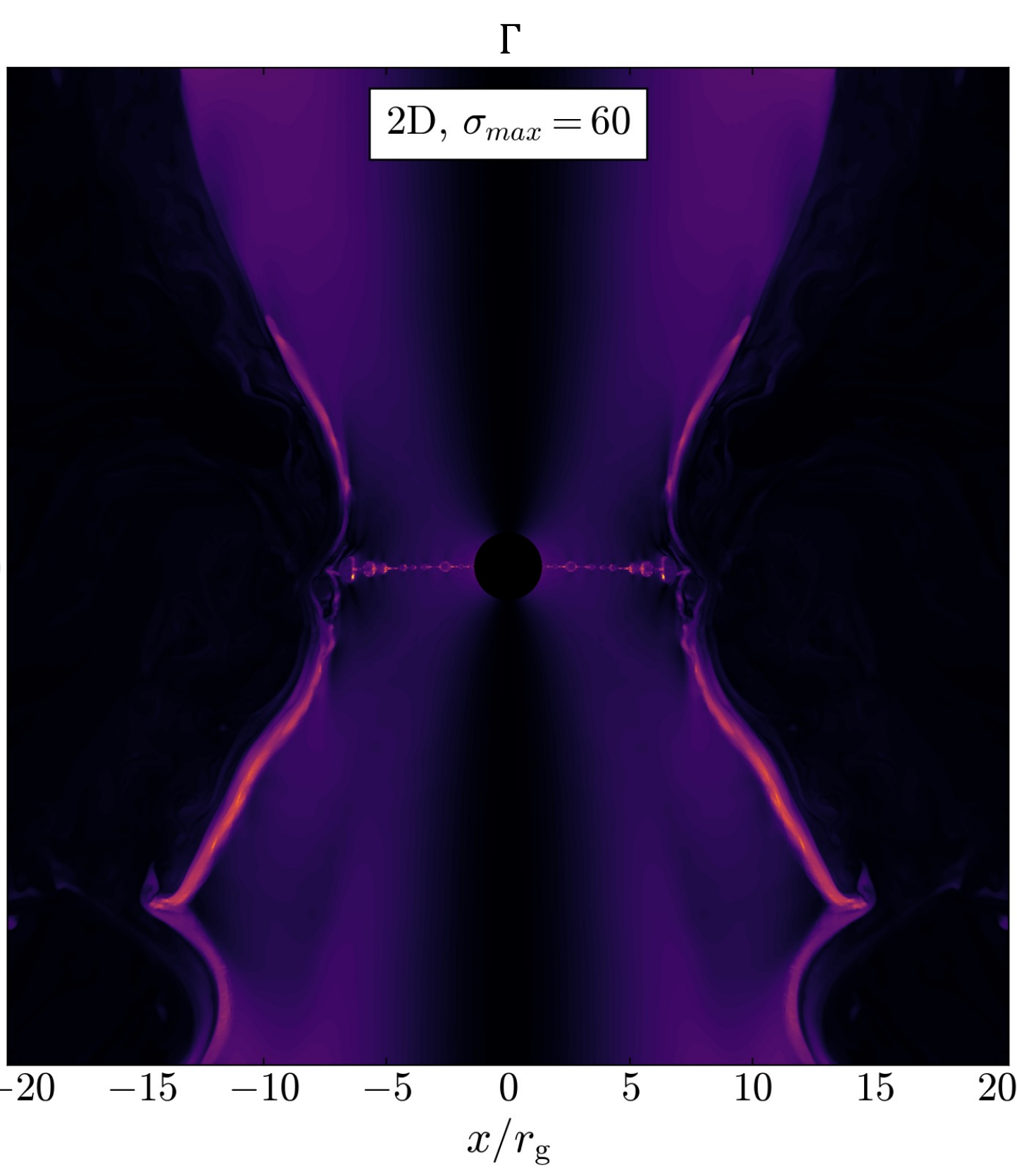} 
    \includegraphics[width=0.35\textwidth,trim= 2.8cm 0.785cm 13.4cm 1.3cm, clip=true]{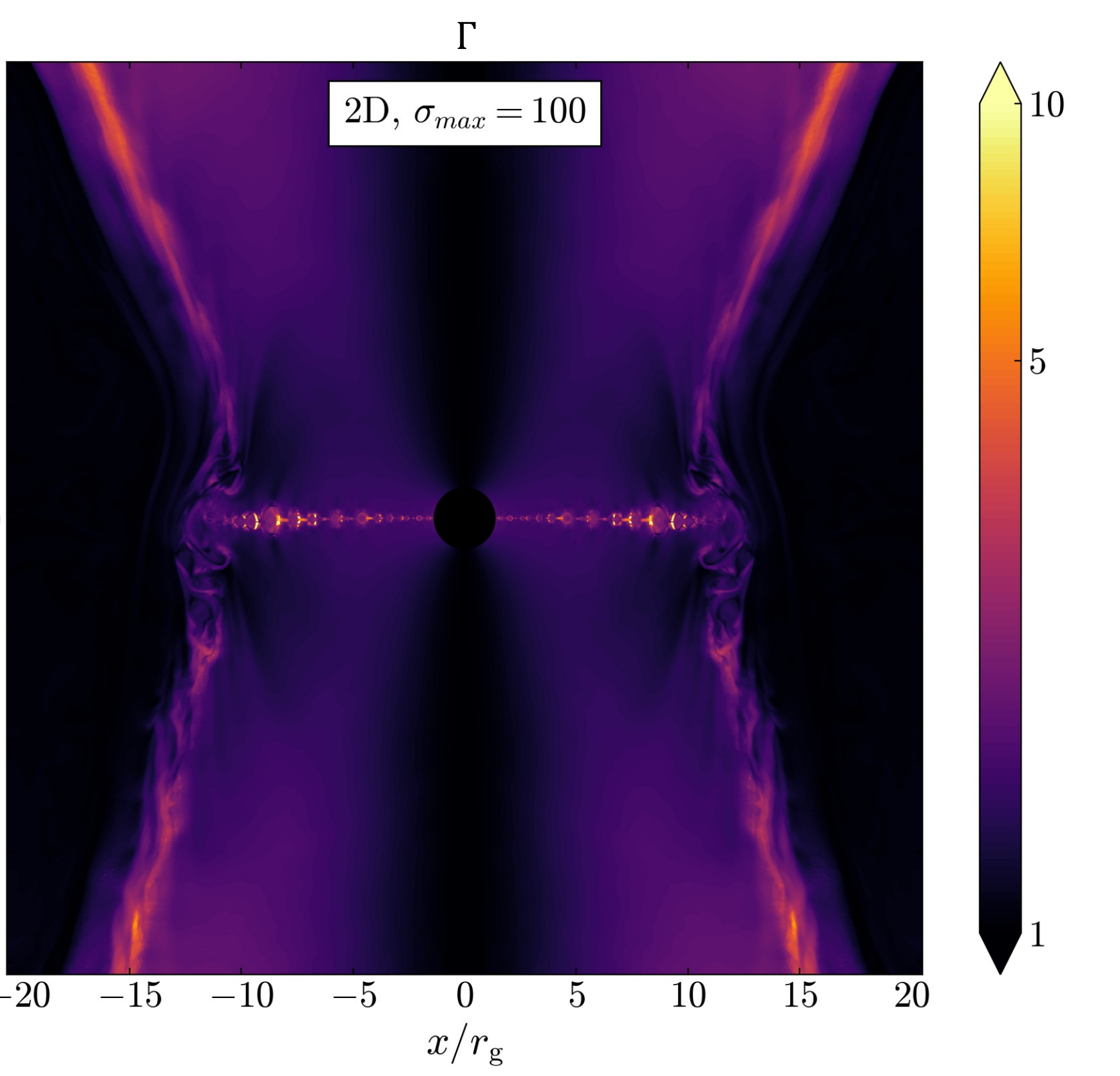}     
    
    \includegraphics[width=0.329\textwidth,trim= 1cm 1cm 7cm 1cm, clip=true]{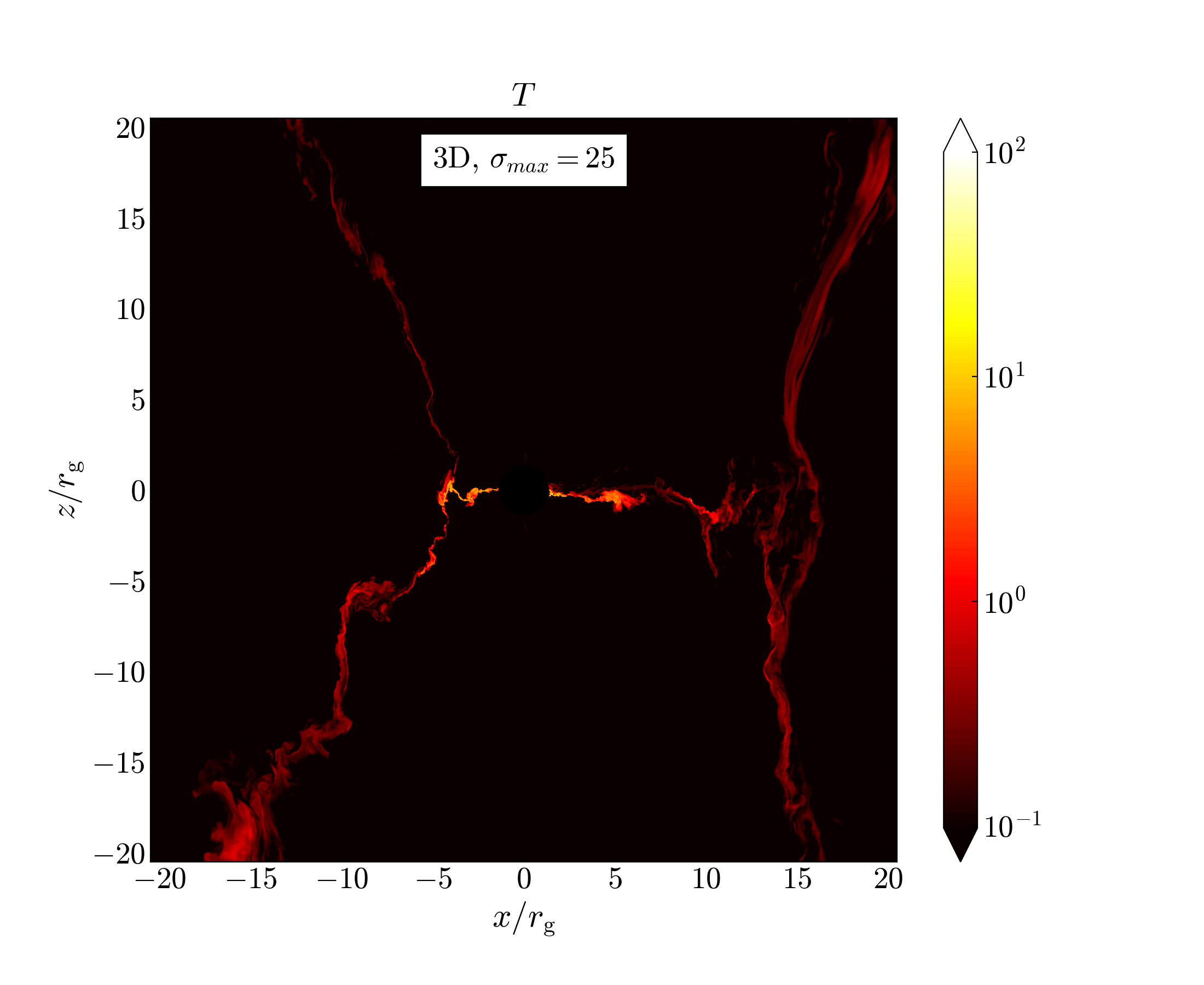} 
    \includegraphics[width=0.29\textwidth,trim= 3.35cm 1cm 7cm 1cm, clip=true]{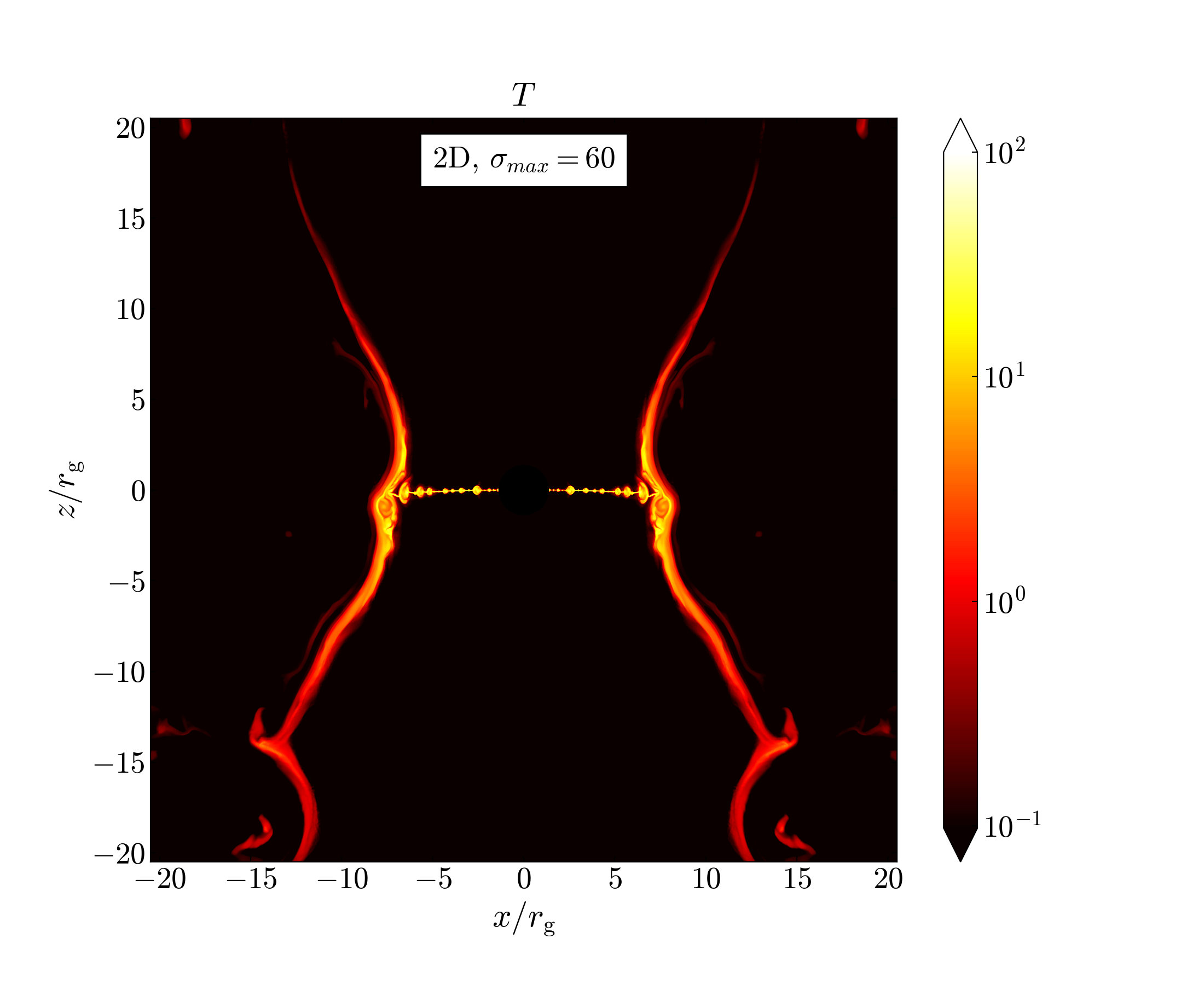} 
    \includegraphics[width=0.35\textwidth,trim= 3.35cm 1cm 3.35cm 1cm, clip=true]{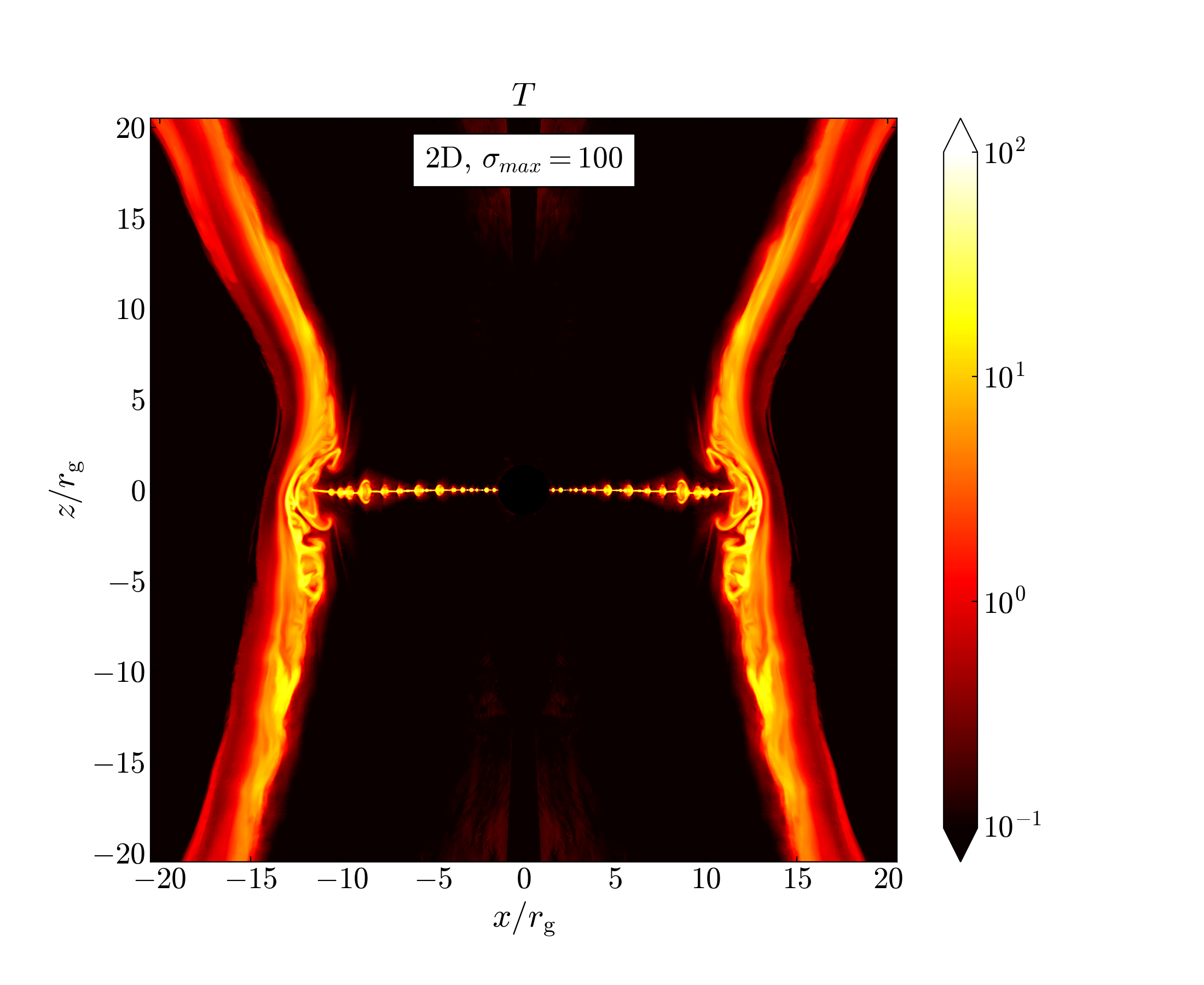} 
    \caption{Lorentz factor $\Gamma$ (top rows) and temperature $T=p/\rho$ for the extreme resolution 3D run with density floor $\sigma_{\rm max}=25$ (left), and two supplementary 2D runs with density floors $\sigma_{\rm max}=60$ (middle) and $\sigma_{\rm max}=100$ (right). Reconnection heats up the plasma in the equatorial current sheet to $T\sim \sigma_{\rm max}$ and $\Gamma \sim \sqrt{\sigma_{\rm max}}$. The hot reconnection exhaust heats up the jet sheath up to temperatures proportional to the magnetization in the jet. We observe limb-brightened jets up to at least $20 r_{\rm g}$.}
    \label{fig:gammacomparison}
\end{figure*}

We performed two additional 2D GRMHD simulations to show that reconnection heats the plasma to $T \sim \sigma_{\rm max}$ and $\Gamma \sim \sqrt{\sigma_{\rm max}}$, for the floor magnetizations in the jet $\sigma_{\rm max}=25, 60, 100$. Figure \ref{fig:gammacomparison} shows both the Lorentz factor $\gamma$ (top row) and temperature $T$ (bottom row) for the three values of $\sigma_{\rm max}$. The plasma in the current sheath is indeed heated to $T \approx \sigma_{\rm max}$ and $\Gamma \approx \sqrt{\sigma_{\rm max}}$ and the reconnection exhaust deposits hot plasma in the jet sheath up to large distances.

\section*{{Appendix C.} Resolution study of the reconnection rate and magnetic flux decay}
\begin{figure*}
    \centering
    \includegraphics[width=0.327\textwidth, trim= 0.21cm 10.5cm 14.7cm 1.15cm, clip=true]{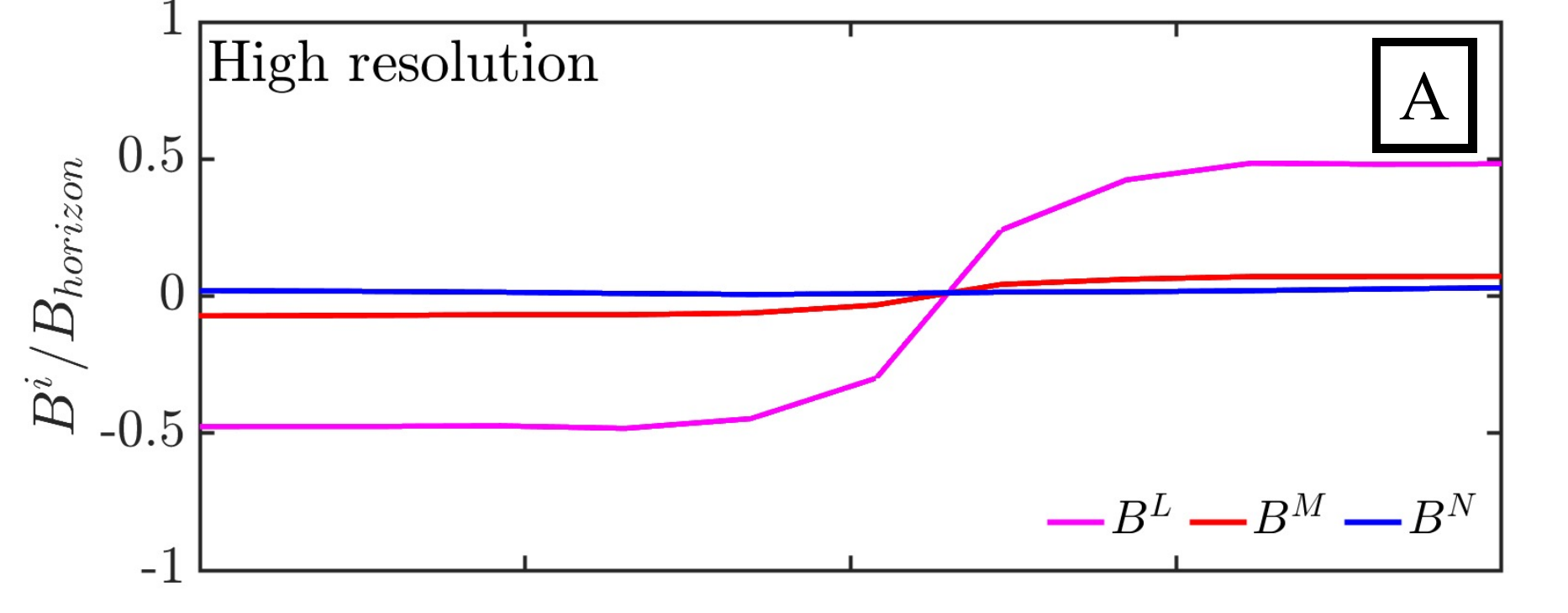}
    \includegraphics[width=0.327\textwidth, trim= 0.21cm 10.5cm 14.7cm 1.15cm, clip=true]{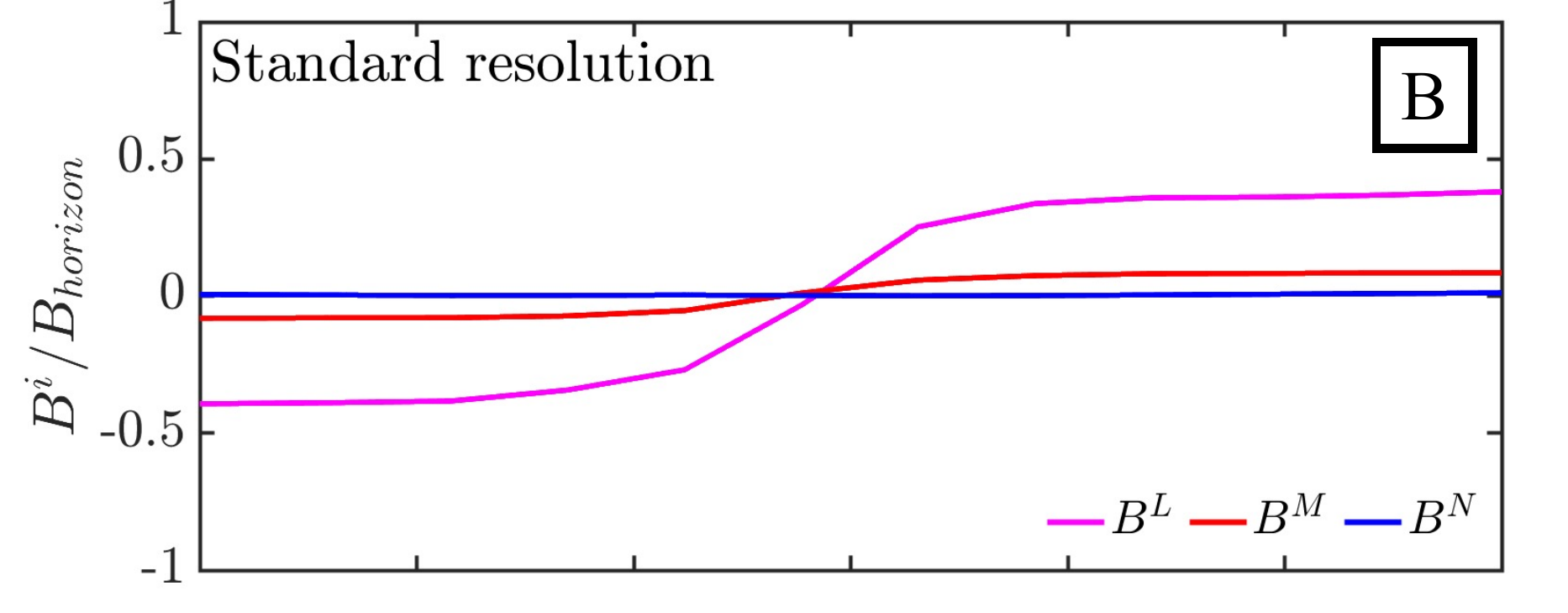} 
    \includegraphics[width=0.327\textwidth, trim= 0.21cm 10.5cm 14.7cm 1.15cm, clip=true]{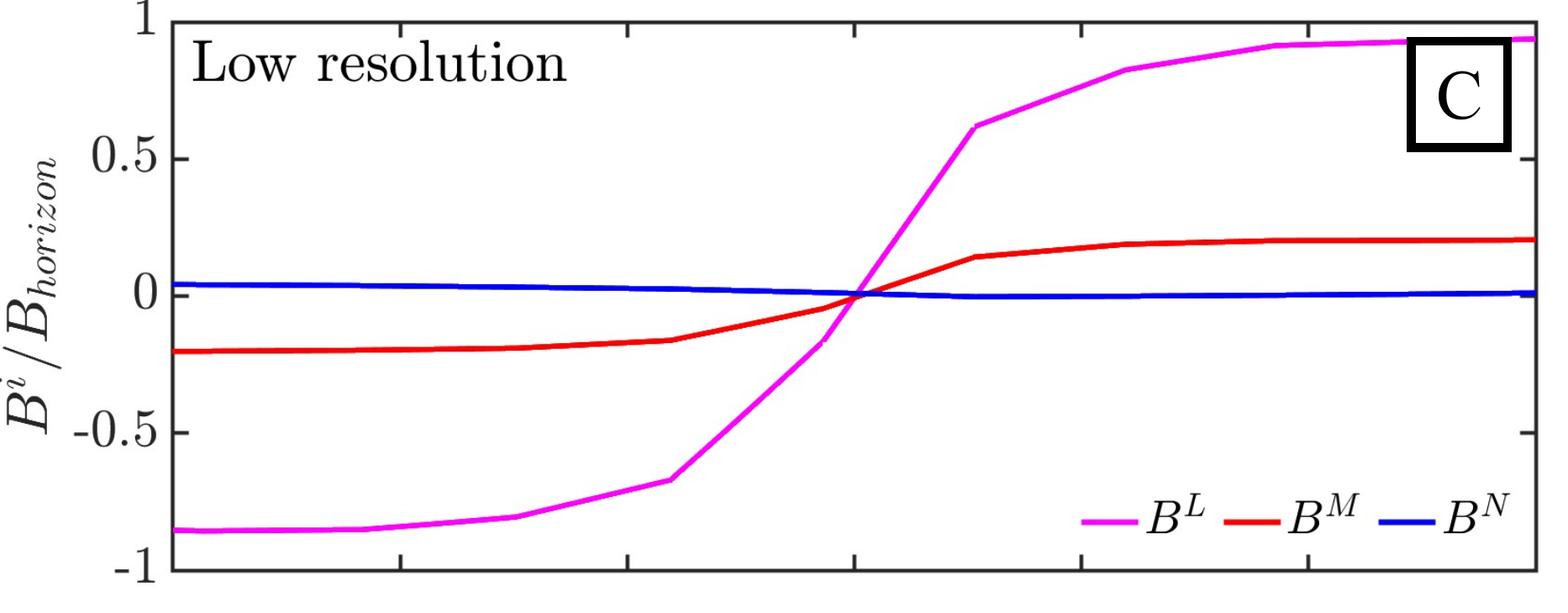} 
    
    \includegraphics[width=0.327\textwidth,  trim= 0.21cm 0.3cm 14.7cm 9.9cm, clip=true]{recrate_1024.pdf}
    \includegraphics[width=0.327\textwidth,  trim= 0.21cm 0.3cm 14.7cm 9.9cm, clip=true]{recrate_512.pdf}
    \includegraphics[width=0.327\textwidth,  trim= 0.21cm 0.3cm 14.7cm 9.9cm, clip=true]{recrate_256.pdf}
    \caption{Top row: Profiles of the magnetic field in minimum variance coordinates in a current sheet in the high (A), standard (B), and low (C) resolution runs, as in Figure \ref{fig:recrate}. Bottom row: Profile of the inflow speed into the current sheet, showing a reconnection rate of $0.01c$ for high resolution (D), and enhanced reconnection rates of $0.02c$ (standard resolution, E) and $0.025c$ (low resolution, F) as a result of a larger numerical resistivity and a broader current sheet (top panels).}
    \label{fig:recrate1024}
\end{figure*}

We analyze the effect of the resolution on the reconnection rate by showing the magnetic field components in minimum variance coordinates for the high (left panels), standard (middle panels) and low (right panels) resolution runs in Figure \ref{fig:recrate1024}, similar to Figure \ref{fig:recrate}.
One can see that the reconnection layer broadens for decreasing resolutions. This results in an increased reconnection rate, in accordance with an increasing numerical resistivity due to the decreasing resolution. 
In the high resolution run we detect plasmoids, and hence the numerical Lundquist number is still above the threshold $S_{\rm crit}=10^4$, confirmed by the measured reconnection rate of $v_{\rm rec} \approx 0.01c$. For the standard resolution run, the reconnection rate increases to $v_{\rm rec} \approx 0.02c$ and for the low resolution run $v_{\rm rec}\approx0.025c$, such that the Lundquist number is of the order $S=(v_{\rm rec}/c)^{-2} \sim \mathcal{O}(10^3)$. No plasmoids are detected in the low and standard resolution simulations. Note that the typical resolutions used in \cite{EHTpaper5,dexter2020sgr,Porth2020flares,Scepi2021} are below our standard resolution. 

\begin{figure*}
    \centering
    \includegraphics[width=0.5\textwidth, trim= 0.3cm 10.5cm 14.3cm 0.4cm, clip=true]{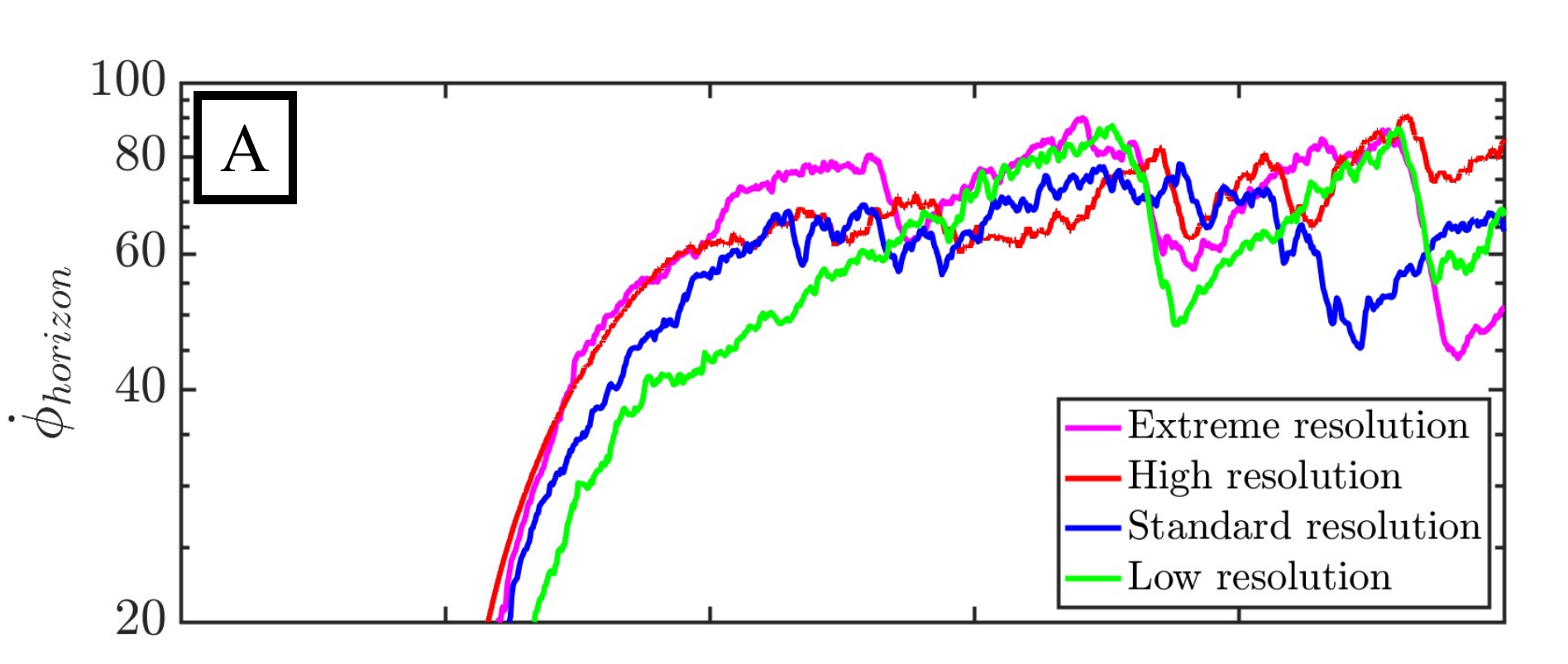}
    \includegraphics[width=0.485\textwidth, trim= 0.3cm 10.5cm 14.9cm 0.4cm, clip=true]{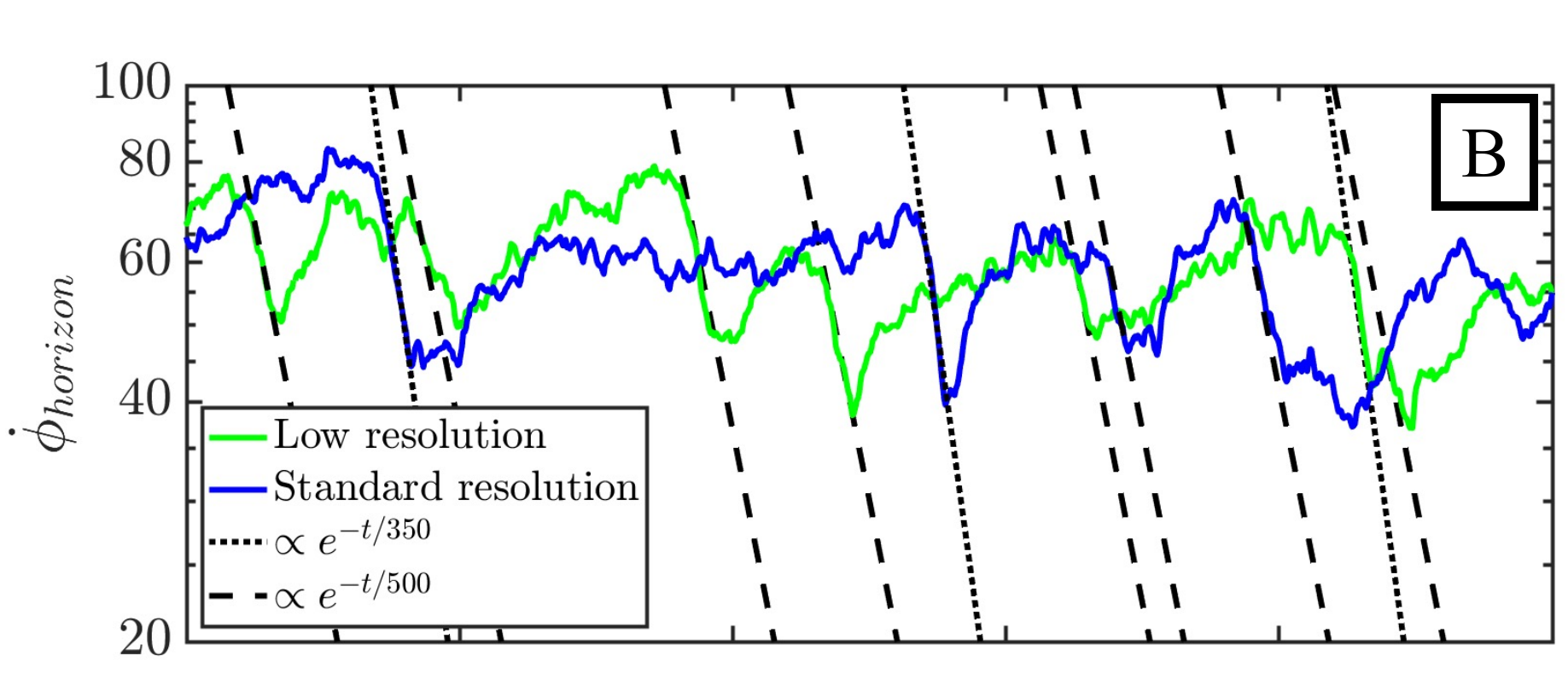}
  
    \includegraphics[width=0.5\textwidth, trim= 0.3cm 0.3cm 14.3cm 9.9cm, clip=true]{mdot_phidot_appendix2.pdf}  
    \includegraphics[width=0.485\textwidth, trim= 0.3cm 0.3cm 14.9cm 9.9cm, clip=true]{mdot_phidot_appendix.pdf}
    \caption{Magnetic flux on the horizon (top) and mass accretion rate at $5 r_{\rm g}$ (bottom) for low (green line), standard (blue line), high (red line), and extreme (magenta line) resolution. Large flares, accompanied by a significant drop of an order of magnitude in mass accretion rate, show a faster flux decay time $\propto e^{-t/350}$ than the flux decay at higher resolution $\propto e^{-t/500}$ due to the enhanced reconnection rate at lower resolutions. mini-flares, not accompanied by a significant flux drop, follow the flux decay rate $\propto e^{-t/500}$ because the current sheet is shorter and therefore better resolved due to the spherically logarithmic grid close to the event horizon.}
    \label{fig:mdot}
\end{figure*}

The increased reconnection rate enhances dissipation of magnetic energy and directly affects the flux decay rate that governs the flare time scale. {In Figure \ref{fig:mdot}B} we show this effect for the low (green line) and standard (blue line) resolution runs. For large flares, accompanied by a drop in mass accretion rate (\ref{fig:mdot}D) the flux decays at a rate $\propto e^{-t/350}$ (indicated by the dotted black lines, e.g., at $t\approx 11300 r_{\rm g}/c$, $15500 r_{\rm g}/c$ and $18500 r_{\rm g}/c$) instead of the converged $\propto e^{-t/500}$. For mini-flares, the current sheet extends less far from the event horizon and is naturally captured by cells of smaller volume due to the spherically logarithmic grid. Additionally, in ideal GRMHD simulations (i.e., relying on numerical resistivity) the reconnection rate is a function of numerical resolution and, hence, of the radius due to the spherical grid. Reconnection close to the horizon, e.g., in a mini-flare, will therefore occur closer to the asymptotic value of $v_{\rm rec} \sim 0.01c$ than at larger radii.
The mini-flares are not accompanied by a significant drop in mass accretion rate, and show a decay rate $e^{-t/500}$ (indicated by black dashed lines) that is similar to the high and extreme resolution simulations. Mini-flares occur more often at low resolutions than at high resolutions due to the larger numerical viscosity; whereas large flares occur less frequently because the funnel region is not cleared out due to diffusion.

{The left panels show the magnetic flux (\ref{fig:recrate1024}A) and mass accretion rate (\ref{fig:recrate1024}C) for the initial $10000 r_{\rm g}/c$ for all resolutions, showing that the simulations are in the quasi-steady state of MAD accretion after $\sim 5000 r_{\rm g}/c$.}

\section*{{Appendix D.} Flares at low numerical resolution}
\begin{figure*}
    \centering
    \includegraphics[width=0.353\textwidth,trim= 0.85cm 2.3cm 13.4cm 1.3cm, clip=true]{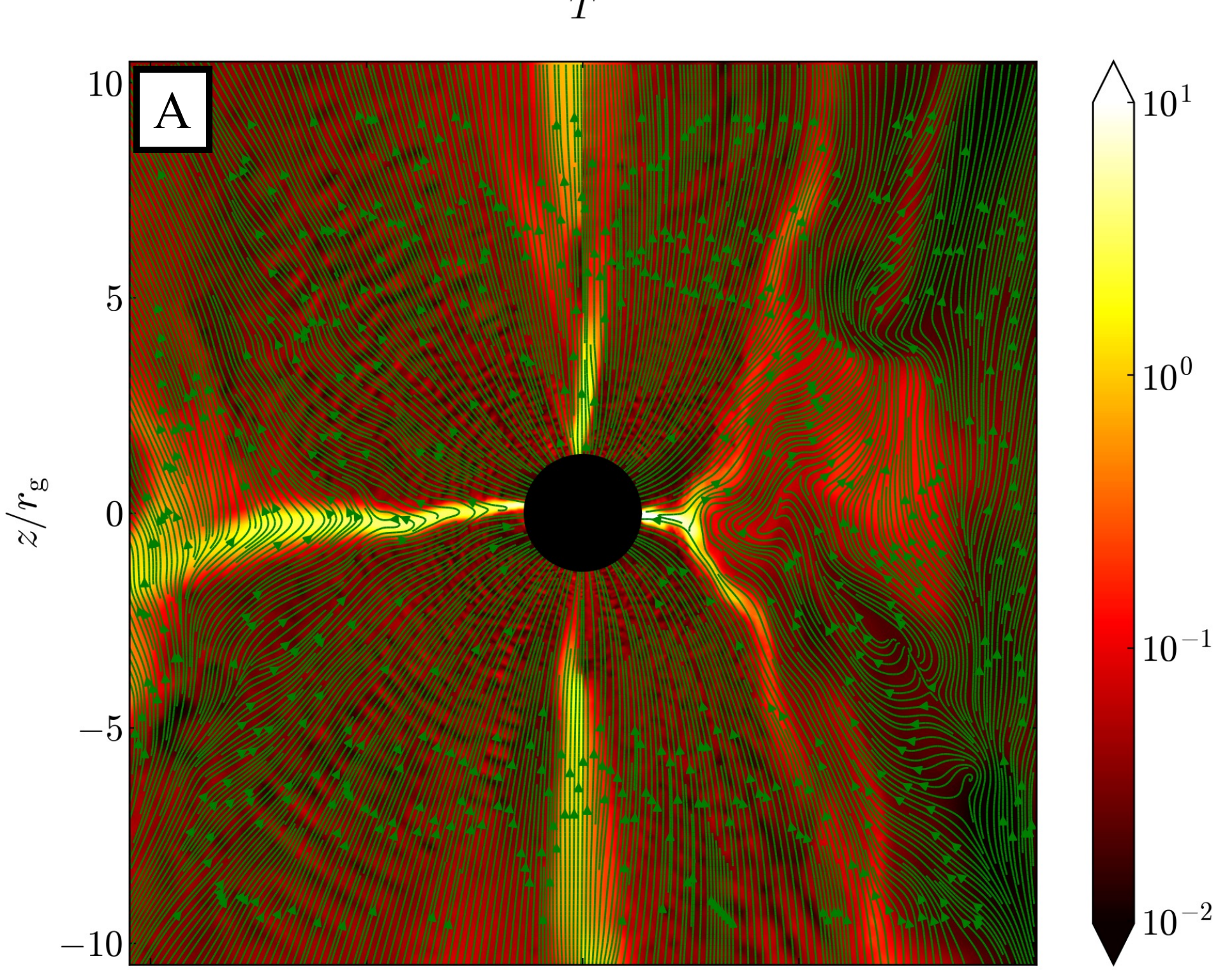}
    \includegraphics[width=0.318\textwidth,trim= 2.8cm 2.3cm 13.4cm 1.3cm, clip=true]{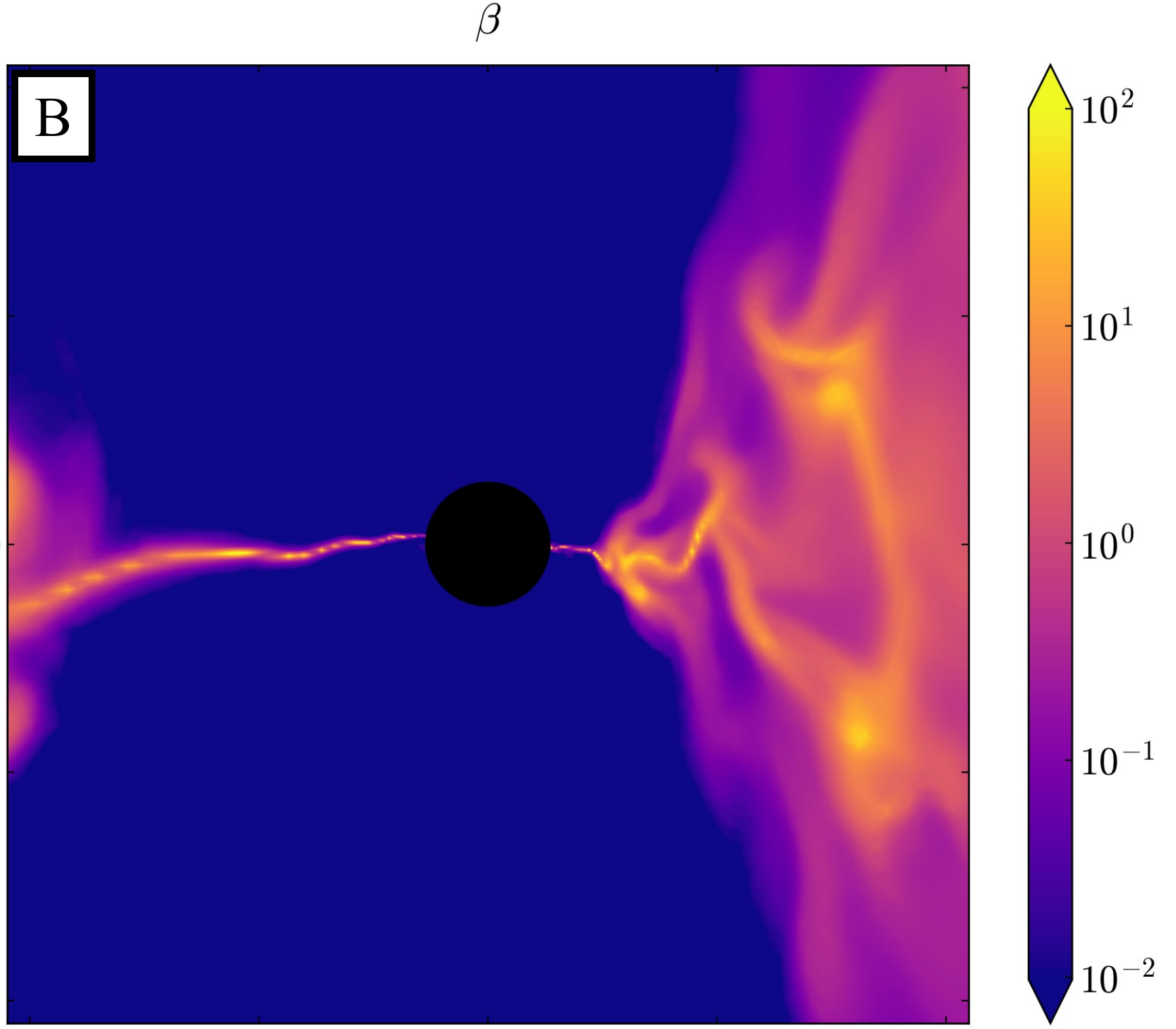}
    \includegraphics[width=0.318\textwidth,trim= 2.8cm 2.3cm 13.4cm 1.3cm, clip=true]{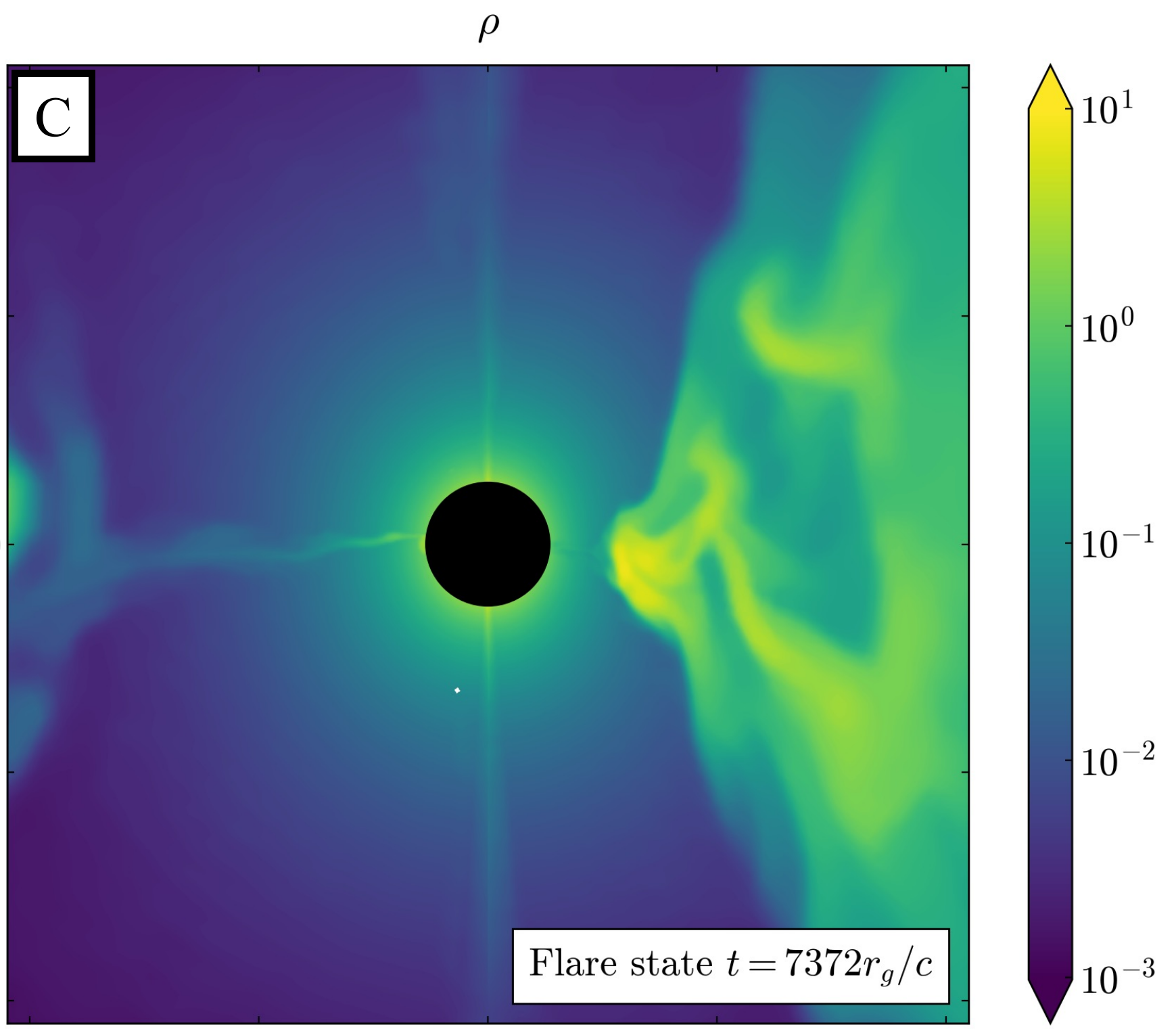}
    
    \includegraphics[width=0.353\textwidth,trim= 0.85cm 0.785cm 13.4cm 1.95cm, clip=true]{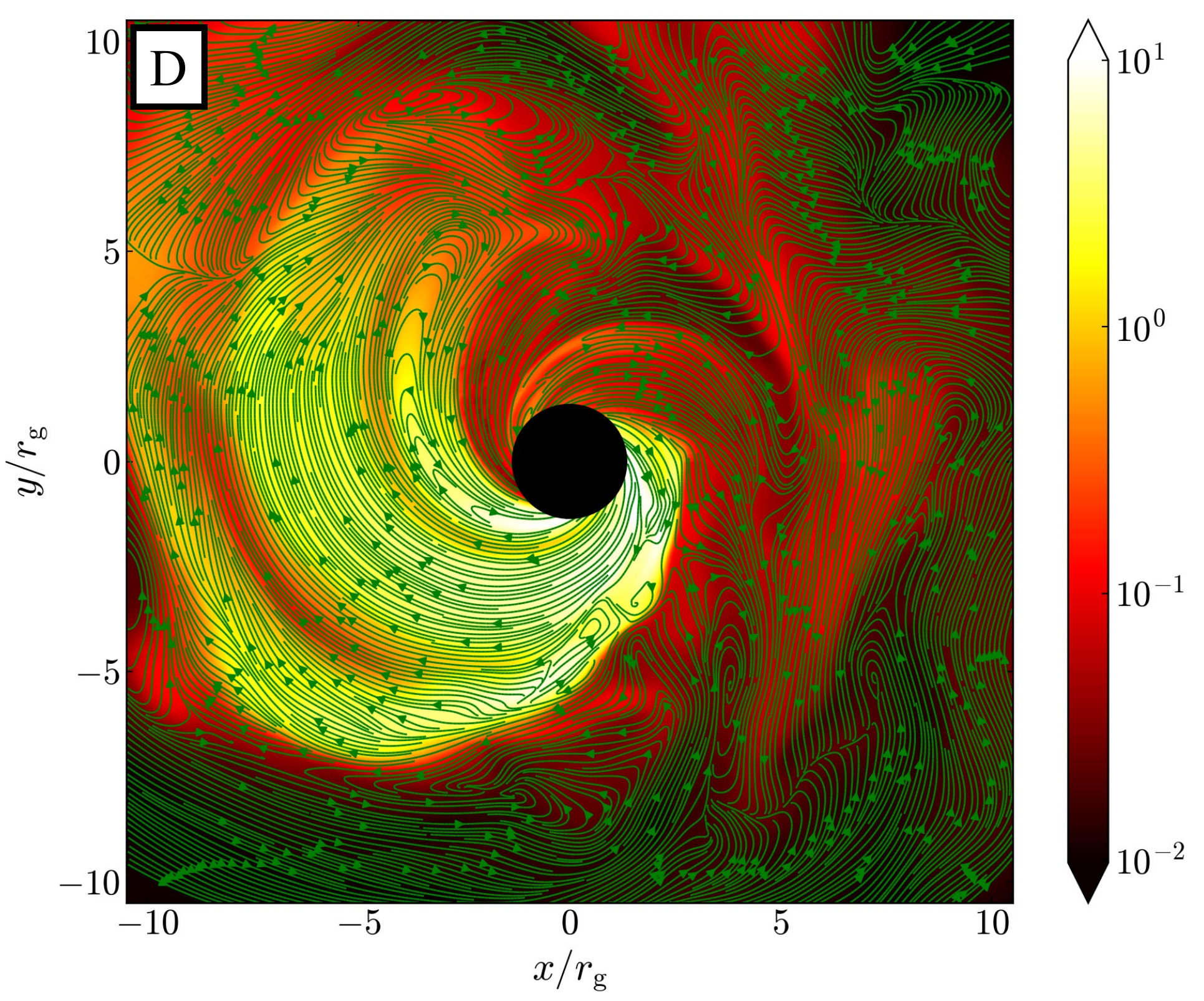}
    \includegraphics[width=0.318\textwidth,trim= 2.8cm 0.785cm 13.4cm 1.95cm, clip=true]{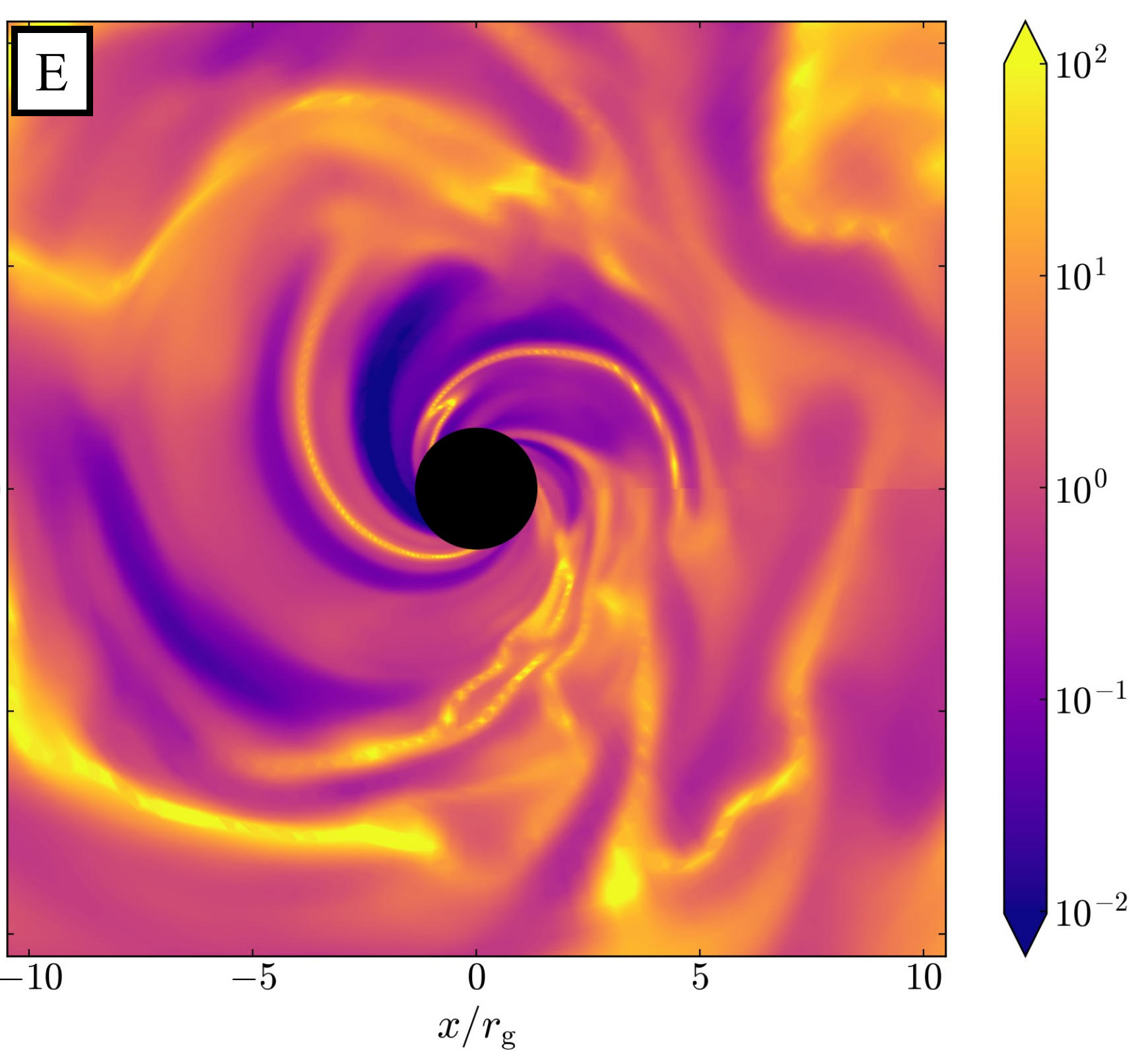}
    \includegraphics[width=0.318\textwidth,trim= 2.8cm 0.785cm 13.4cm 1.95cm, clip=true]{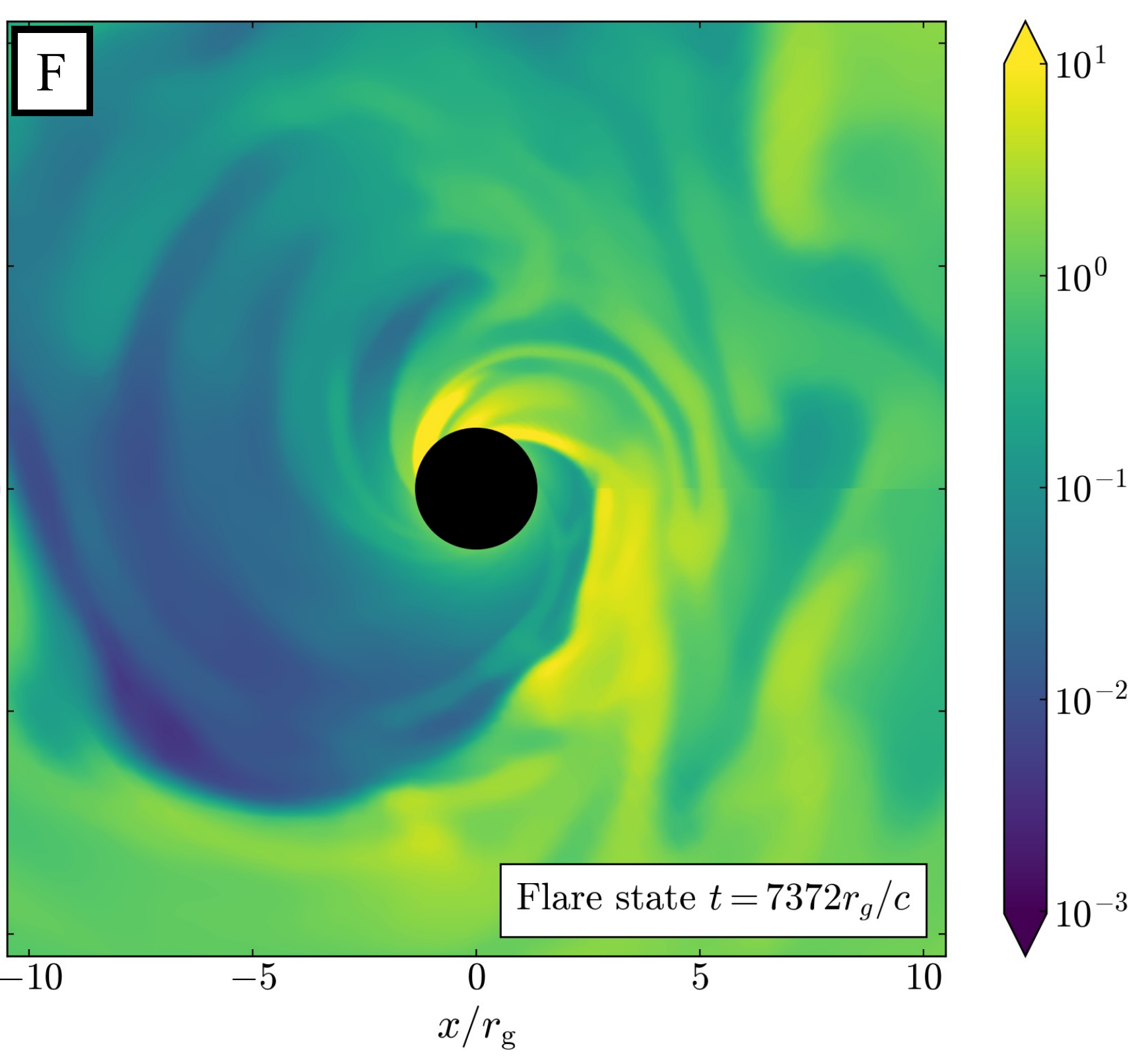} 
    
    \includegraphics[width=0.353\textwidth,trim= 0.85cm 0.785cm 13.4cm 2.15cm, clip=true]{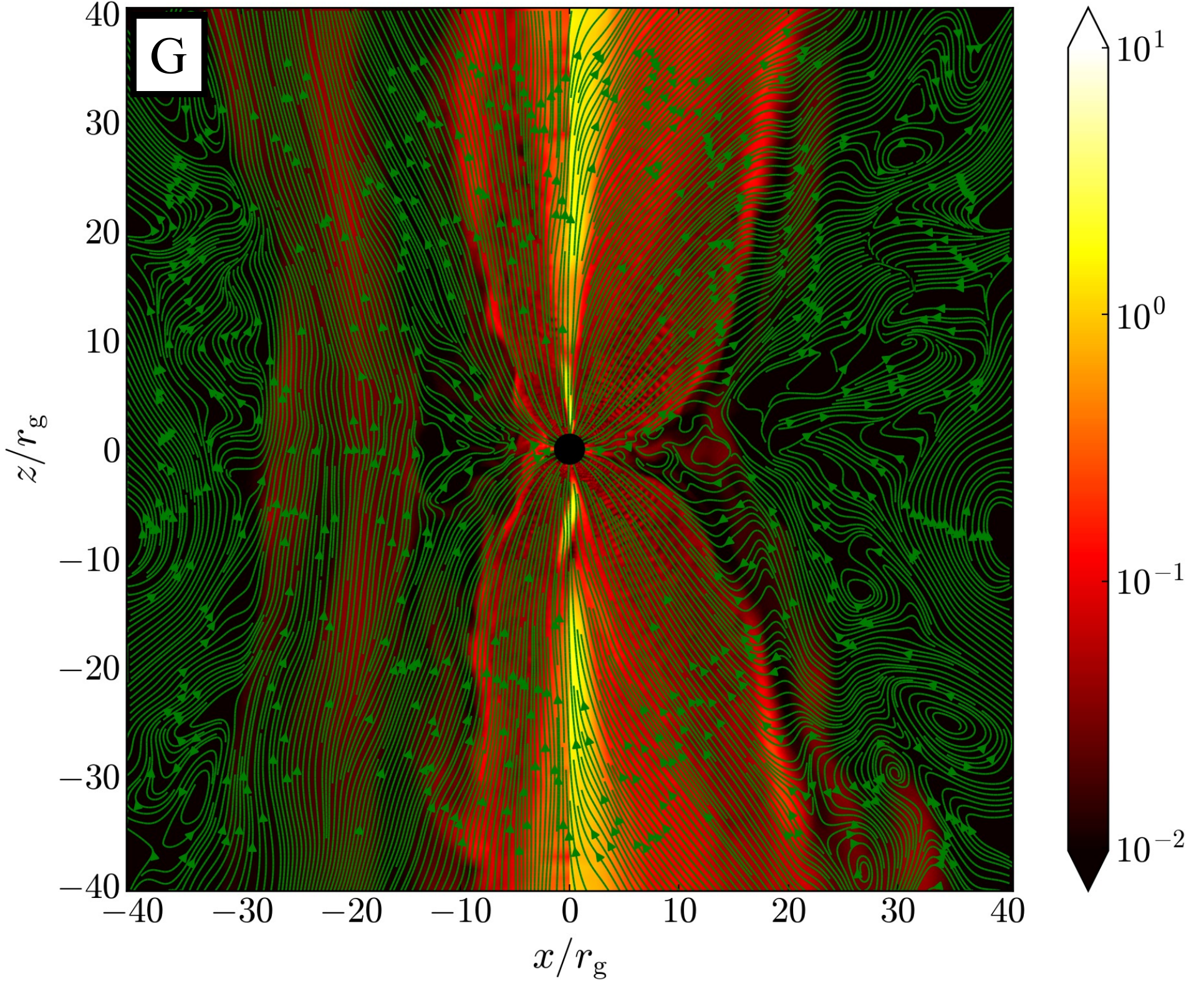}
    \includegraphics[width=0.318\textwidth,trim=  2.8cm 0.785cm 13.4cm 2.15cm, clip=true]{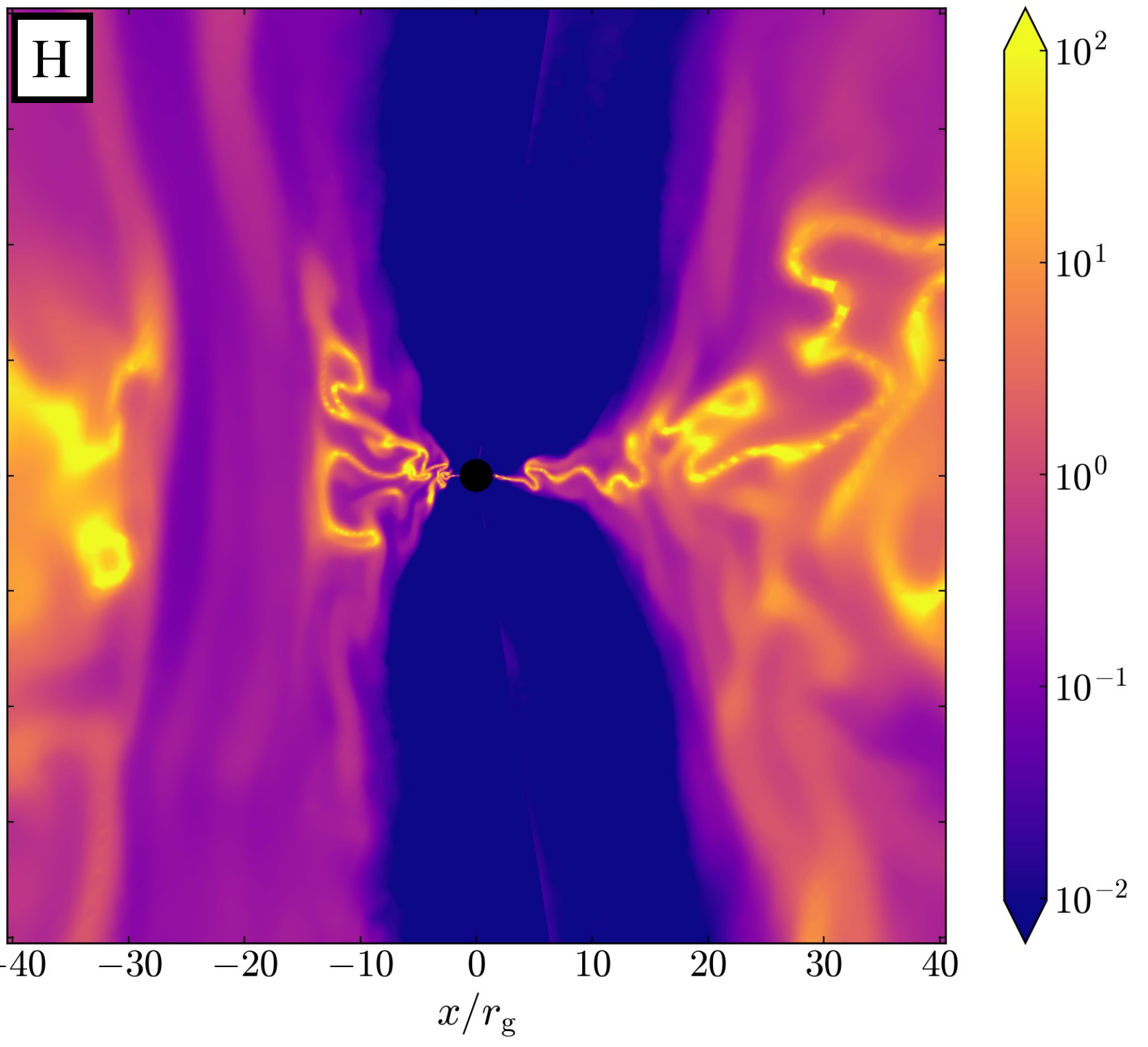}
    \includegraphics[width=0.318\textwidth,trim= 2.8cm 0.785cm 13.4cm 2.15cm, clip=true]{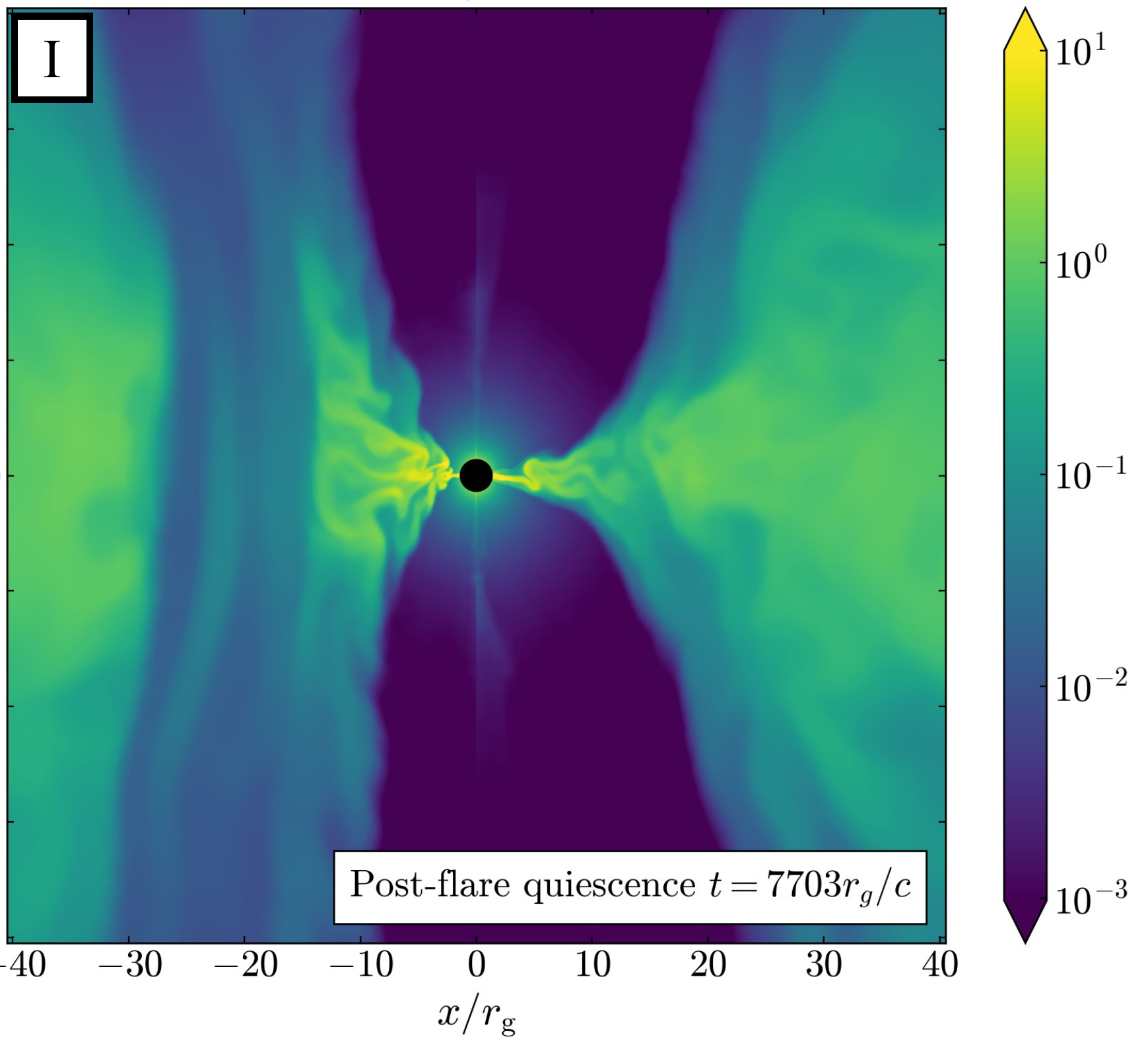}
    
    \includegraphics[width=0.353\textwidth,trim=  0.85cm 0.785cm 13.4cm 1.95cm clip=true]{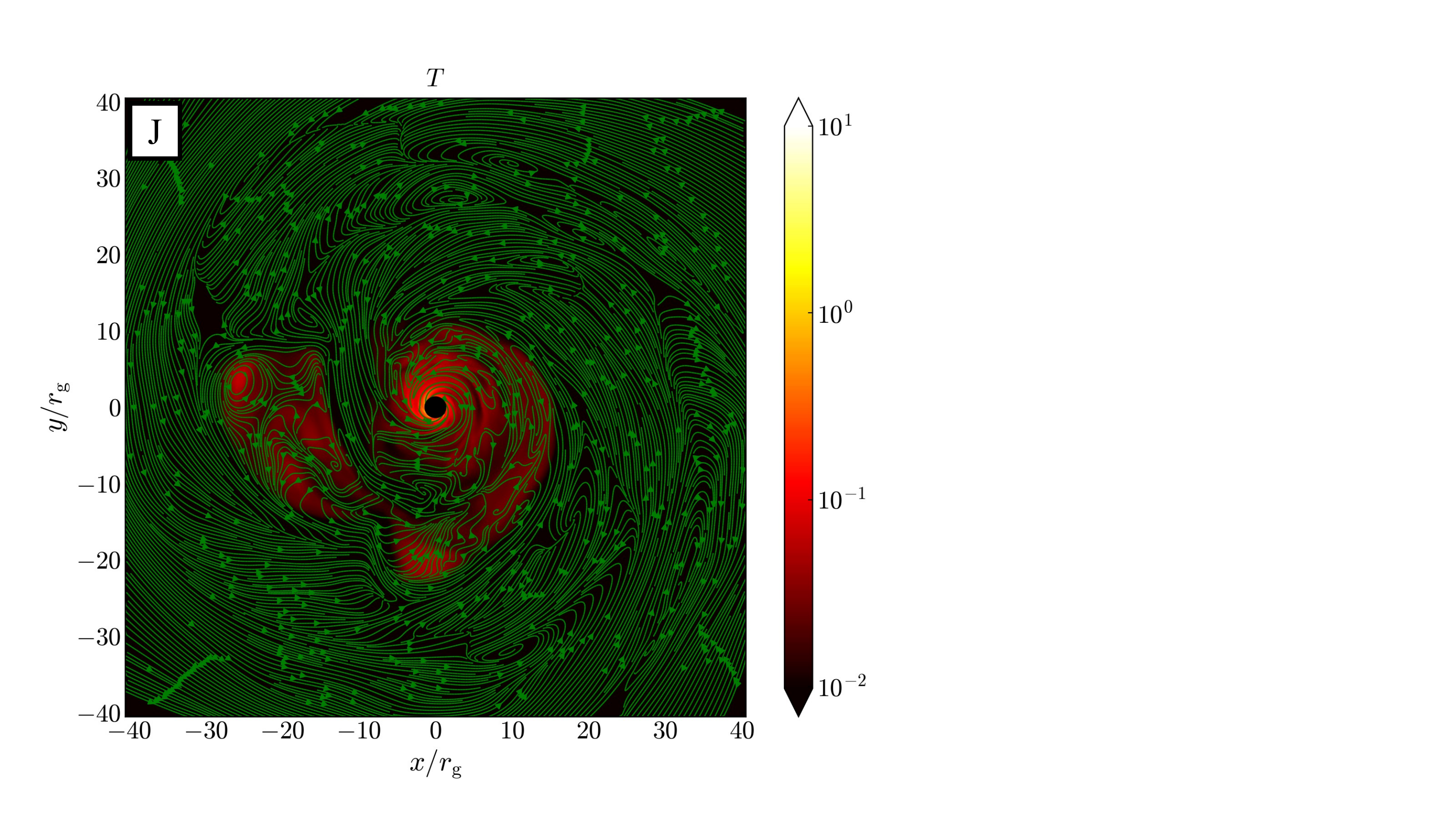}
    \includegraphics[width=0.318\textwidth,trim=  2.8cm 0.785cm 13.4cm 1.95cm, clip=true]{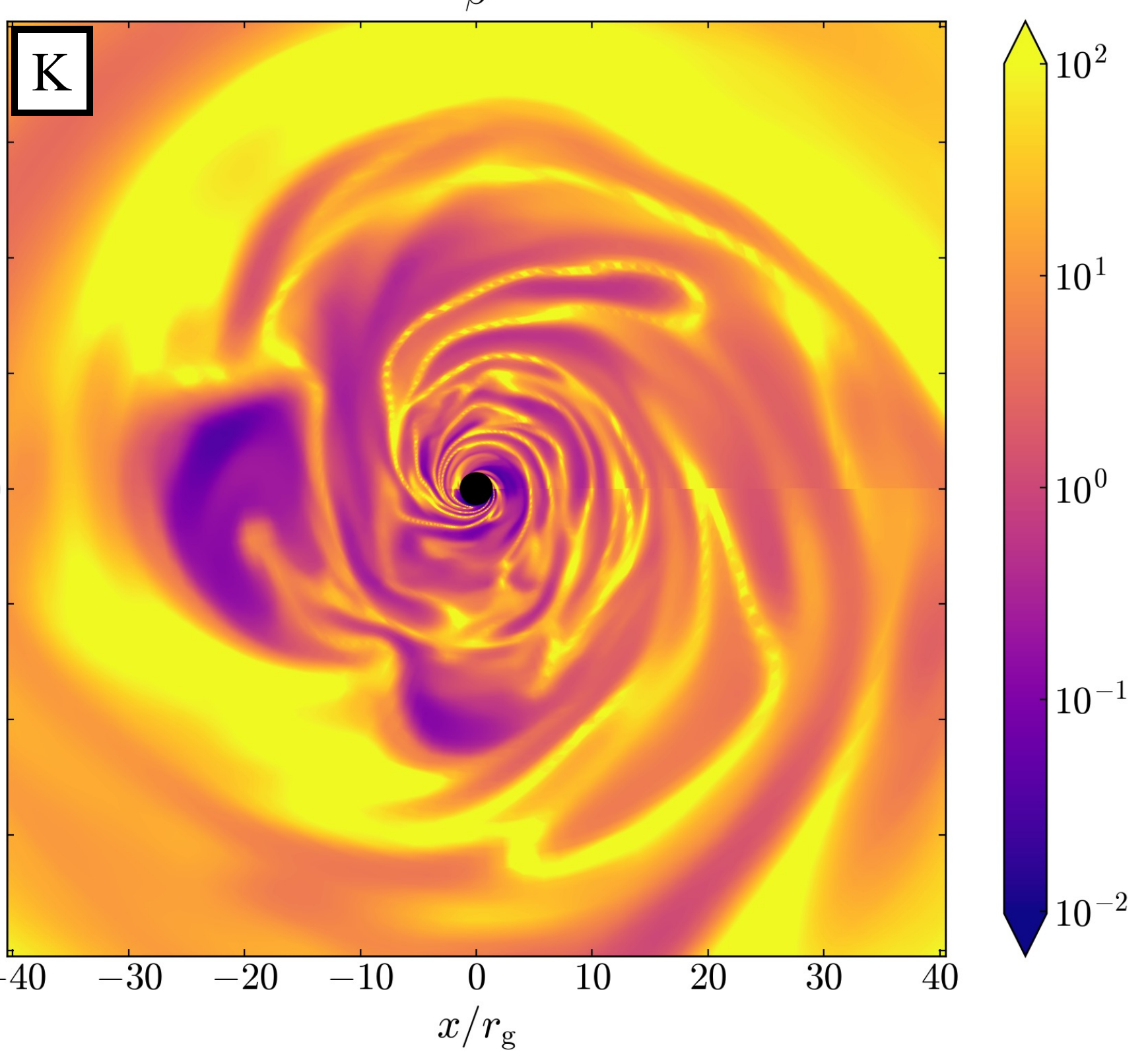}
    \includegraphics[width=0.318\textwidth,trim= 2.8cm 0.785cm 13.4cm 1.95cm, clip=true]{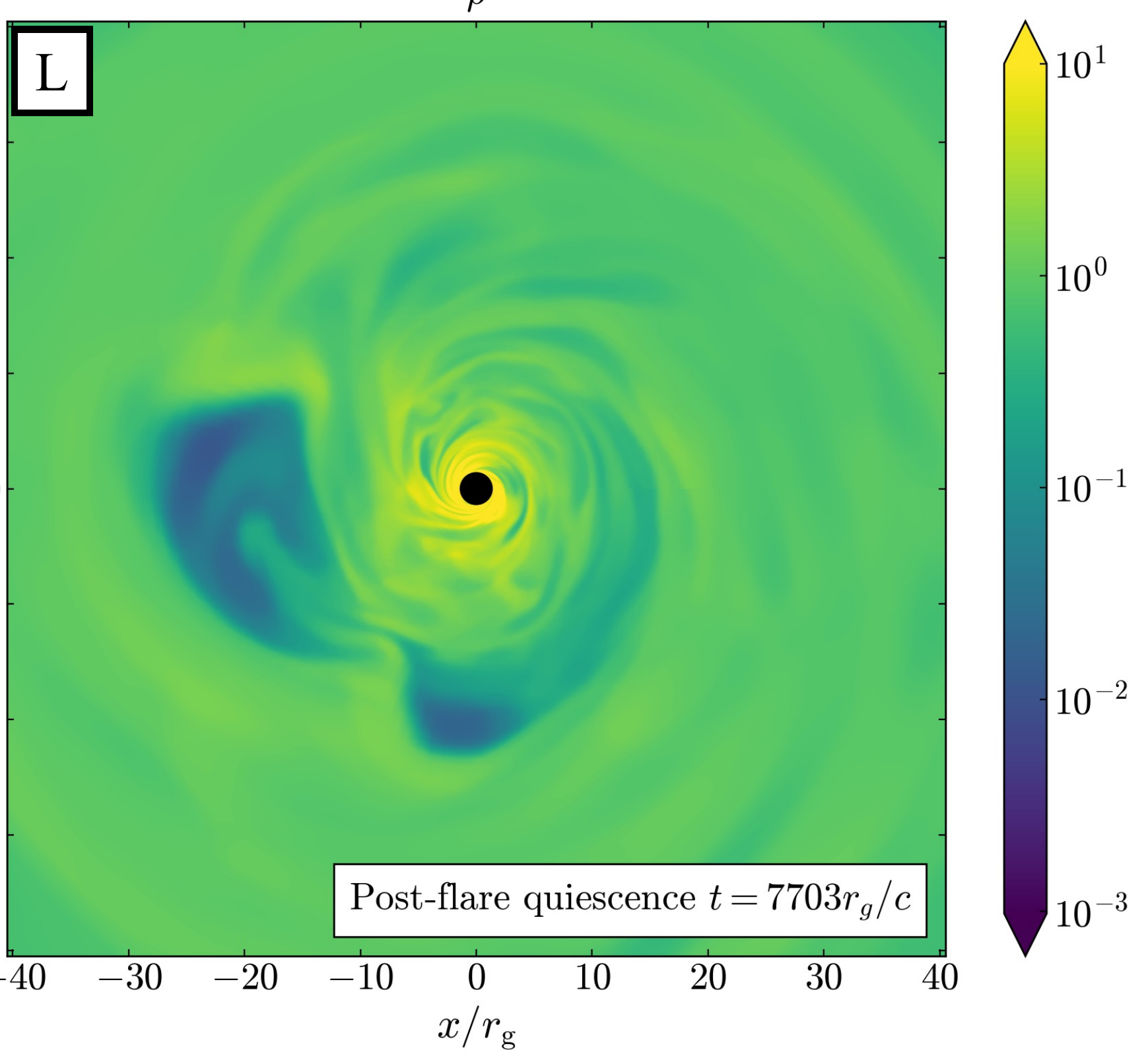} 
    \caption{Dimensionless temperature $T=p/\rho$ (left), plasma-$\beta$ (middle), and density $\rho$ (right) in the meridional plane (first and third rows) and equatorial plane (second and fourth rows) during a {magnetic flux eruption in the inner $10 r_{\rm g}$ (first and second row) and during quiescence} in the inner $40 r_{\rm g}$ (third and fourth row).} 
    \label{fig:panelXZlowres}
\end{figure*}

Figure \ref{fig:panelXZlowres} shows a large flare that forms during a magnetic flux decay at an enhanced rate $\propto e^{-t/350}$, and a significant drop in mass accretion rate, at low resolution. During the large flare a very broad current sheet forms, indicated by the high plasma-$\beta$ in panel (B). The magnetic field lines (green lines, left panels) diffuse through the equatorial sheet without reconnecting.
Due to the applied floors and the large numerical diffusion at low resolution the temperature in the jet region is particularly unreliable, showing a wavy pattern that is absent at high and extreme resolution. The area that is heated due to reconnection is broader due to numerical diffusion, and due to the large cells that increase in volume with radius. This results in a large $T>1$ area that lies fully in the equatorial plane (D) because of the thickness of the sheet in the $z$-direction. The heated area does not correspond to a clear floored region, which is visible in the middle and right plots of $\beta$ and density $\rho$. After the flare (bottom two rows), the inner $10 r_{\rm g}$ is in the quiescent state and an ejected low density flux tube with vertical field orbits at $x \approx -25 r_{\rm g}$, $y\approx 0 r_{\rm g}$

During mini-flares, a short and broad current sheet forms close to the horizon, but there is no clear expulsion of the accretion disk. Therefore it is hard to distinguish the flare state from the quiescent state. The mini-flare shows a clear magnetic flux decay at a rate $\propto e^{-t/500}$ in accordance with a reconnection rate $\sim 0.01c$. There is no clear drop in the mass accretion rate.

\end{document}